\documentclass[12pt]{article}
\usepackage{amssymb,amsmath}
\usepackage{color,graphicx}
\usepackage{cite}
\usepackage{yfonts}
\AtBeginDocument{
  \addtocontents{toc}{\small}
  \addtocontents{lof}{\small}
}

\newcommand\fverb{\setbox\pippobox=\hbox\bgroup\verb}
\newcommand\fverbdo{\egroup\medskip\noindent%
                              \fbox{\unhbox\pippobox}\ }
\newcommand\fverbit{\egroup\item[\fbox{\unhbox\pippobox}]}
\newbox\pippobox

\newcommand{\be} {\begin{equation}}
\newcommand{\ee} {\end{equation}}
\newcommand{\beq} {\begin{equation}}
\newcommand{\eeq} {\end{equation}}
\newcommand{\bea} {\begin{eqnarray}}
\newcommand{\eea} {\end{eqnarray}}
\newcommand{\bear}{\begin{eqnarray}}
\newcommand{\eear}{\end{eqnarray}}

\newcommand{\rc}{\nonumber\\}
\def\tr{{\rm Tr}}
\def\ie{{\em i.e.}}

\begin{document}
 
\begin{flushright}
HIP-2016-04/TH
\end{flushright}

\begin{center}

\centerline{\Large {\bf Collective excitations of massive flavor branes}}

\vspace{8mm}

\renewcommand\thefootnote{\mbox{$\fnsymbol{footnote}$}}
Georgios Itsios${}^{1,4,5}$\footnote{itsiosgeorgios@uniovi.es},
Niko Jokela${}^{2,3}$\footnote{niko.jokela@helsinki.fi},
and Alfonso V. Ramallo${}^{4,5}$\footnote{alfonso@fpaxp1.usc.es}

\vspace{2mm}
${}^1${\small \sl Department of Physics, University of Oviedo}  \\
{\small \sl Avda. Calvo Sotelo 18, 33007 Oviedo, Spain }

\vspace{2mm}
${}^2${\small \sl Department of Physics} and ${}^3${\small \sl Helsinki Institute of Physics} \\
{\small \sl P.O.Box 64} \\
{\small \sl FIN-00014 University of Helsinki, Finland} 

\vspace{2mm}
${}^4${\small \sl Departamento de  F\'\i sica de Part\'\i  culas} \\
{\small \sl Universidade de Santiago de Compostela} \\
{\small \sl and} \\
${}^5${\small \sl Instituto Galego de F\'\i sica de Altas Enerx\'\i as (IGFAE)} \\
{\small \sl E-15782 Santiago de Compostela, Spain}

\end{center}

\vspace{6mm}
\numberwithin{equation}{section}
\setcounter{footnote}{0}
\renewcommand\thefootnote{\mbox{\arabic{footnote}}}

\begin{abstract}
\noindent
We study the intersections of two sets of D-branes  of different dimensionalities. This configuration is
dual to a supersymmetric gauge theory with flavor hypermultiplets in the fundamental representation of the gauge group which live on the defect of the unflavored theory determined by the  directions common to the two types of branes.  One set of branes is dual to the color degrees of freedom, while the other set adds flavor to the system. We work in the quenched approximation, \ie, where the flavor branes are considered as probes, and focus specifically on the case in which the quarks are massive. We study the thermodynamics and the speeds of first and zero sound at zero temperature and non-vanishing chemical potential. We show that the system undergoes a quantum phase transition when the chemical potential approaches its minimal value and we obtain the corresponding non-relativistic critical exponents that characterize its critical behavior. In the case of $(2+1)$-dimensional intersections, we further study alternative quantization and the zero sound of the resulting anyonic fluid. We finally extend these results to non-zero temperature and magnetic field and compute  the diffusion constant in the hydrodynamic regime. The numerical results we find match the predictions by the Einstein relation.

\end{abstract}

\newpage
\tableofcontents
\addtocontents{toc}{\vspace{-2\baselineskip}}

\renewcommand{\theequation}{{\rm\thesection.\arabic{equation}}}


\section{Introduction}

There is hope that the gauge/gravity holographic duality  could serve to characterize new types of compressible states of matter, \ie, states with non-zero charge density which vary continuously with the chemical potential. Indeed, holography provides gravitational descriptions of strongly interacting systems without long-lived quasiparticles, situations which cannot be accommodated within the standard  Landau's Fermi liquid theory. Although the field theories with known holographic dual are very different from those found so far in Nature, there are good reasons to believe that these studies could reveal generic universal features of strongly interacting quantum systems \cite{AdS_CFT_reviews}. 

In this paper we approach this problem in a top-down model of intersecting branes of different dimensionalities. We will consider a stack of $N_c$ color D$p$-branes which intersect $N_f$ flavor D$q$-branes ($q\ge p)$ along $n$ common directions. This configuration, which we will denote by $(n\, |\, p\perp q)$, is dual to a $(p+1)$-dimensional $SU(N_c)$ gauge theory with $N_f$ fundamental hypermultiplets (quarks) living on a $(n+1)$-dimensional defect \cite{Karch:2002sh}. In the context of holography, we will work in the large $N_c$ 't Hooft limit with $N_f\ll N_c$. In this limit the quarks are quenched and the D$q$-branes can be treated as probes, whose action is the Dirac-Born-Infeld (DBI) action,  in the gravitational background created by the D$p$-branes. The embedding of the flavor branes is parameterized by a function which measures the distance between the two types of branes. The field theory dual of this distance is the mass of the hypermultiplet.  Moreover, in order to engineer a system with non-zero baryonic charge density, we must switch on a suitable gauge field on the worldvolume of the flavor brane \cite{Kobayashi:2006sb}. We will also study the influence of a magnetic field directed along two of the spatial directions of the worldvolume. 
 
In \cite{Jokela:2015aha} we studied the collective excitations of generic brane intersections corresponding to massless quarks and we uncovered a certain universal structure. The purpose of this article is to extend the results of \cite{Jokela:2015aha} to the case in which the quarks have a non-zero mass.  We will study first the system at zero temperature and non-zero chemical potential. This is the so-called collisionless quantum regime, in which the dynamics is dominated by the zero sound mode. This mode is a collective excitation, first found in the holographic context in  \cite{Karch:2008fa,Karch:2009zz}. These results were generalized to non-zero temperature in \cite{Bergman:2011rf, Davison:2011ek} and to  non-vanishing magnetic field in \cite{Jokela:2012vn,Brattan:2012nb} (see \cite{Kulaxizi:2008kv,Kulaxizi:2008jx,Kim:2008bv,Hung:2009qk,Edalati:2010pn,Lee:2010uy,HoyosBadajoz:2010kd,Lee:2010ez,Ammon:2011hz,Goykhman:2012vy,Gorsky:2012gi,Jokela:2012se,Davison:2013bxa,Dey:2013vja,Edalati:2013tma,Davison:2013uha,DiNunno:2014bxa,Itsios:2015kja} for studies on different aspects of the holographic zero sound). In \cite{Jokela:2015aha} we developed a  general formalism which included all possible intersections  $(n\, |\, p\perp q)$  and, in particular, we found an index $\lambda$ (depending on $n$, $p$, and $q$) which determines the speed of zero sound for massless quarks. This is intimately related with the fact that $\lambda$ determines the scaling dimension of the charge density or to put it slightly differently, $\lambda$ acts as the polytropic index in the equation of state for the holographic matter.

In the case of massive quarks the embedding of the D$q$-brane is non-trivial and must be determined in order to extract the different physical properties.  When the charge density is non-vanishing, the brane reaches the horizon of the geometry, \ie, we have a black hole embedding. This embedding depends on a function which parameterizes the shape of the flavor brane in the background geometry and, in general, must be found by numerical integration of the equations of motion of the probe.  However, in the case of intersections  $(n\, |\, p\perp q)$ which preserve some amount of supersymmetry at zero temperature $T$ and chemical potential $\mu$ some remarkable simplification occurs.  Indeed, as shown in \cite{Karch:2007br}, in these intersections one can choose a system of coordinates such that the embedding function is a cyclic variable of the DBI Lagrangian when $T=0$ and $\mu\not=0$. As a consequence, the embedding function and the physical properties of the configuration, can be found analytically. In particular, one can study the zero temperature thermodynamics of these systems  and find the speed of first sound. This was done in refs. \cite{Kulaxizi:2008jx,Kim:2008bv} for the D3-D$q$ intersections $(3\, |\, 3\perp 7)$, $(2\, |\, 3\perp 5)$, and $(1\, |\, 3\perp 3)$. Moreover, by studying the quasinormal  fluctuation modes of the probe, one can also compute analytically the speed of zero sound which, non-trivially, equals that of the first sound \cite{Kulaxizi:2008kv, Davison:2011ek,Ammon:2012je}. 

In this paper we generalize these results for any $(n\, |\, p\perp q)$ intersection with \#ND=4, \ie, when $n=(p+q-4)/2$. These cases correspond to those brane intersections which are supersymmetric in flat space at low energies as the gravitational and Ramond-Ramond forces cancel out. Here the index $\lambda$ can only take three different values $\lambda=2,4,6$, corresponding to codimension 2 (D$p$-D$p$), codimension 1 (D$p$-D$(p+2)$), and codimension 0 (D$p$-D$(p+4)$) intersections, respectively.
As in the conformal D3-background, the speeds of first and zero sounds coincide. Moreover, we find the same kind of universality as in the massless case: the speed is the same for those intersections which have the same $\lambda$ index, or codimension. However, in the massive case the speed of sound depends continuously on the chemical potential, \ie, on the charge density, and vanishes when the chemical potential reaches its minimal value, which corresponds to a vanishing charge density $d$. Actually, as argued in \cite{Ammon:2012je} for the D3-D7 and D3-D5 intersections, there is a quantum phase transition as $d\to 0$ which exhibits a non-relativistic scaling behavior with hyperscaling violation. At the transition point the black hole embeddings with $d\not=0$ degenerate into a Minkowski embedding with zero charge density. Here we will find the critical exponents for the general \#ND=4 intersections, generalizing the results of \cite{Ammon:2012je}. 

When the number $n$ of common dimensions of the color and flavor branes is equal to two, the matter hypermultiplets live on a (2+1)-dimensional theory. In this case one can perform an alternative quantization of the quasinormal modes, which consists in imposing mixed Dirichlet-Neumann boundary conditions at the UV. As shown in \cite{Jokela:2013hta}, this alternative quantization amounts to transforming the charged excitations into particles of fractional statistics, \ie, anyons (see also \cite{Jokela:2014wsa,Brattan:2013wya,Brattan:2014moa,Itsios:2015kja} for the analysis of different aspects of the holographic anyonic systems). In \cite{Jokela:2015aha} we studied the zero sound mode as a function of the constant that measures the degree of mixing the UV  boundary conditions.  We found that the anyonic zero sound is generically gapped and that this gap can be fine-tuned to zero if a suitable magnetic field is switched on. This choice corresponds to the case, where the anyons experience no effective magnetic field. In this paper we generalize these results to the case in which the quarks are massive. 

In this article we also study the hydrodynamic regime that is reached when the temperature is high enough. The dominant collective mode in this regime is a diffusion mode, which has a purely imaginary dispersion relation characterized by a diffusion constant $D$.  When the temperature is non-zero the embedding function is no more a cyclic coordinate of the DBI action and cannot therefore be found analytically. Thus, we study this $T\not= 0$ case by using numerical methods, after performing a convenient change of variables.  Moreover, this numerical analysis allow us to check the analytic results found at zero temperature, by taking the $T\to 0$ limit. We also study numerically the system in the presence of a magnetic field $B$. We compare the results for the diffusion constant obtained from the fluctuation analysis at $T\not=0$ with the ones predicted by the Einstein relation, which gives $D$ in terms of the DC conductivity $\sigma$ and the charge susceptibility $\chi$. Both $\sigma$ and $\chi$ can be obtained from the embedding function. We find a very good agreement between the numerical results for $D$ and the value given by the Einstein relation. 

The rest of this paper is organized as follows. In section \ref{zero_temperature_embeddings} we formulate our top-down holographic model, solve the equations of motion of the probe at $T=0$ and $\mu\not=0$, and study the thermodynamics at zero temperature. In particular, in this section we find the speed of first sound and compute the charge susceptibility at $T=0$. In section \ref{zeroT_fluct} we write the equations of motion for the fluctuations of the probe at zero temperature. In section \ref{zero_sound_section} we analyze the zero sound and find analytically the dispersion relation of this collective mode.  Section \ref{critical} is devoted to the study of the scaling behavior near the quantum critical point. In section \ref{alternative_quantization} we study the zero sound mode in an anyonic fluid. Section \ref{finiteT} contains our results at non-zero temperature and magnetic field.  We summarize  our results and discuss some possible future research directions in section \ref{conclusions}. 

We complement and give further details of our analysis in several appendices.  Appendix \ref{appendixA} contains a detailed derivation of the Lagrangian of the fluctuations at zero temperature which is used in section \ref{zeroT_fluct}. In appendix \ref{appendixB} we work out the equations of motion of the fluctuations at $T\not=0$. In appendix \ref{appendixC} we find the correlator of two transverse currents and extract  the DC conductivity in the absence of magnetic field. Finally, in appendix \ref{appendixD} we provide an alternative derivation of the conductivity, valid also when $B\not=0$.


\section{Massive D$p$-D$q$ systems with charge}
\label{zero_temperature_embeddings}

Let us begin our analysis by introducing our setup and studying its properties at zero temperature and magnetic field.  We will consider a generic D$p$-brane metric at zero temperature of the type:
\beq
ds_{10}^2\,=\,g_{tt}(r)\,dt^2\,+\,g_{xx}(r)\,
\big[(dx^1)^2\,+\,\cdots +(dx^p)^2\big]\,+\,g_{rr}(r)\, d\vec y\cdot  d\vec y
\,\,,
\eeq
where $\vec y\,=\,(y^1, \ldots, y^{9-p})$ are the coordinates transverse to the D$p$-brane and the functions $g_{tt}$, $g_{xx}$, and 
$g_{rr}$ depend on the transverse radial direction $r=\sqrt{\vec y\cdot \vec y}$. We now embed $N_f$ D$q$-brane probes, with $N_f\ll N_c$, extended along the directions 
\beq
(t, x^1,\ldots, x^n,y^1,\ldots, y^{q-n})\,\,.
\eeq
We will refer to this configuration as a $(n\, |\, p\perp q)$ intersection ($n$ is the number of common spatial directions of the D$p$ and D$q$).  This intersection is represented by the array:
\beq
\begin{array}{cccccccccccccl}
 &x^1&\cdots&x^n&x^{n+1}& \cdots& x^p&y^1 &\cdots&y^{q-n}&y^{q-n+1}&\cdots &y^{9-p}& \nonumber \\
Dp: & \times &\cdots &\times&\hspace{-0.2truein}\times&\cdots &\times &-&\cdots &-&\hspace{-0.2truein}-&\cdots  &\hspace{-0.2truein}-    \nonumber \\
Dq: &\times&\cdots&\times&\hspace{-0.2truein}-&\cdots&-&\times&\cdots&\times&\hspace{-0.2truein}-&\cdots&\hspace{-0.2truein}-&
\end{array}
\label{DpDqintersection}
\eeq
We shall denote by $\vec z$ the coordinates $\vec y$ transverse to the D$q$-brane:
\beq
\vec z\,=\,(z^1,\ldots,z^{9+n-p-q})\,\,,
\eeq
with $z^m\,=\,y^{q-n+m}$ for $m=1,\ldots, 9+n-p-q$. Moreover, we define $\rho$ as the radial coordinate for the subspace spanned by $(y^1,\ldots, y^{q-n})$:
\beq
\rho^2\,=\,(y^1)^2+\cdots+(y^{q-n})^2\,\,.
\eeq

Let us make a short comment on the global symmetries. The original D$p$-background has a rotational symmetry in the $y^i$ directions, this corresponds to the $SO(9-p)$ R-symmetry. When we add $N_f$ coincident probe D$q$-branes we introduce $U(N_f)$ flavor symmetry. The D$p$-D$q$-intersection $(n\, |\, p\perp q)$ breaks the original R-symmetry, which can be easily read off from the isometries. We end up with the global symmetry $SO(n,1)\times U(N_f)\times SO(p-n)_p\times SO(q-n)_q\times SO(9+n-p-q)$. The last group will be further broken when we consider massive D$q$-brane embeddings.

Since,
\beq
d\vec y^{\,\,2}\,=\,d\rho^2\,+\,\rho^2\,d\Omega^2_{q-n-1}\,+\,d\vec z^{\,\,2},
\eeq
the background metric in these coordinates can be written as:
\bear
&&ds_{10}^2\,=\,g_{tt}(r)\,dt^2\,+\,g_{xx}(r)\,\Big[
(dx^1)^2\,+\,\cdots +(dx^n)^2\,+\,
(dx^{n+1})^2\,+\,\cdots +(dx^p)^2\,\Big]\nonumber \\
&&\qquad\qquad\qquad\qquad\qquad\qquad
+\,g_{rr}(r)\,\Big[\,d\rho^2\,+\,\rho^2\,d\Omega^2_{q-n-1}\,+\,d\vec z^{\,\,2}\,\Big]\,\,.
\eear
Let us consider a stack of D$q$-branes with a non-trivial profile in the transverse space. We will choose our transverse coordinates in such a way that this profile can be parameterized as $\vec z=(z^1(\rho), 0,\ldots, 0)$. In what follows we just write $z(\rho)$ instead of 
 $z^1(\rho)$ and we will denote by $r=r(\rho)$ the function:
 \beq
 r(\rho)\,=\,\sqrt{\rho^2+z(\rho)^2}\,\,.
 \eeq
The induced metric on the D$q$-brane worldvolume at zero temperature is:
\beq
ds_{q+1}^2=g_{tt}(\rho)\,dt^2+g_{xx}(\rho)\,\big[
(dx^1)^2+\cdots +(dx^n)^2\,\big]
+g_{rr}(\rho)\big[(1+z'^{\,2})\,d\rho^2
+\rho^2\,d\Omega^2_{q-n-1}\,\big]\,\,,
\eeq
with $z'=dz/d\rho$. Let us compute the DBI action of the D$q$-brane in the case in which there is a worldvolume gauge field $F$  with components $\rho t$. Thus, we will take $F$ to be given by:
\beq
F\,=\,A_t'\, d\rho\wedge dt  \,\,.
\label{F_massive}
\eeq
where $A_t'=\partial_{\rho} A_t$ and we have chosen a gauge for $A$ such that $A_{\rho}=0$.  This means that we aim to study holographic matter at non-zero baryon charge density by introducing a chemical potential for the diagonal $U(1)\subset U(N_f)$.
The DBI action becomes:
\beq
S_{Dq}\,=-\,N_f T_{Dq}\,\int d^{q+1}\xi \,e^{-\phi}\,
\sqrt{-\det(g+2\pi\alpha'F)}\,=\,\int dt\,d^n x\,d\rho\,{\cal L}_{DBI}\,\,,
\eeq
with the Lagrangian density ${\cal L}_{DBI}$ given by:
\beq
{\cal L}_{DBI}\,=\,-{\cal N}\,e^{-\phi}\,\rho^{q-n-1}\,
g_{xx}^{\frac{n}{2}}\,g_{rr}^{{q-n-1\over 2}}\,
\sqrt{g_{rr}|g_{tt}|\,(1+z'^{\,2})\,-(2\pi\alpha')^2\,A_t'^2}\,\,,
\eeq
where  ${\cal N}$ is the normalization factor
\beq
{\cal N}\,=\, N_f T_{Dq}\,{\rm Vol}({\mathbb S}^{q-n-1})\,\,,
\label{calN}
\eeq
and where the tension of the D$q$-brane and the volume of the unit sphere are
\be
 T_{Dq} = \frac{1}{(2\pi)^q \sqrt{\alpha'}^{q+1} g_s} \ , \ \ {\rm Vol}({\mathbb S}^{q-n-1})=\frac{2\pi^{\frac{q-n}{2}}}{\Gamma\left(\frac{q-n}{2}\right)} \ .
\ee
For a D$p$-brane background at zero temperature, the metric and the dilaton are given by:
\beq
-g_{tt}\,=\,g_{xx}\,=\,\Big({r\over R}\Big)^{{7-p\over 2}}\,\,,
\qquad
g_{rr}\,=\,\Big({R\over r}\Big)^{{7-p\over 2}}\,\,,
\qquad
e^{-2\phi}\,=\,\Big({R\over r}\Big)^{{(7-p)(p-3)\over 2}}\,\,.
\label{Dp-metric_dilaton}
\eeq
This background  satisfies $g_{rr}|g_{tt}|=1$ and the Lagrangian density ${\cal L}_{DBI}$ can be written as:
\beq
{\cal L}_{DBI}\,=\,-{\cal N}\,\rho^{q-n-1}\,
\Big({r\over R}\Big)^{{(2n-p-q+4)(7-p)\over 4}}
\sqrt{1+z'^{\,2}\,-(2\pi\alpha')^2\,A_t'^2}\,\,.
\label{DBI_density_zeroT}
\eeq
In the following we will scale out the constant $R$, \ie, we will take directly $R=1$. To avoid clutter, we also redefine the gauge field by absorbing the factors of the string length $2\pi\alpha'A_\mu\to A_\mu$.
Moreover, we will restrict ourselves to the case in which the embedding function $z(\rho)$ is a cyclic variable, \ie, when ${\cal L}_{DBI}$ depends on $z'$ and not on $z$. The only dependence on $z$ in (\ref{DBI_density_zeroT}) is the one induced by the power of $r$ multiplying the DBI square root. Therefore $z(\rho)$ is cyclic only when the following condition  between $n$, $p$, and $q$  is satisfied:
\beq
n\,=\,{p+q-4\over 2}\,\,.
\label{n_p_q}
\eeq
One can check that this happens only in the supersymmetric intersections with \#ND=4: $(p\, |\, p\perp  p+4)$, $(p-1\, |\, p\perp  p+2)$, and $(p-2\, |\, p\perp  p)$. In the following we will restrict ourselves to these cases. Let us define $\lambda$ as:
\beq
\lambda\,=\,2(q-n-1)\,=\,q-p+2\,\,.
\label{lambda_SUSY}
\eeq
Notice that $\lambda=6,4,2$ for the intersections D$p$-D$(p+4)$, D$p$-D$(p+2)$, and D$p$-D$p$, respectively.
We can then write the Lagrangian density as:
\beq
{\cal L}_{DBI}\,=\,-{\cal N}\,\rho^{{\lambda\over 2}}\,
\sqrt{1+z'^{\,2}\,-\,A_t'^2}\,\,.
\eeq
The cyclic nature of $z$ and $A_t$ implies the following conservation laws:
\bear
&&{1\over {\cal N}}{\partial {\cal L}_{DBI}\over \partial z'}\,=\,-
{\rho^{{\lambda\over 2}}\,z'\over \sqrt{1+z'^{\,2}\,-\,A_t'^2}}\,\equiv -c \nonumber \\
&&{1\over {\cal N}}{\partial {\cal L}_{DBI}\over \partial A_t'}\,=\,
{\rho^{{\lambda\over 2}}\,A_t'\over \sqrt{1+z'^{\,2}\,-\,A_t'^2}}\,\equiv d\,\,,
\eear
with $c$ and $d$ being constants of integration. 
These relations can be inverted as:
\beq
z'\,=\,{c\over \sqrt{\rho^{\lambda}\,+\,d^2-c^2}}\,\,,
\qquad\qquad\qquad
A_t'\,=\,{d\over \sqrt{\rho^{\lambda}\,+\,d^2-c^2}}\,\,.
\label{zprime-Atprime}
\eeq
When $c=d=0$, both $z(\rho)$ and $A_t(\rho)$ are constant and we have a Minkowski embedding. Let us suppose that $c$ does not vanish. Then,  it follows from (\ref{zprime-Atprime}) that $A_t'$ and $z'$ are related as:
\beq
A_t'\,=\,{d\over c}\,z'\,\,.
\eeq
When $c^2=d^2\not=0$ both $z(\rho)$ and $A_t(\rho)$ diverge at $\rho=0$. Therefore, we discard this configuration 
and we will assume in the following that $d^2>c^2$. In this case, 
from the expression of $z'$ and $A_t'$ written in  (\ref{zprime-Atprime}) it is easy to conclude that   the point $\rho=0$ is reached. In what follows we will assume that this condition holds.  We will integrate the equation for $A_t(\rho)$ by imposing that $A_t(0)=0$. We have:
\beq
A_t(\rho)\,=\,d\,\int_{0}^{\rho}\,{d\bar \rho\over \sqrt{\bar\rho^{\lambda}+d^2-c^2}}\,\,.
\eeq
This integral can be computed analytically and expressed 
in terms of the hypergeometric function as:
\beq
A_t(\rho)\,=\,{d\over \big(d^2-c^2\big)^{{1\over 2}-{1\over \lambda}}}\,
{\rho\over \big[ \rho^{\lambda}+d^2-c^2\big]^{{1\over \lambda}}}\,\,
F\Big(\,{1\over \lambda}, {1\over 2}+{1\over\lambda}; 1\,+\,{1\over\lambda};
{\rho^{\lambda}\over \rho^{\lambda}+d^2-c^2}\,\Big)\,\,.
\label{At_rho_massive}
\eeq
Similarly, the embedding function $z(\rho)$ can be written as:
\beq
z(\rho)\,=\,{c\over \big(d^2-c^2\big)^{{1\over 2}-{1\over \lambda}}}\,
{\rho\over \big[ \rho^{\lambda}+d^2-c^2\big]^{{1\over \lambda}}}\,\,
F\Big(\,{1\over \lambda}, {1\over 2}+{1\over\lambda}; 1\,+\,{1\over\lambda};
{\rho^{\lambda}\over \rho^{\lambda}+d^2-c^2}\,\Big)\,\,.
\label{z_rho_massive}
\eeq
Notice that when $d^2>c^2$ the brane reaches the Poincar\'e horizon of the metric at $\rho=z=0$ and we have a black hole embedding. The two constants $d$ and $c$ are related to the charge density and condensate of the dual theory, respectively.

\subsection{Zero temperature thermodynamics}

Let us first consider the intersections with $\lambda >2$. We also restrict to $T=0$, as at non-zero temperature not much can be said analytically. In this case the functions $A_t(\rho)$ and $z(\rho)$ in (\ref{At_rho_massive})  and (\ref{z_rho_massive}) 
approach a constant value in the UV region $\rho\to\infty$. 
According to the standard AdS/CFT dictionary,  the  flavor chemical potential $\mu$ is the UV value of $A_t$:
\beq
\mu\,=\,A_t(\rho\to\infty)\,=\,
{d\over \big(d^2-c^2\big)^{{1\over 2}-{1\over \lambda}}}\,
F\Big(\,{1\over \lambda}, {1\over 2}+{1\over\lambda}; 1\,+\,{1\over\lambda};1\Big)=\,{d\over \Big(d^2-c^2\Big)^{{1\over 2}-{1\over \lambda}}}\,\gamma \ ,
\label{chemical-pot-zeroT}
\eeq
where $\gamma$ is the constant
\beq
\gamma\,=\,{1\over \sqrt{\pi}}\,\Gamma\Big({1\over 2}-{1\over \lambda}\Big)\,
\Gamma\Big(1+{1\over \lambda}\Big)
\label{gamma_def}
\eeq 
and we used the identity $F(A,B;C;1)\,=\,{\Gamma(C)\,\Gamma(C-A-B)\over \Gamma(C-A)\,\Gamma(C-B)}$.

\begin{figure}[ht]
\center
 \includegraphics[width=0.7\textwidth]{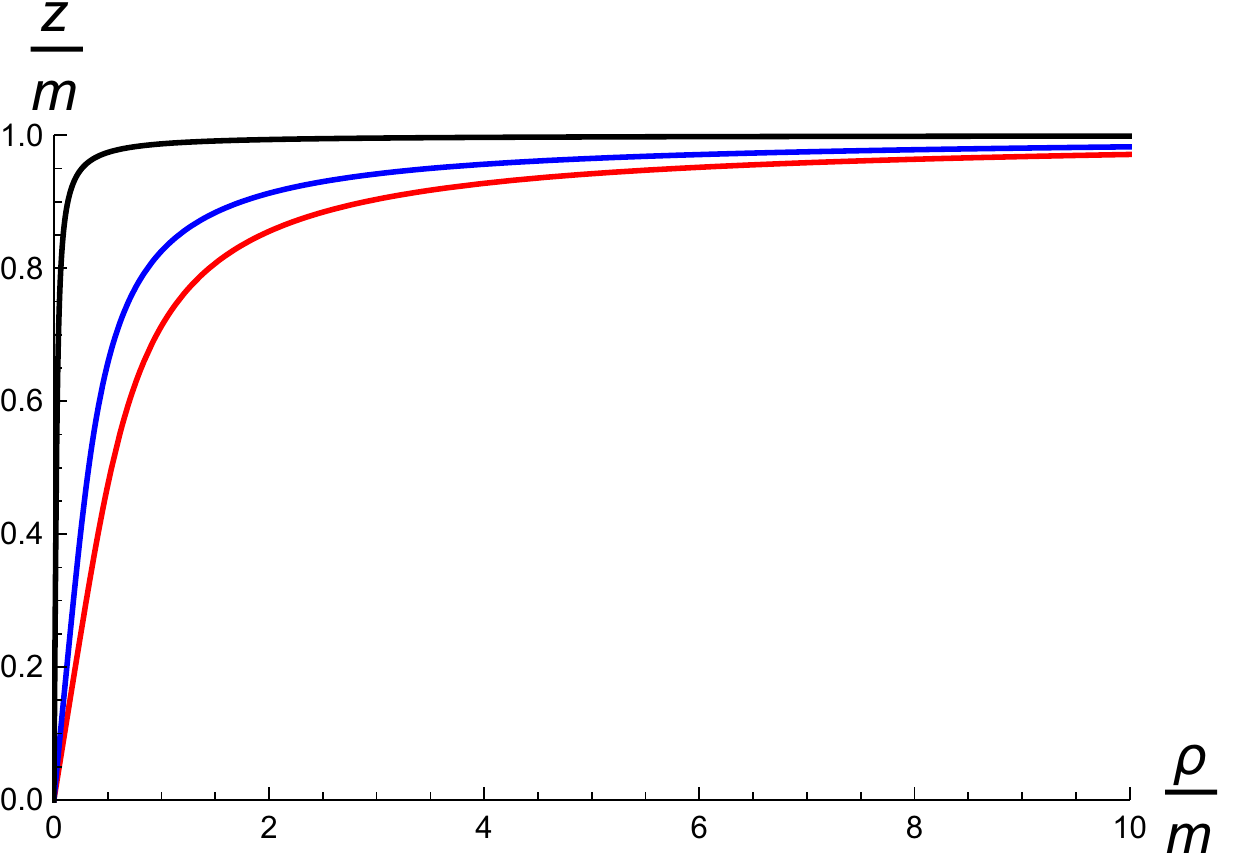}
\caption{In this figure we plot the different embeddings for $\lambda=4$ and $m/\mu=0.1, \, 0.8, \,0.999$ (bottom-up). The Minkowski embeddings at zero density and $m/\mu= 1$ would correspond to the  constant horizontal line $z/m=1$.    }
 \label{embeddings}
\end{figure}

The mass parameter $m$ of the embedding is defined as $m=z(\rho\to\infty)$. It follows from
(\ref{z_rho_massive}) that:
\beq
m\,=\,{c\over \Big(d^2-c^2\Big)^{{1\over 2}-{1\over \lambda}}}\,\gamma\,\,.
\label{mass_massive}
\eeq
Let us invert (\ref{chemical-pot-zeroT}) and (\ref{mass_massive}) and compute $c$ and $d$ in terms of $\mu$ and $m$. First, we notice that:
\beq
\mu^2\,-\,m^2\,=\,\Big(d^2-c^2\Big)^{{2\over \lambda}}\,\gamma^2\,\,.
\label{mu-m}
\eeq
Since  $d^2\ge c^2$, eq. (\ref{mu-m}) implies that  $\mu\ge m$ for the embeddings we are considering. Moreover, from (\ref{mu-m}) we  get $d^2-c^2$ as a function of $\mu$ and $m$ and,  using this result in (\ref{chemical-pot-zeroT}) and (\ref{mass_massive}), we  obtain
\beq
c\,=\,m\,\gamma^{-{\lambda\over 2}}\,
\Big(\mu^2-m^2\Big)^{{\lambda-2\over 4}}\,\,,
\qquad\qquad\qquad
d\,=\,\mu\,\gamma^{-{\lambda\over 2}}\,
\Big(\mu^2-m^2\Big)^{{\lambda-2\over 4}}\,\,.
\label{c_d_massive}
\eeq
When $\lambda>2$,   $\mu=m$  in  (\ref{c_d_massive}) corresponds to $c=d=0$, \ie, to the Minkowski embeddings with vanishing density discussed above. Actually, as illustrated in Fig.~\ref{embeddings}, the topology of the embeddings changes when $m\to\mu$, where a quantum phase transition takes place. The order parameter of this transition is the charge density (see \cite{Karch:2007br} for further details).

Let us now evaluate the on-shell action of the probe. Using
\beq
\sqrt{1+z'^{\,2}\,-\,A_t'^2}\Big|_{on-shell}\,=\,{\rho^{{\lambda\over 2}}\over 
 \sqrt{\rho^{\lambda}\,+\,d^2-c^2}}\,\,,
 \eeq
we find
\beq
S_{on-shell}\,=\,-{\cal N}\,\,\int_{0}^{\infty}\,{\rho^{\lambda}\over 
 \sqrt{\rho^{\lambda}\,+\,d^2-c^2}}\,d\rho\,\,,
\eeq
which is divergent and must be regulated. We will do it by subtracting  the same integral with the integrand evaluated at the UV ($\rho\to\infty$). We arrive at
\beq
S_{on-shell}^{reg}\,=\,-{\cal N}\,\,\int_{0}^{\infty}\,
\rho^{{\lambda\over 2}}\,\Bigg[\,
{\rho^{{\lambda\over 2}}\over  \sqrt{\rho^{\lambda}\,+\,d^2-c^2}}\,-\,1
\Bigg]\,d\rho\,=\,{2\,{\cal N}\over \lambda+2}\,
\Big(d^2-c^2\Big)^{{1\over \lambda}+{1\over 2}}\,\gamma
\,\,.
\label{S_on-shell}
\eeq
The zero-temperature grand canonical potential $\Omega$  is given by minus the regulated on-shell action:
\beq
\Omega\,=\,-S_{on-shell}^{reg}\,=\,-
{2\,{\cal N}\over \lambda+2}\,
\Big(d^2-c^2\Big)^{{1\over \lambda}+{1\over 2}}\,\gamma\,\,.
\eeq
In terms of $m$ and $\mu$ the grand canonical potential can be written as:
\beq
\Omega\,=\,-{2\,{\cal N}\over \lambda+2}\,
\gamma^{-{\lambda\over 2}}\,
\Big(\mu^2-m^2\Big)^{{\lambda+2\over 4}}\,\,,
\label{Omega_m_mu_massive}
\eeq
where we have used (\ref{mu-m}). Moreover,  the charge density is:
\beq
\rho_{ch}\,=\,-{\partial\Omega\over \partial\mu}\,=\,\mu{\cal N}\gamma^{-{\lambda\over 2}}\,
\Big(\mu^2-m^2\Big)^{{\lambda-2\over 4}}\,=\,{\cal N}\,d\,\,,
\label{rho-d}
\eeq
which confirms our identification of the constant $d$. It is worth noting that the formulas that we will write down do not have the factor of the (infinite) volume of the gauge theory directions $V_{\mathbb{R}^n}$, rather all thermodynamic quantities are densities per unit volume. Next, we compute the energy density as:
\beq
\epsilon\,=\,\Omega+\mu\,\rho_{ch}\,=\,{{\cal N}\over (\lambda+2)}\,
\gamma^{-{\lambda\over 2}}\,
\Big(\mu^2-m^2\Big)^{{\lambda-2\over 4}}\,(\lambda\mu^2+2m^2)\,\,.
\label{epsilon_massive}
\eeq
To calculate the speed of first sound $u_s$  we make use of the equation
\beq
u_s^2\,=\,{\partial P\over \partial \epsilon}\,=\,
{\partial P\over \partial \mu}\Big( {\partial \epsilon\over \partial \mu}\Big)^{-1}\,\,,
\eeq
where $P$ is the pressure. 
Let us first compute the derivative appearing in the numerator. Since $P=-\Omega$, we get from (\ref{Omega_m_mu_massive}):
\beq
{\partial P\over \partial \mu}\,=\,\mu\,{\cal N}\,\gamma^{-{\lambda\over 2}}\,
\Big(\mu^2-m^2\Big)^{{\lambda-2\over 4}}\,\,.
\eeq
Moreover, from (\ref{epsilon_massive}) we have:
\beq
{\partial \epsilon\over \partial \mu}\,=\,\mu\,{\cal N}\,\gamma^{-{\lambda\over 2}}\,
\Big(\mu^2-m^2\Big)^{{\lambda-6\over 4}}\,
\Big({\lambda\over 2}\,\mu^2\,-\,m^2\Big)\,\,.
\eeq
These yield
\beq
u_s^2\,=\,2\,{\mu^2-m^2\over \lambda\,\mu^2\,-\,2\,m^2}\,\,,
\label{speed_sound_massive}
\eeq
which is the result we were looking for. As a check notice that (\ref{speed_sound_massive})  gives 
$u_s^2=2/\lambda$ for $m=0$, which is the universal result found in \cite{Jokela:2015aha}.\footnote{For the massless \#ND=4 intersections one can rewrite the global symmetry in a suggestive form: $SO(n,1)\times SU(N_f)\times U(1)\times SO(3-\lambda/2)_p\times SO(1+\lambda/2)_q\times SO(5-n)$. The $SO(1+\lambda/2)_q$ part rotates a sphere of $\lambda/2$ dimensions, which curiously coincides with the value for the speed of sound (\ref{speed_sound_massive}) for $m=0$.}  Moreover, the speed of sound (\ref{speed_sound_massive}) depends on the integers $(n,p,q)$ through the combination $\lambda$, \ie, $u_s$ is the same for conformal and non-conformal brane backgrounds with the same index $\lambda$. 
In particular, for the D3-D7 and D3-D5 supersymmetric intersections we have:
\bear
&& u_s^2\,=\,{\mu^2-m^2\over 3\,\mu^2\,-\,m^2}\,\,,
\qquad\qquad\qquad {\rm for\,\,D3-D7}\,\,,\rc\rc
&& u_s^2\,=\,{\mu^2-m^2\over 2\,\mu^2\,-\,m^2}\,\,,
\qquad\qquad\qquad {\rm for\,\,D3-D5}\,\,.
\eear
These  results agree with the calculation in \cite{Kulaxizi:2008jx,Kim:2008bv}. Notice that the speed of sound vanishes in the zero density limit with $\mu=m$, which is a clear sign of a quantum phase transition. 

Let us now consider the case $\lambda=2$, which corresponds to the $(p-2|p\perp p)$ intersections. In these systems $A_t(\rho)$ and $z(\rho)$ grow logarithmically when $\rho\to\infty$ and the AdS/CFT dictionary must be adapted accordingly. Indeed, in this case the chemical potential and the mass are obtained from the subleading terms of $A_t$ and $z$ in the UV . Moreover, the on-shell action has additional logarithmic divergences, which must be eliminated with new counterterms \cite{Karch:2005ms,Benincasa:2009ze}. As the result of this analysis one gets that the grand canonical potential for black hole embeddings takes the form $\Omega=-a\, (\mu^2-m^2)$, where $a$ is a positive constant \cite{Benincasa:2009be}. Repeating the calculation  of $u_s$ performed above, it is straightforward to verify that $u_s^2=1$ in this $\lambda=2$ case. Notice that this value is exactly the one obtained by taking $\lambda=2$ in (\ref{speed_sound_massive}).

\section{Fluctuations}
\label{zeroT_fluct}

We now allow fluctuations of both the gauge field along the Minkowski directions of the intersection and of the scalar function in the form:
\beq
A_\nu\,=\,A_\nu^{(0)}\,+\,a_{\nu} (\rho, x^{\mu})\,\,,
\qquad\qquad
z\,=\,z_0(\rho)\,+\,\xi(\rho, x^{\mu})\,\,,
\label{zeroT-fluct}
\eeq
where $A^{(0)}=A_\nu^{(0)}\,dx^{\nu}=A_t\,dt$ is the one-form for the unperturbed gauge field  (\ref{At_rho_massive}) and $z_0$ is the embedding  function  written in (\ref{z_rho_massive}). The total gauge field strength is:
\beq
F\,=\,F^{(0)}\,+\,f\,\,,
\label{zeroT-fluct-F}
\eeq
with  $F^{(0)}=dA^{(0)}$ and $f=da$.  The dynamics of the fluctuations is determined by the Lagrangian density ${\cal L}$ that results after expanding the DBI action to second order in the perturbations $a_\mu$ and $\xi$. The detailed calculation of ${\cal L}$ is performed in appendix \ref{appendixA}. 
The Lagrangian can be neatly written in terms of open string metric ${\cal G}^{ab}$, which is symmetric and has the following non-vanishing components:
\bear
{\cal G}^{tt} & = & -{\rho^{\lambda}+d^2\over (\rho^2+z_0^2)^{{7-p\over 4}}\,\rho^{\lambda}} \nonumber \\
{\cal G}^{\rho\rho} & = & (\rho^2+z_0^2)^{{7-p\over 4}}\,\,
 {\rho^{\lambda}+d^2-c^2\over \rho^{\lambda}} \nonumber \\
 {\cal G}^{x^i\,x^j} & = & {\delta^{ij}\over 
 (\rho^2+z_0^2)^{{7-p\over 4}} }\,\,.
 \label{os-metric-zeroT}
\eear
The Lagrangian density for the fluctuations can be written as:
\bear
&&{\cal L}\,=\,-{\cal N}\,{\rho^{\lambda}\over \sqrt{\rho^{\lambda}+d^2-c^2}}
\Bigg[
{1\over 4}\,{\cal G}^{ac}\,{\cal G}^{bd}\,f_{cd}\,f_{ab}\,+\,
{1\over 2 r_0^{{7-p\over 2}}}\,
\Big(1-{c^2\over \rho^{\lambda}}\Big)\,{\cal G}^{a b}\,
\partial_{a}\xi\,\partial_{b}\xi\,\qquad\qquad\rc\rc
&&\qquad\qquad\qquad\qquad\qquad\qquad\qquad
-{c^2\,d^2\over 2\, r_0^{7-p}\rho^{2\lambda}}\, 
(\partial_{t}\xi)^2\,-\,
{cd\over  r_0^{{7-p\over 2}}\,\rho^{\lambda}}\,
{\cal G}^{ab}\,\partial_{a}\xi\,f_{tb}\,\,\Bigg]\,\,,\qquad
\label{massive_Lagrangian}
\eear
where we have defined  $r_0=r_0(\rho)$ as:
\beq
r_0(\rho)\,=\,\sqrt{\rho^2+z_0(\rho)^2}\,\,.
\eeq
In (\ref{massive_Lagrangian}) the tensor indices $a,b,c,d$ run over the directions $(\rho, x^{\mu})$.

Let us now explicitly write down the equations of motion derived from the Lagrangian (\ref{massive_Lagrangian}).  We will choose the gauge in which $a_{\rho}=0$. Moreover, we will consider fluctuation fields  $a_{\nu}$ which depend on $\rho$, $t$, and $x^1$. Then, it is possible to restrict to the case in which $a_{\nu}\not=0$ only when $\nu=t, x^1\equiv x$, and $ x^2\equiv y$. The equation of motion for $a_{\rho}$ when $a_{\rho}=0$ leads to the following transversality condition:
\beq
\Big(1+{d^2\over \rho^{\lambda}}\Big)\,\partial_t\,a_t'\,-\,{c\,d\over \rho^{\lambda}}\,
\partial_t\,\xi'\,-\,\partial_x\,a_x'\,=\,0\,\,.
\label{trans-coordinates}
\eeq
Let us Fourier transform  the gauge and scalar fields   to momentum space as:
\bear
a_\nu(\rho, t, x) & = & \int {d\omega\,dk\over (2\pi)^2}\,
a_\nu(\rho, \omega, k)\,e^{-i\omega\,t\,+\,i k x}\nonumber \\
\xi(\rho, t, x) &  = & \int {d\omega\,dk\over (2\pi)^2}\,
\xi(\rho, \omega, k)\,e^{-i\omega\,t\,+\,i k x}\,\,.
\eear
In momentum space the transversality condition (\ref{trans-coordinates})  takes the form:
\beq
\Big(1+{d^2\over \rho^{\lambda}}\Big)\,\omega\,a_t'\,-\,{c\,d\over \rho^{\lambda}}\,
\omega\,\xi'\,+\,k\,a_x'\,=\,0\,\,.
\label{trans-momentum}
\eeq
Let us now   define the electric field $E$ as the gauge-invariant combination:
\beq
E\,=\,k\,a_t\,+\,\omega\,a_x\,\,.
\label{E_at_ax}
\eeq
Using (\ref{trans-momentum})  we can obtain $a_t'$ and $a_x'$ in terms of $E'$ and $\xi'$ as follows:
\beq
a_t'\,=\,{k\,\rho^{\lambda}\,E'\,-\,c\,d\,\omega^2\,\xi'\over
\rho^{\lambda}\,k^2\,-\,(\rho^{\lambda}+d^2)\,\omega^2}\,\,,
\qquad\qquad
a_x'\,=\,{\omega\over
\rho^{\lambda}\,k^2\,-\,(\rho^{\lambda}+d^2)\,\omega^2}\,
\big[\,c\,d\,k\xi'\,-\,(\rho^{\lambda}+d^2)\,E'\big]\,\,.
\label{massive_transversality}
\eeq
The equation of motion for $a_t$ derived from (\ref{massive_Lagrangian}) is:
\bear
&&\partial_{\rho}\Bigg[\,{\sqrt{\rho^{\lambda}+d^2-c^2}\over \rho^{\lambda}}
\Big[(\rho^{\lambda}+d^2)a_t'\,-\,c\,d\,\xi'\Big]\Bigg]\nonumber \\
&&\qquad\qquad-
{1\over r_0^{7-p}\sqrt{\rho^{\lambda}+d^2-c^2}}
\Big[(\rho^{\lambda}+d^2)\partial_x\,(\partial_t a_x-\partial_x a_t)\,+\,c\,d\,
\partial^2_{x}\xi\,\Big]\,=\,0\,\,.\qquad\qquad
\label{eom_at}
\eear
The equation for $a_x$ is:
\bear
\partial_{\rho}\Big(\sqrt{\rho^{\lambda}+d^2-c^2}a_x'\Big)-{1\over r_0^{7-p}\sqrt{\rho^{\lambda}+d^2-c^2}}\big[(\rho^{\lambda}+d^2)\partial_t(\partial_t a_x-\partial_x a_t)+c d\partial_{t}\partial_{x}\xi\big]=0 \ . \label{eom_ax}
\eear
By using  (\ref{massive_transversality}), eqs. 
(\ref{eom_at}) and (\ref{eom_ax})  reduce (in momentum space) to the following equation in terms of the electric field $E$:
\beq
\partial_{\rho}\Big[{\sqrt{\rho^{\lambda}+d^2-c^2}\over 
(\omega^2-k^2)\rho^{\lambda}+\omega^2 d^2}\,
\big[(\rho^{\lambda}+d^2)E'\,-\,cd\,k\,\xi'\big]\Big]+
{(\rho^{\lambda}+d^2)E-cd\,k\,\xi\over r_0^{7-p}\sqrt{\rho^{\lambda}+d^2-c^2}}
=0\,\,.
\label{eom_massive_E}
\eeq
The equation for $a_y$ in momentum space  is:
\beq
\partial_{\rho}\,\Big(\sqrt{\rho^{\lambda}+d^2-c^2}\,a_y'\Big)+
{(\rho^{\lambda}+d^2)\,\omega^2 -\rho^{\lambda}\,k^2 
\over r_0^{7-p}\sqrt{\rho^{\lambda}+d^2-c^2}}\,a_y\,=\,0\,\,.
\label{eom_massive_ay}
\eeq
Finally, the equation for the scalar $\xi$ in momentum space can be written as:
\bear
&&\partial_{\rho}\Bigg[\,{\sqrt{\rho^{\lambda}+d^2-c^2}\over \rho^{\lambda}}
\Big[(\rho^{\lambda}-c^2)\,\xi'+ c d\,a_t'
\Big]\Bigg] \nonumber \\
&&\qquad\qquad\qquad\qquad\qquad\qquad+
{[(c^2-\rho^{\lambda})k^2\,+\,(\rho^{\lambda}+d^2-c^2)\omega^2]\xi-cd\,k\,E
\over r_0^{7-p}\sqrt{\rho^{\lambda}+d^2-c^2}}\,=\,0\,\,.\qquad\qquad
\eear
By using (\ref{massive_transversality}) we can rewrite this equation in terms of the electric field $E$:
\bear
&&\partial_{\rho}\Bigg[\,{\sqrt{\rho^{\lambda}+d^2-c^2}\over 
(\omega^2-k^2)\rho^{\lambda}+\omega^2\,d^2
}
\Big[
[(c^2-\rho^{\lambda})k^2\,+\,(\rho^{\lambda}+d^2-c^2)\omega^2]\xi'-cd\,k\,E'
\Big]\Bigg]\nonumber \\
&&\qquad\qquad\qquad\qquad\qquad\qquad+
{[(c^2-\rho^{\lambda})k^2\,+\,(\rho^{\lambda}+d^2-c^2)\omega^2]\xi-cd\,k\,E
\over r_0^{7-p}\sqrt{\rho^{\lambda}+d^2-c^2}}\,=\,0\,\,.\qquad\qquad
\label{eom_massive_xi}
\eear
In the next section we study these equations of motion in the regime in which the frequency $\omega$ and the momentum $k$ are small and of the same order. We will find a sound mode, the zero sound, and we will be able to determine analytically its dispersion relation following the matching technique introduced in \cite{Karch:2008fa}.

\section{Zero sound}
\label{zero_sound_section}

We now study the zero sound of the massive embeddings by matching the near-horizon and low frequency behavior of the fluctuations. The technique we employ consists in performing these two limits in different order \cite{Karch:2008fa}.

\subsection{Near-horizon analysis}
Let us first consider the equations of motion (\ref{eom_massive_E}) and (\ref{eom_massive_xi}) near the Poincar\'e horizon $\rho\approx 0$. To perform this analysis we define the functions $\chi_1$ and $\chi_2$ as:
\bear
\chi_1 & = & (\rho^{\lambda}+d^2)E-cd\,k\,\xi\nonumber \\
\chi_2 & = & [(c^2-\rho^{\lambda})k^2\,+\,(\rho^{\lambda}+d^2-c^2)\omega^2]\xi-cd\,k\,E
\,\,.
\label{chis_def}
\eear
For small $\rho$ we just neglect the terms containing $\rho^{\lambda}$ in the $\chi_i$'s. Then, these functions take the form:
\beq
\chi_1\,\approx \,d^2\,E-cd\,k\,\xi\,\,,
\qquad\qquad
\chi_2\,\approx \,
[c^2\,k^2\,+\,(d^2-c^2)\omega^2]\xi-cd\,k\,E\,\,,
\label{chis_nh}
\eeq
and, therefore, are related to $E$ and $\xi$ by linear combinations with constant coefficients. In order to write the near-horizon equations for  $\chi_1$ and $\chi_2$, let us study the behavior of the embedding function $z_0(\rho)$ for small $\rho$. From (\ref{z_rho_massive}) we easily obtain:
\beq
z_0\,\approx {c\over \sqrt{d^2-c^2}}\,\rho\,\,.
\eeq
It follows that $r_0(\rho)$ behaves near $\rho\approx 0$ as:
\beq
r_0\,\approx {d\over \sqrt{d^2-c^2}}\,\rho\,\,.
\eeq
Using these results it is straightforward to demonstrate that, for small $\rho$, the 
$\chi_i$'s satisfy the equation:
\beq
\chi_i''\,+\,{\Lambda^2\over \rho^{7-p}}\,\chi_i\,=\,0\,\,,
\label{chi_i_eq}
\eeq
where $\Lambda$ is the following rescaled frequency:
\beq
\Lambda^{2}= \Big({d^2-c^2\over d^2}\Big)^{{5-p\over 2}}\,\omega^2\,\,.
\eeq
Eq. (\ref{chi_i_eq}) is the same equation as in the massless case (with $\omega\to \Lambda$). 
When $p<5$, the solution of this equation with incoming boundary condition at the horizon  is given by  the following Hankel function:
\beq
\chi_i(\rho)\,=\,\rho^{{1\over 2}}\,H_{{1\over 5-p}}^{(1)}\,
\Big({2\,\Lambda\over 5-p}\,\rho^{{p-5\over 2}}\Big)\,\,,
\qquad\qquad (p<5)\,\,.
\label{Hankel_nh}
\eeq
For $d\not =c$ the equations in (\ref{chis_nh}) can be inverted and one can obtain $E$ and $\xi$ as linear combinations (with constant coefficients) of $\chi_1$ and $\chi_2$. Thus, $E$ and $\xi$ behave as in (\ref{Hankel_nh}).  Moreover, when $p<4$ and $\omega$ is small, we have:
\beq
E(\rho)\,=\,A\,\rho+A\,c_p\,\Lambda^{{2\over 5-p}}\,+\,\cdots\,\,,
\qquad\qquad
\xi(\rho)\,=\,B\,\rho+B\,c_p\,\Lambda^{{2\over 5-p}}\,+\,\cdots\,\,,
\qquad\qquad
(p<4)\,\,,
\label{E-chi_nh_low}
\eeq
where $A$ and $B$ are  constants and the coefficient $c_p$ is:
\beq
c_p\,=\,\pi\,{(5-p)^{{p-3\over 5-p}}
\over 
\,\Big[\Gamma\Big({1\over 5-p}\Big)\Big]^2}\,\Big[
i\,-\,
\cot\Big({\pi\over 5-p}\Big)\Big]\,\,,
\qquad\qquad 
(p<4)\,\,.
\label{cp}
\eeq

\subsection{Low frequency analysis}
Let us now start by taking the low frequency limit of the fluctuation equations (\ref{eom_massive_E}) and (\ref{eom_massive_xi}). One can show that in this limit one can neglect the terms without derivatives. Then, the fluctuation equations reduce to:
\bear
\partial_{\rho}\Big[{\sqrt{\rho^{\lambda}+d^2-c^2}\over 
(\omega^2-k^2)\rho^{\lambda}+\omega^2 d^2}\,
\big[(\rho^{\lambda}+d^2)E'\,-\,cd\,k\,\xi'\big]\Big] & = & 0\nonumber \\
\partial_{\rho}\Bigg[\,{\sqrt{\rho^{\lambda}+d^2-c^2}\over 
(\omega^2-k^2)\rho^{\lambda}+\omega^2\,d^2}\Big[[(c^2-\rho^{\lambda})k^2\,+\,(\rho^{\lambda}+d^2-c^2)\omega^2]\xi'-cd\,k\,E'\Big]\Bigg] & = & 0\,\,.\qquad\qquad
\eear
These equations can be immediately integrated once to give:
\bear
(\rho^{\lambda}+d^2)E'\,-\,cd\,k\,\xi' & = & C_1\,
{(\omega^2-k^2)\rho^{\lambda}+\omega^2 d^2\over 
\sqrt{\rho^{\lambda}+d^2-c^2}}\nonumber \\
c\,d\,k\,E'-[(c^2-\rho^{\lambda})k^2\,+\,(\rho^{\lambda}+d^2-c^2)\omega^2]\xi'& = & C_2\,
{(\omega^2-k^2)\rho^{\lambda}+\omega^2 d^2\over 
\sqrt{\rho^{\lambda}+d^2-c^2}}\,\,,
\eear
where $C_1$ and $C_2$ are integration constants. Solving for $E'$ and $\xi'$, we get:
\bear
 E' & = & {[(\omega^2-k^2)\,\rho^{\lambda}\,+\,(k^2-\omega^2)\,c^2\,+\,\omega^2\,d^2]C_1\,-\,cd\,k\,C_2\over (\rho^{\lambda}+d^2-c^2)^{{3\over 2}}}\nonumber \\
\xi' & = & {cd\,k\,C_1\,-\,(\rho^{\lambda}+d^2)\,C_2\over (\rho^{\lambda}+d^2-c^2)^{{3\over 2}}}\,\,.
\label{E-xi-prime_zeroT}
\eear
In order to perform a further integration, let us define the following functions:
\beq
{\cal J}_1(\rho)\,\equiv\,\int_{\rho}^{\infty}\,
{\bar\rho^{\lambda}\over
(\bar\rho^{\lambda}+d^2-c^2)^{{3\over 2}}}\,d\bar\rho\,\,,
\qquad\qquad
{\cal J}_2(\rho)\,\equiv\,\int_{\rho}^{\infty}\,
{d\bar\rho\over
(\bar\rho^{\lambda}+d^2-c^2)^{{3\over 2}}}\,\,.
\label{calJ1_calJ2_def}
\eeq
For $\lambda>2$ these integrals are convergent and can be computed analytically:
\bear
&&{\cal J}_1(\rho)\,=\,{2\over \lambda-2}\,\rho^{1-{\lambda\over 2}}\,
F\Big(\,{3\over 2}, {1\over 2}-{1\over\lambda}; {3\over 2}\,-\,{1\over\lambda};
- {d^2-c^2\over \rho^{\lambda}}\,\Big)\nonumber \\
&&{\cal J}_2(\rho)\,=\,{2\over 3\lambda-2}\,\rho^{1-{3\lambda\over 2}}\,
F\Big(\,{3\over 2}, {3\over 2}-{1\over\lambda}; {5\over 2}\,-\,{1\over\lambda};
- {d^2-c^2\over \rho^{\lambda}}\,\Big)\,\,.
\eear
Moreover, by construction ${\cal J}_1(\rho\to\infty)={\cal J}_2(\rho\to\infty)=0$. It follows that:
\bear
E(\rho) & = & E^{(0)}\,-\,(\omega^2-k^2)\,C_1\,{\cal J}_1(\rho)\,-\,
\Big[[(k^2-\omega^2)\,c^2+\omega^2\,d^2\,]C_1-cd\,k\,C_2\Big]\,{\cal J}_2(\rho)\nonumber \\
\xi(\rho) & = & \xi^{(0)}\,+\,C_2\,{\cal J}_1(\rho)\,+\,
d\big[d\,C_2-c\,k\,C_1\big]\,{\cal J}_2(\rho)\,\,,\label{E_xi_low_freq}
\eear
where $E^{(0)}$  and $\xi^{(0)}$ are the values of $E$ and $\xi$ at the boundary $\rho\to\infty$. Let us now expand $E(\rho)$ and $\xi(\rho)$  near $\rho\approx 0$. With this purpose it is better to deal directly with the integrals defining ${\cal J}_1$ and ${\cal J}_2$. One can easily prove that:
\bear
{\cal J}_1(\rho) & = & {2\over \lambda}\,\gamma\,(d^2-c^2)^{{1\over \lambda}-{1\over 2}}\,+\,{\mathcal O}(\rho^2)\nonumber \\
{\cal J}_2(\rho) & = & {\lambda-2\over \lambda}\,\gamma\,(d^2-c^2)^{{1\over \lambda}-{3\over 2}}\,-\,
{\rho\over (d^2-c^2)^{{3\over 2}}}\,+\,{\mathcal O}(\rho^2)\,\,,
\eear
where $\gamma$ is the constant defined in (\ref{gamma_def}). Using these expansions we can represent $E(\rho)$ near the horizon as:
\beq
E(\rho)\,=\,E^{(0)}\,+\,b_1\,C_1\,+b_2\,C_2\,+\,(a_1\,C_1\,+\,a_2\,C_2)\,\rho\,+\,\cdots
\,\,,
\label{E_low_freq_nh}
\eeq
where the coefficients $b_i$ and $a_i$ are given by:
\bear
 b_1 & = & -{\gamma\over \lambda}\,(d^2-c^2)^{{1\over \lambda}-{3\over 2}}\,\Big[\,(\lambda c^2\,-\,2\,d^2)\,k^2\,+\,\lambda\,(d^2-c^2)\,\omega^2\,\Big]\nonumber \\
 b_2 & = & {\lambda-2\over \lambda}\,\gamma\,(d^2-c^2)^{{1\over \lambda}-{3\over 2}}\,c\,d\,k\nonumber \\
 a_1 & = & {(d^2-c^2)\omega^2\,+\,c^2\,k^2\over (d^2-c^2)^{{3\over 2}}}\nonumber \\
 a_2 & = &-{c\,d\,k\over (d^2-c^2)^{{3\over 2}}}\,\,.
\label{b_a_coeff}
\eear
Similarly, $\xi(\rho)$ can be expanded as:
\beq
\xi(\rho)\,=\,\xi^{(0)}\,+\,\tilde b_1\,C_1\,+ \tilde  b_2\,C_2\,+\,
(\tilde  a_1\,C_1\,+\, \tilde  a_2\,C_2)\,\rho\,+\,\cdots\,\,,
\label{xi_low_freq_nh}
\eeq
where the different coefficients are:
\bear
 \tilde b_1 & = & -{\lambda-2\over \lambda}\,\gamma\,(d^2-c^2)^{{1\over \lambda}-{3\over 2}}\,c\,d\,k\,=\,-b_2\nonumber \\
 \tilde b_2 & = & {\gamma\over \lambda}\,(d^2-c^2)^{{1\over \lambda}-{3\over 2}}\,(\lambda\,d^2\,-\,2\, c^2)\nonumber \\
 \tilde a_1 & = & {c\,d\,k\over (d^2-c^2)^{{3\over 2}}}\,=\,-a_2\nonumber \\
 \tilde a_2 & = & -{d^2\over (d^2-c^2)^{{3\over 2}}}\,\,.
\label{tilde_b_a_coeff}
\eear

\subsection{Matching}
We now match (\ref{E_low_freq_nh}) and (\ref{xi_low_freq_nh}) with (\ref{E-chi_nh_low}). By identifying the terms linear in $\rho$, we can write the constants $A$ and $B$ of (\ref{E-chi_nh_low}) in terms of the coefficients of (\ref{b_a_coeff}) and (\ref{tilde_b_a_coeff}):
\bear
 A & = & a_1\,C_1\,+\,a_2\,C_2\,=\,{\big[(d^2-c^2)\,\omega^2\,+\,c^2\,k^2\big]\,C_1\,-\,c\,d\,k\,C_2\over (d^2-c^2)^{{3\over 2}}}\nonumber \\
 B & = & \tilde a_1\,C_1\,+\, \tilde  a_2\,C_2\,=\,{c\,d\,k\,C_1\,-\,d^2\,C_2\over (d^2-c^2)^{{3\over 2}}}\,\,.
\label{linear_terms_matching}
\eear
By eliminating $A$ and $B$ and comparing the constant terms in (\ref{E-chi_nh_low}), (\ref{E_low_freq_nh}), and (\ref{xi_low_freq_nh}) we get the boundary values of $E$ and $\xi$ as functions of $C_1$ and $C_2$:
\beq
\begin{pmatrix}
E^{(0)}\\ \\ \xi^{(0)}
\end{pmatrix}
\,=\,
\begin{pmatrix}
 \Lambda^{{2\over 5-p}}\,c_p\,a_1\,-\,b_1
 &&&
 \Lambda^{{2\over 5-p}}\,c_p\,a_2\,-\,b_2
  \\
  {} & {} \\
 \Lambda^{{2\over 5-p}}\,c_p\,\tilde a_1\,-\, \tilde  b_1 &&& 
  \Lambda^{{2\over 5-p}}\,c_p\, \tilde  a_2\,-\, \tilde  b_2
  \end{pmatrix}\,
 \begin{pmatrix}
C_1\\ \\ C_2
\end{pmatrix}\,\,.
\label{E0_xi0}
\eeq
We now require the vanishing of the sources $E^{(0)}$ and $\xi^{(0)}$, which only happens non-trivially if the determinant of the matrix written in (\ref{E0_xi0}) is zero. This leads to the following relation:
\beq
(\,a_1 \,\tilde a_2+a_2^2\,)\, \Lambda^{{4\over 5-p}}\,c_p^2\,-\,
(a_1\,\tilde b_2\,+\,\tilde a_2\,b_1\,+\,2\,a_2\,b_2\,)\,\Lambda^{{2\over 5-p}}\,c_p\,+\,
b_1\,\tilde b_2\,+\,b_2^2\,=\,0\,\,.
\label{zero-det}
\eeq
From (\ref{zero-det}) we can find the dispersion relation of the zero sound modes. Indeed, let us assume that $\omega\sim k\sim  {\mathcal O}(\epsilon)$. Then, 
$\Lambda\sim {\mathcal O}(\epsilon)$ and the orders of the different coefficients in (\ref{zero-det}) are:
\beq
b_1\sim a_1\sim {\mathcal O}(\epsilon^2)\,\,,\qquad
b_2\sim a_2\sim {\mathcal O}(\epsilon)\,\,,\qquad
\tilde b_1\sim \tilde  a_1\sim {\mathcal O}(\epsilon)\,\,,\qquad
\tilde  b_2\sim \tilde  a_2\sim {\mathcal O}(\epsilon^0)\,\,.
\eeq
At leading order the only contribution comes from the last two terms in (\ref{zero-det}), which  therefore reduces to:
\beq
b_1\,\tilde b_2\,+\,b_2^2\,=\,0\,\,.
\label{leading_zero_det}
\eeq
By using the values of the constants $b_1$, $b_2$, and $\tilde b_2$ from (\ref{b_a_coeff}) and (\ref{tilde_b_a_coeff}) it is straightforward to verify that (\ref{leading_zero_det}) leads to the following dispersion relation:
\beq
\omega^2\,=\,\omega^2_0\,=\,{2(d^2-c^2)\over \lambda\, d^2-2\,c^2}\,k^2\,\,.
\eeq
Let us write this result in terms of the reduced mass parameter ${\bf m}$, defined as:
\beq
{\bf m}\,=\,{m\over \mu}\,\,.
\label{reduced_mass}
\eeq
One easily checks that:
\beq
{c\over d}\,=\,{\bf m}\,\,.
\eeq
and the leading dispersion relation can be written as:
\beq
\omega^2_0\,=\,c_s^2\,k^2\,\,,
\label{leading_disp_relation}
\eeq
where $c_s$ is the speed of zero sound, given by
\beq
c_s^2\,=\,2\,\,{1-{\bf m}^2\over \lambda\, -2\,{\bf m}^2}\,\,.
\label{speed_zero_sound}
\eeq
Notice that, non-trivially, $c_s$ is equal to the speed of first sound written in (\ref{speed_sound_massive}). 

Let us now compute the next order in the dispersion relation. We write:
\beq
\omega\,=\,\omega_0\,+\,\delta\omega\,\,.
\eeq
At first-order in $\delta\omega$, we get:
\beq
\delta\omega\,=\,-\,2^{{p-3\over 2(5-p)}}\,c_p\,{\lambda\,d\over \gamma}\,
{(d^2-c^2)^{{6-p\over 5-p}-{1\over \lambda}}\over (\lambda\,d^2-2 c^2)
^{{7-p\over 2(5-p)}+1}}
\,k^{{7-p\over 5-p}}\,\,.
\label{next_leading_disp_rel}
\eeq
In terms of ${\bf m}$ this expression becomes:
\beq
\delta\omega\,=\,-\,
2^{{p-3\over 2(5-p)}}\,c_p\,{\lambda\,\over \mu}\,
{(1-{\bf m}^2)^{{6-p\over 5-p}-{1\over 2}}\over (\lambda-2 {\bf m}^2)
^{{7-p\over 2(5-p)}+1}}
\,\,k^{{7-p\over 5-p}}\,\,,
\eeq
where we used the following relation of  $\mu$, $d$, and $m$:
\beq
\mu\,=\,\gamma\,d^{2\over \lambda}\,(1-{\bf m}^2)^{{1\over \lambda}-{1\over 2}}\,\,.
\label{mu_d_massive}
\eeq
Let us use the expression of $c_p$ in (\ref{cp}) and separate the imaginary and real parts:
\bear
{\rm Im} \,\delta  \omega & = &-{\pi\,\lambda\over \mu}\,
{(5-p)^{{p-3\over 5-p}}\over \Big[\Gamma\Big({1\over 5-p}\Big)\Big]^2}\,
2^{{p-3\over 2(5-p)}}\,
{(1-{\bf m}^2)^{{6-p\over 5-p}-{1\over 2}}\over (\lambda-2 {\bf m}^2)
^{{7-p\over 2(5-p)}+1}}
\,\,k^{{7-p\over 5-p}}\nonumber \\
{\rm Re} \,\delta\omega & = & {\pi\,\lambda\over \mu}\,
{(5-p)^{{p-3\over 5-p}}\over \Big[\Gamma\Big({1\over 5-p}\Big)\Big]^2}\,
\cot\Big({\pi\over 5-p}\Big)\,
2^{{p-3\over 2(5-p)}}\,
{(1-{\bf m}^2)^{{6-p\over 5-p}-{1\over 2}}\over (\lambda-2 {\bf m}^2)
^{{7-p\over 2(5-p)}+1}}
\,\,k^{{7-p\over 5-p}}\,\,.
\label{Im_Re_delta_omega}
\eear
In particular, for $p=3$ the real part of ${\rm Re} \,\delta\omega$ vanishes at the order we are working in (\ref{Im_Re_delta_omega}) and the complete dispersion relation is given by:
\beq
\omega_{p=3}\,=\,\pm\,\sqrt{2}\,
\Bigg[{1-{\bf m}^2\over \lambda-2{\bf m}^2}\Bigg]^{{1\over 2}}\,k\,\,-\,
{i\,\lambda\over \mu}\,
{1-{\bf m}^2\over (\lambda-2 {\bf m}^2)^{2}}\,\,k^2\,\,.
\eeq
In order to compare with the results in \cite{Kulaxizi:2008kv,Davison:2011ek}, let us substitute $\mu$ by its expression in terms of the density $d$ (eq. (\ref{mu_d_massive})). We find
\beq
\omega_{p=3}\,=\,\pm\,\sqrt{2}\,
\Bigg[{1-{\bf m}^2\over \lambda-2{\bf m}^2}\Bigg]^{{1\over 2}}\,k\,-\,
i\,{ \lambda^2\over d^{{2\over \lambda}}}
\,{\Gamma\big({1\over 2}\big)\over 
\Gamma\Big({1\over 2}-{1\over \lambda}\Big)\,
\Gamma\Big({1\over \lambda}\Big)
}\,
{(1-{\bf m}^2)^{{3\over 2}-{1\over \lambda}}\over (\lambda-2 {\bf m}^2)
^{2}}\,\,k^2\,\,.
\eeq
In particular, for the D3-D5 system we take $\lambda=4$ and arrive at the following dispersion relation:
\beq
\omega_{D3-D5}\,=\,\pm\,
\Bigg[{1-{\bf m}^2\over 2-{\bf m}^2}\Bigg]^{{1\over 2}}\,k\,-\,
i\,\,{4\over d^{{1\over 2}}}
\,{\Gamma\big({1\over 2}\big)\over 
\Big[\Gamma\Big({1\over 4}\Big)\Big]^2\,
}\,
{(1-{\bf m}^2)^{{5\over 4}}\over (2-{\bf m}^2)
^{2}}\,\,k^2\,\,.
\label{disp_zero_sound_D3D5}
\eeq
\begin{figure}[ht]
\center
 \includegraphics[width=0.45\textwidth]{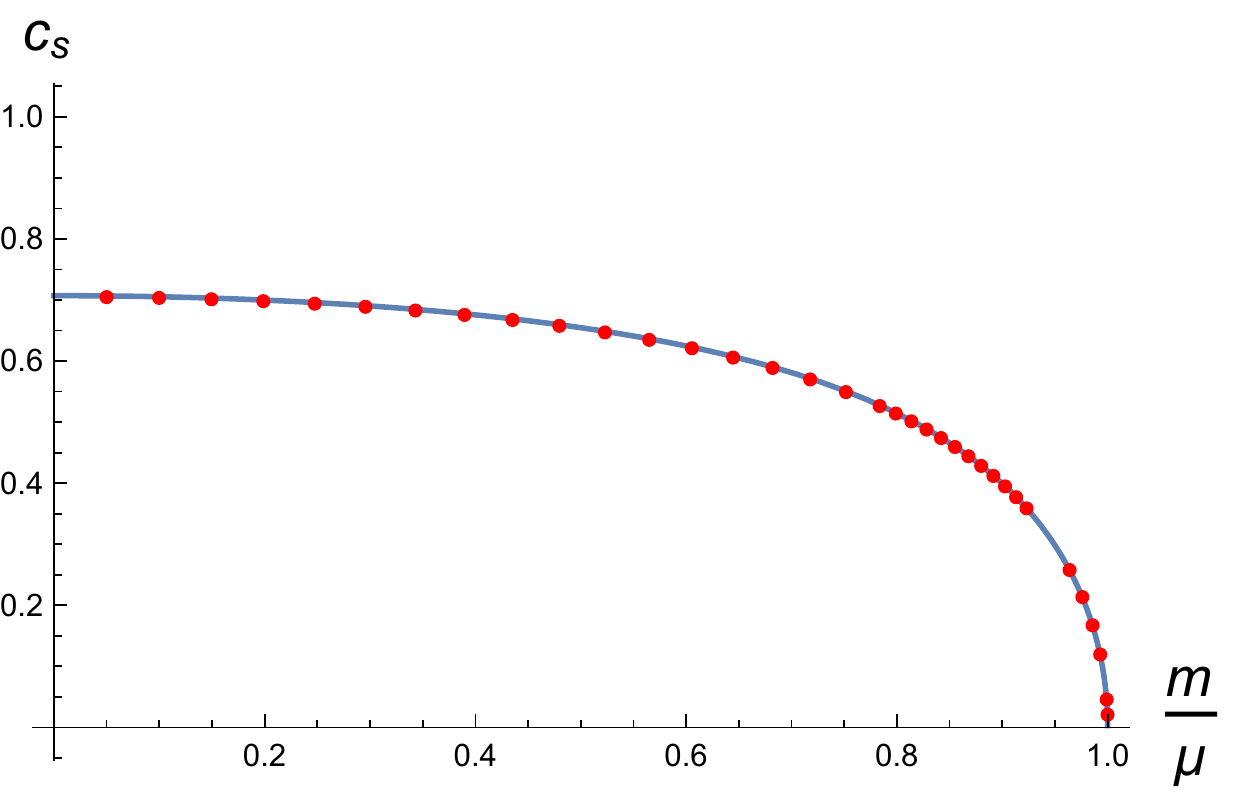}
\includegraphics[width=0.45\textwidth]{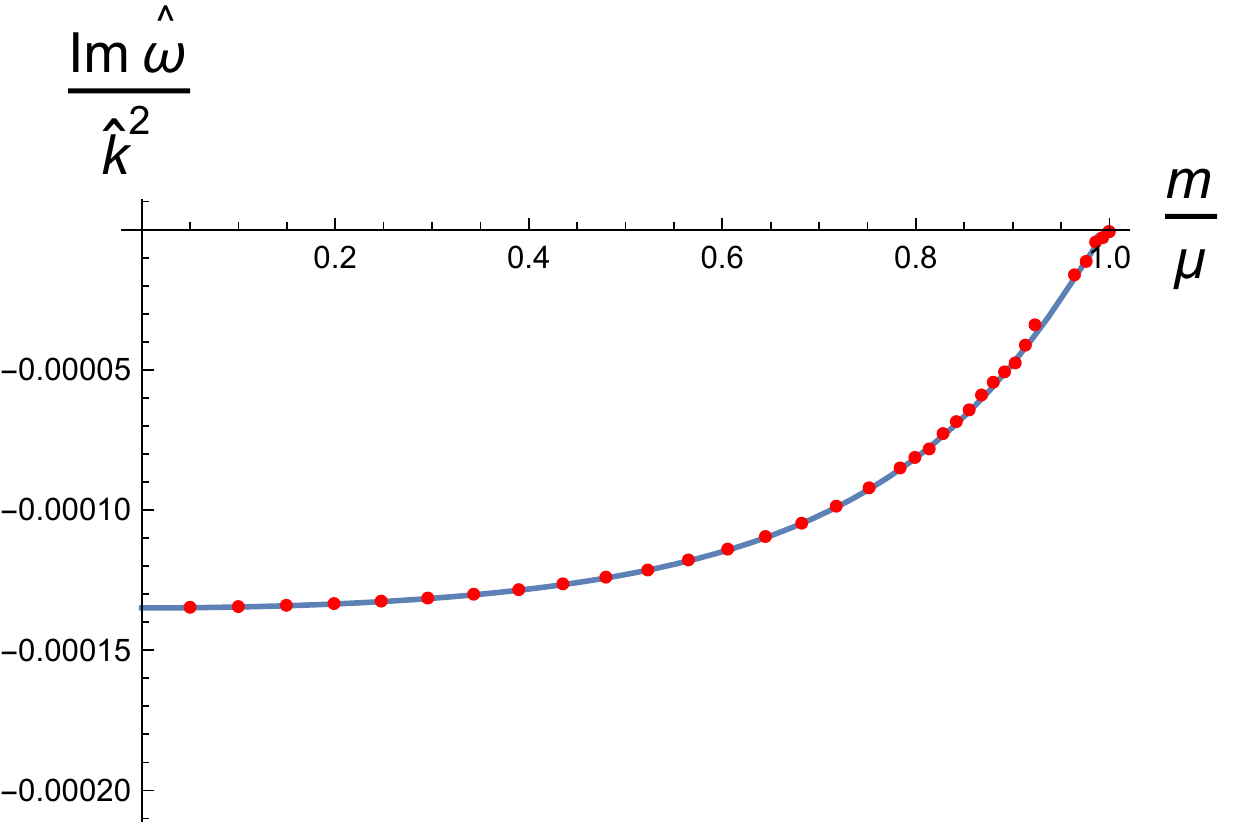}
\caption{We depict the speed of zero sound  (left) and the attenuation divided by momentum squared (right) for the D3-D5 intersection with $\hat d=10^6$ ($\hat d$, $\hat \omega$, and $\hat k$ are defined in (\ref{hat_d_hat_B_def}) and  (\ref{hat_omega_k})). The  dots have been obtained by integrating numerically the fluctuation equations (\ref{E_eom_momentum_B}) and (\ref{zeta_eom_momentum_B}).  The continuous line corresponds to the analytic expression (\ref{disp_zero_sound_D3D5}).}
 \label{zero_sound}
\end{figure}
In Fig.~\ref{zero_sound}  we check (\ref{disp_zero_sound_D3D5})  by comparing it with the results obtained by numerical integration at non-zero (but small) temperature. As it can be appreciated in this figure, the agreement is very good, both for the speed of zero sound $c_s$ and for  the attenuation  (\ie, the imaginary part of $\omega$).

\subsection{The $p=4$ case}
\label{p4_zero_sound}
As pointed out around (\ref{E-chi_nh_low}), the $p=4$ case is special and we have to modify our analysis. Indeed, the expansion of the Hankel function $H_1^{(1)}(x)$ near $x=0$ contains logarithmic terms, which implies that $E(\rho)$ and $\xi(\rho)$ behave near the horizon at low frequency  as:
\bear
 E(\rho) & = & A\,\rho+A\,c_4\,\Lambda^2\,+\,A\,\Lambda^2\,\log \big({\rho\over \Lambda^2}\big)\,+\,\cdots\nonumber \\
 \xi(\rho) & = & B\,\rho+B\,c_4\,\Lambda^2\,+\,B\,\Lambda^2\,\log \big({\rho\over \Lambda^2}\big)\,+\,\cdots\,\,,
\label{E-chi_nh_low_p4}
\eear
where $c_4$ is the constant:
\beq
c_4=i\,\pi+1-2\gamma_{E}\,\,.
\label{c4}
\eeq
In (\ref{c4})  $\gamma_E=0.577\cdots$ is the Euler-Mascheroni constant. Let us now try to obtain the expansion (\ref{E-chi_nh_low_p4}) by performing the limits in the opposite order. As in \cite{Kulaxizi:2008jx}, we have to compute the next correction to (\ref{E_low_freq_nh}) and (\ref{xi_low_freq_nh}) near the horizon.  First we notice that the equations satisfied by $E(\rho)$ and $\xi(\rho)$ near $\rho=0$ are just obtained by taking $p=4$ in (\ref{chi_i_eq}):
\beq
E''\,=\,-{\Lambda^2\over \rho^3}\,\,E\,\,,\qquad\qquad
\xi''\,=\,-{\Lambda^2\over \rho^3}\,\xi\,\,.
\label{E_xi_nh_p4}
\eeq
Neglecting the right-hand side in (\ref{E_xi_nh_p4}) and integrating twice, we arrive at a linear solution as in 
(\ref{E_low_freq_nh}) and (\ref{xi_low_freq_nh}). To go beyond this approximation we plug the values of $E$ and $\xi$ into the right-hand side of (\ref{E_xi_nh_p4}) and perform the integration. In the low-frequency limit $\omega^2\ll \rho$, we have:
\bear
E(\rho) & = & E^{(0)}+b_1C_1+b_2C_2+(a_1C_1+a_2C_2)\rho+
(a_1C_1+a_2C_2)\Lambda^2\log\rho +\cdots\nonumber \\
\xi(\rho) & = & \ \xi^{(0)}+\tilde b_1C_1+ \tilde  b_2C_2+
(\tilde  a_1C_1+\tilde  a_2C_2)\rho+(\tilde  a_1C_1+\tilde  a_2C_2)\Lambda^2\log\rho +\cdots\,\,.
\label{E_xi_low_freq_nh_p4}
\eear
Let us now match (\ref{E-chi_nh_low_p4}) and (\ref{E_xi_low_freq_nh_p4}). By comparing the linear and logarithmic terms of these equations we arrive at the same values of $A$ and $B$ as those written in  (\ref{linear_terms_matching}).  Moreover, using these values of $A$ and $B$ and identifying the constant terms, we find the following matrix relation between 
$(E^{(0)}, \xi^{(0)})$ and $(C_1, C_2)$:
\beq
\begin{pmatrix}
E^{(0)}\\ \\ \xi^{(0)}
\end{pmatrix}
\,=\,
\begin{pmatrix}
 \Lambda^{2}\,(c_4-\log \Lambda^2)\,a_1\,-\,b_1
 &&&
 \Lambda^{2}\,(c_4-\log \Lambda^2)\,a_2\,-\,b_2
  \\
  {} & {} \\
 \Lambda^{2}\,(c_4-\log \Lambda^2)\,\tilde a_1\,-\, \tilde  b_1 &&& 
  \Lambda^{2}\,(c_4-\log \Lambda^2)\, \tilde  a_2\,-\, \tilde  b_2
  \end{pmatrix}\,
 \begin{pmatrix}
C_1\\ \\ C_2
\end{pmatrix}\,\,.
\label{E0_xi0_p4}
\eeq
As in the $p<4$ case, the  sources vanish non-trivially when the determinant of the matrix written in (\ref{E0_xi0_p4})
is zero, namely:
\bear
&&(\,a_1 \,\tilde a_2+a_2^2\,)\, \Lambda^{2}\,(c_4-\log \Lambda^2)^2\,-\,
(a_1\,\tilde b_2\,+\,\tilde a_2\,b_1\,+\,2\,a_2\,b_2\,)\,\Lambda^{2}\,(c_4-\log \Lambda^2)\nonumber \\
&&\qquad\qquad\qquad\qquad
+\,b_1\,\tilde b_2\,+\,b_2^2\,=\,0\,\,.
\label{zero-det_p4}
\eear
Notice that (\ref{zero-det_p4})  is obtained from  (\ref{zero-det}) by taking $p=4$ and changing $c_p\to c_4-\log \Lambda^2$ on the latter.  Using this observation it is straightforward to find the dispersion relation encoded in (\ref{zero-det_p4}). At leading order in $\omega\sim k$ (\ref{zero-det_p4}) reduces to (\ref{leading_zero_det}), which means that the leading dispersion relation is just given by (\ref{leading_disp_relation}) and (\ref{speed_zero_sound}). Moreover, the next-to-leading contribution $\delta \omega$ is:
\beq
\delta\omega\,=\,-\,{\sqrt{2}\,\,\lambda\over  \mu}\,\,(c_4-\log \Lambda^2)\,
{(1-{\bf m}^2)^{{3\over 2}}\over (\lambda-2 {\bf m}^2)
^{{5\over 2}}}
\,\,k^{3}\,\,.
\label{next_leading_disp_rel_p4}
\eeq
The imaginary part of  $\delta \omega$ is easily deduced from (\ref{next_leading_disp_rel_p4}):
\beq
{\rm Im}\,\delta\omega\,=\,-
{\pi\,\sqrt{2}\,\lambda\over  \mu}\,
{(1-{\bf m}^2)^{{3\over 2}}\over (\lambda-2 {\bf m}^2)
^{{5\over 2}}}
\,\,k^{3}\,\,.
\label{im_delta_omega_p4}
\eeq
Notice that (\ref{im_delta_omega_p4}) is the same as in the first equation in (\ref{Im_Re_delta_omega}) for $p=4$. Similarly, the real part of $\delta\omega$ can be written as:
\beq
{\rm Re}\,\delta\omega\,=\,{\sqrt{2}\,\lambda\over \mu}\,
\Bigg[2\gamma_E-1+\log\Bigg(2\,
{(1-{\bf m}^2)^{{3\over 2}}\over \lambda-2 {\bf m}^2}\,k^2\Bigg)\Bigg]\,
{(1-{\bf m}^2)^{{3\over 2}}\over (\lambda-2 {\bf m}^2)^{{5\over 2}}}
\,\,k^{3}\,\,.
\label{re_delta_omega_p4}
\eeq

\subsection{The $\lambda=2$ case}

For $\lambda=2$ the integral ${\cal J}_1(\rho)$, defined in (\ref{calJ1_calJ2_def}), is not convergent and, therefore, the expressions  written in (\ref{E_xi_low_freq})  for  $E(\rho)$ and $\xi(\rho)$ at low frequency  are not  correct. In order to obtain the solution of (\ref{E-xi-prime_zeroT}) for $\lambda=2$, let us define the integral $\bar{\cal J}_1(\rho)$ as:
\beq
\bar{\cal J}_1(\rho)\,\equiv\,\int_{\rho}^{\infty}\,d\bar \rho\Big[
{\bar\rho^2\over (\bar\rho^2+d^2-c^2)^{{3\over 2}}}-{1\over \bar\rho}\Big]=
{\rho\over \sqrt{\rho^2+d^2-c^2}}-1\,+\,\log{2\rho\over \sqrt{\rho^2+d^2-c^2}+\rho}\,\,.
\eeq
Then, (\ref{E-xi-prime_zeroT}) for $\lambda=2$ can be integrated as:
\bear
 E(\rho) &  = & E^{(0)}\,-\,(\omega^2-k^2)\,C_1\,[\bar {\cal J}_1(\rho)-\log\rho]
\,-\,
\Big[[(k^2-\omega^2)\,c^2+\omega^2\,d^2\,]C_1-cd\,k\,C_2\Big]\,{\cal J}_2(\rho)\nonumber \\
\xi(\rho) & = & \xi^{(0)}\,+\,C_2\,[\bar {\cal J}_1(\rho)-\log\rho]\,+\,
d\big[d\,C_2-c\,k\,C_1\big]\,{\cal J}_2(\rho)\,\,,
\label{E_xi_low_freq_lambda2}
\eear
where $E^{(0)}$ and $\xi^{(0)}$ are constants. When $\rho$ is very large the integrals 
$\bar {\cal J}_1(\rho)$ and ${\cal J}_2(\rho)$ vanish by construction and thus $E(\rho)$ and $\xi(\rho)$ behave at the UV as:
\bear
 E(\rho) & = & E^{(0)}+(\omega^2-k^2)\,C_1\,\log\rho\,+\,\cdots\nonumber \\
 \xi(\rho) & = & \xi^{(0)}-C_2\,\log\rho\,+\cdots\,\,,
\qquad\qquad (\rho\to\infty)\,\,.
\label{UV_lambda2}
\eear
As argued in \cite{Karch:2005ms}, when the logarithmic behavior displayed in (\ref{UV_lambda2}) is present, the sources are identified with the coefficients of the logarithms, which should vanish. It is clear from the behavior of $\xi(\rho)$ in 
(\ref{UV_lambda2}) that we must require that $C_2=0$. Moreover, the logarithmic term in $E(\rho)$ is absent either when $C_1=0$ or when:
\beq
\omega=\pm k\,\,.
\label{disp_rel_lambda2}
\eeq
If  $C_1=C_2=0$ it follows from (\ref{E_xi_low_freq_lambda2}) that  the functions $E(\rho)$ and $\xi(\rho)$  are constant and also the matching with the near-horizon results in (\ref{E-chi_nh_low}) imply that both $E$ and $\xi$ must vanish. Therefore, the only non-trivial solution is given by the dispersion relation (\ref{disp_rel_lambda2}), which corresponds to a zero sound mode without dissipation and  speed $c_s^2=1$. Notice that this result coincides with the value of the speed of first sound in (\ref{speed_sound_massive}) for $\lambda=2$. Moreover, when $C_2=0$ and $\omega^2=k^2$, eq. (\ref{E_xi_low_freq_lambda2}) reduces to:
\beq
E(\rho)\,=\,E^{(0)}\,-\,\omega^2\,d^2\,C_1\,{\cal J}_2(\rho)\,\,,
\qquad\qquad
\xi(\rho)\,=\,\xi^{(0)}\,-\,c\,d\,k\,C_1\,{\cal J}_2(\rho)\,\,.
\label{E_xi_lambda2}
\eeq
Taking $\rho\to 0$ in (\ref{E_xi_lambda2}) we can match this result with (\ref{E-chi_nh_low}) and, as a consequence, we can show that $E^{(0)}$ and $\xi^{(0)}$ are related to the constant $C_1$ as:
\bear
 E^{(0)} & = & C_1\,{d^2\over d^2-c^2}\,\omega^2\,+\,
c_p\, C_1\,{d\over d^2-c^2}\,\omega^{{2(6-p)\over 5-p}}\nonumber \\
 \xi^{(0)} & = & C_1\,{c\,d\over d^2-c^2}\,k\,+\,
c_p\, C_1\,{c\over d^2-c^2}\,k\,\omega^{{2\over 5-p}}\,\,.
\label{E0_xi0_lambda2}
\eear
Notice that (\ref{E0_xi0_lambda2}) coincides with (\ref{E0_xi0}) when $\lambda=2$, $C_2=0$ and $\omega^2=k^2$. In particular, these relations imply that the ratio of $E^{(0)}$ and $\xi^{(0)}$ is given by:
\beq
{E^{(0)}\over \xi^{(0)}}\,=\,{d\over c}\,k\,\,.
\eeq
The analysis performed so far in this section is valid for $p<4$. When $p=4$ we have to go beyond the leading term in $\omega$, as in section \ref{p4_zero_sound}, in order to match the logarithmic terms in the near-horizon expansion. It is easy to check that the $\lambda=2$ solution written above can be corrected to match the $\rho\to 0$  expansion in (\ref{E-chi_nh_low_p4}).  The dispersion relation is still given by (\ref{disp_rel_lambda2}) and (\ref{E0_xi0_lambda2}) continues to hold in this case.

\section{Hyperscaling violation near the critical point}
\label{critical}
As  already mentioned, the probe D-brane systems analyzed above undergo a quantum phase transition as $\mu\to m$ and the density $d$ vanishes. It was shown in \cite{Ammon:2012je} that the critical points of the D3-D7 and D3-D5 intersections are described by a non-relativistic scale invariant field theory exhibiting hyperscaling violation. In this section we extend these results to the case of non-conformal backgrounds (\ie, for $p\not=3$) and we compute the corresponding critical exponents.

Let us  thus follow the approach of \cite{Ammon:2012je} and study the behavior of the system near the quantum critical point at $\mu=m$. Accordingly, we consider a chemical potential of the form:
\beq
\mu\,=\,m+\,\bar \mu\,\,,
\eeq
where $\bar \mu$ is considered to be small.  At leading order in $\bar \mu$ we can expand the different thermodynamic functions of (\ref{Omega_m_mu_massive}), (\ref{epsilon_massive}), and (\ref{c_d_massive}) as:
\bear
\Omega & = & -P \,\approx\,-{2^{{\lambda+6\over 4}}\over \lambda+2}\,
\gamma^{-{\lambda\over 2}}\,{\cal N}\,\big(m\,\bar \mu)^{{\lambda +2\over 4}}\nonumber \\
\epsilon & =& f\, \approx\,2^{{\lambda-2\over 4}}\,
\gamma^{-{\lambda\over 2}}\,{\cal N}\,\,m^{{\lambda+6\over 4}}\,\,
\bar \mu^{{\lambda-2\over 4}}\nonumber \\
d & \approx & 2^{{\lambda-2\over 4}}\,
\gamma^{-{\lambda\over 2}}\,m^{{\lambda+2\over 4}}\,\,
\bar \mu^{{\lambda-2\over 4}}\,\,.
\label{P_f_d_critical}
\eear
where $f$ is the free energy density. The non-relativistic energy density $e$ is defined as in \cite{Ammon:2012je}:
\beq
e\,=\,\epsilon\,-\, \rho_{ch}\,m\,=\,\epsilon\,-\,{\cal N}\, d\,m\,\,,
\eeq
where $ \rho_{ch}\,=\,{\cal N}\,d$ is the physical charge density. By using (\ref{epsilon_massive}) and (\ref{c_d_massive}) we get:
\beq
e\,=\,{\cal{N}}\, \gamma^{-{\lambda\over 2}}\,\big(\mu^2\,-\,m^2\big)^{{\lambda-2\over 4}}\,
\Big[{\lambda\mu^2+2 m^2\over \lambda+2}\,-\,\mu\,m\,\Big]\,\,.
\eeq
Expanding at leading order in $\bar \mu$, we arrive at:
\beq
e\,\approx\,2^{{\lambda-2\over 4}}\,{\lambda-2\over \lambda+2}\,
{\cal N}\, \gamma^{-{\lambda\over 2}}\,
\big(m\bar\mu\big)^{{\lambda+2\over 4}}\,\,.
\eeq
Comparing this result  with the one for the pressure in (\ref{P_f_d_critical}), we obtain the following relation between $e$ and $P$:
\beq
e\,=\,{\lambda-2\over 4}\,\,P\,\,.
\label{e_p_critical}
\eeq
According to the analysis in \cite{Ammon:2012je}, the relation between $e$ and $P$ at zero temperature  near the 
quantum critical point is:
\beq
e\,=\,{n-\theta\over z}\,P\,\,,
\label{e_p_critical_general}
\eeq
where $\theta$ is the hyperscaling violation exponent and $z$ is the dynamical critical exponent.  Eq. (\ref{e_p_critical_general})  is a consequence of the scaling dimensions of $e$, $P$, $ \bar\mu$, and $d$, namely:
$[e]=[P]=z+n-\theta$, $[\bar \mu]=z$,  and $[d]=n-\theta$. Thus, in our case we have the following relation between 
$\theta$ and $z$:
\beq
\theta\,=\,n\,-\,{\lambda-2\over 4}\,z\,\,.
\label{theta_z}
\eeq
Notice that the relation (\ref{theta_z}) between $\theta$ and $z$  coincides with the ones found in  \cite{Ammon:2012je} for the D3-D7 system (taking $n=3$ and $\lambda=6$) and for the D3-D5 intersection (taking  $n=2$ and $\lambda=4$). In order to determine $z$ we look at the speed of  sound  (\ref{speed_sound_massive}) for $\mu\approx m$.
At first-order in $\bar \mu$ it is given by:
\beq 
u_s^2\,\approx\,{4\over \lambda-2}\,{\bar\mu\over m}\,\,,
\eeq
and the corresponding dispersion relation is:
\beq
\omega\, \approx\, \sqrt{{4\over \lambda-2}\,{\bar\mu\over m}}\,\,k\,\,.
\label{critical_disp_relation}
\eeq
Matching the scaling dimensions of both sides of (\ref{critical_disp_relation}) 
as in \cite{Ammon:2012je}, using that $[\omega]=z$ and $[k]=1$, we conclude that:
\beq
z\,=\,2\,\,.
\eeq
Therefore $\theta$ takes the value:
\beq
\theta\,=\,n\,-\,{\lambda\over 2}\,+1\,\,.
\eeq
Taking into account that  for the  SUSY D$p$-D$q$ intersections we are considering 
\beq
n\,=\,{p+q-4\over 2}\,\,,\qquad\qquad
\lambda\,=\,q-p+2\,\,,
\label{n_lambda_SUSY}
\eeq
we can rewrite the expression of $\theta$ simply as:
\beq
\theta\,=\,p\,-\,2\,\,.
\label{theta_any_p}
\eeq
Notice that for a D3-D$q$ intersection the previous formula gives $\theta=1$, in agreement with \cite{Ammon:2012je}. Eq. (\ref{theta_any_p}) is the generalization of this result for any $p$.

Let us now consider the system at finite temperature $T$. According to the analysis of
\cite{Karch:2009eb}, when $T$ is small the free energy density can be approximated as:
\beq
f(\mu,m,T)\,=\,f(\mu,m,T=0)\,+\,\pi\, \rho_{ch}\,T\,+\,{\mathcal O}(T^2)\,\,.
\eeq
Then, the non-relativistic free energy  density is given by:
\beq
f_{non-rel}(\mu,m,T)\,=\,f(\mu,m, T)\,-\, \rho_{ch}\,m\,=\,e\,+\,
\pi\, \rho_{ch}\,T\,+\,{\mathcal O}(T^2)\,\,.
\eeq
At leading order in $\bar\mu$ we have:
\beq
f_{non-rel}(\mu,m,T)\,=\,
2^{{\lambda-2\over 4}}\,{\lambda-2\over \lambda+2}\,
{\cal N}\, \gamma^{-{\lambda\over 2}}\,
\big(m\bar\mu\big)^{{\lambda+2\over 4}}\,\Big[
1\,+\,\pi\,{\lambda+2\over \lambda-2}\,{T\over \bar\mu}\,+\,
{\mathcal O}\Big(\Big({T\over \bar \mu}\Big)^2\Big)\Big]\,\,.
\label{f_nonrel}
\eeq
In the quantum critical region  the non-relativistic free energy  density
should scale as:
\beq
f_{non-rel}\sim \big(\bar\mu\big)^{2-\alpha}\, g\Big({T\over \bar\mu ^{\nu z}}\Big)\,\,,
\label{f_nonrel_scaling}
\eeq
where $\alpha$ is the exponent which characterizes the scaling of the specific heat capacity $C$ and $\nu$ is the exponent corresponding to the correlation length $\xi$ (\ie, $C\sim (T-T_c)^{-\alpha}$ and $\xi\sim (T-T_c)^{-\nu}$ near a phase transition at $T=T_c$). Comparing (\ref{f_nonrel_scaling}) and
(\ref{f_nonrel})  it follows that, in our case, we have:
\beq
2-\alpha\,=\,{\lambda+2\over 4}\,\,,
\qquad\qquad
\nu\,z\,=\,1\,\,.
\eeq
Since $z=2$ for our system, the exponents $\alpha$ and $\nu$ are:
\beq
\alpha\,=\,{6-\lambda\over 4}\,\,,
\qquad\qquad
\nu\,=\,{1\over 2}\,\,.
\eeq
Using the expression of $\lambda$ in terms of $p$ and $q$ written in (\ref{n_lambda_SUSY}), we can recast $\alpha$ simply as:
\beq
\alpha\,=\,1\,-\,{q-p\over 4}\,\,.
\label{alpha_p_q}
\eeq
These results again coincide with the ones in \cite{Ammon:2012je} for the D3-D7 and D3-D5 intersections. 
Remarkably, the exponents obtained above satisfy the hyperscaling-violation relation:
\beq
(n+z-\theta)\,\nu\,=\,2-\alpha\,\,.
\eeq

\section{Zero sound in alternative quantization}
\label{alternative_quantization}

In this section we will restrict ourselves to the study of intersections which are $(2+1)$-dimensional. In this case one can impose mixed Dirichlet-Neumann boundary conditions to the fluctuation modes, \ie, one can adopt an alternative quantization scheme \cite{Witten:2003ya,Yee:2004ju}. The equations of motion are the same for different quantizations, only the boundary conditions in the UV are different. On the dual field theory side this corresponds to having an anyonic fluid \cite{Jokela:2013hta,Jokela:2014wsa,Brattan:2013wya,Brattan:2014moa}. 
Let us  impose the following boundary condition at the UV:
\beq
\lim_{\rho\to\infty}\,\Big[\,\textfrak{n}\,\rho^{{\lambda\over 2}}\,f_{\rho\,\mu}\,-\,{1\over 2}\,
\epsilon_{\mu\alpha\beta}\,f^{\alpha\beta}\,\big]\,=\,0\,\,,
\label{bc_with_n}
\eeq 
where  $\textfrak{n}$ is a constant that characterizes the boundary condition (the normal quantization condition considered so far corresponds to $\textfrak{n}=0$). As in \cite{Jokela:2015aha}, it is straightforward to prove that (\ref{bc_with_n}) 
is equivalent to require:
\beq
\lim_{\rho\to\infty}\,E\,=\,-i\,\textfrak{n}\,\lim_{\rho\to\infty}\,\big[\,\rho^{{\lambda\over 2}}\,a_y'\,\big]\,\,,\qquad\qquad
\lim_{\rho\to\infty}\,a_y\,=\,
i\,{\textfrak{n}\over \omega^2-k^2}\,\lim_{\rho\to\infty}\,\big[\,\rho^{{\lambda\over 2}}\,E'\,\big]\,\,.
\label{lim_E_ay}
\eeq
Notice that, even if the equations of motion  (\ref{eom_massive_E}) and   (\ref{eom_massive_ay})  for $E$ and $a_y$ are decoupled,  the mixed boundary conditions (\ref{lim_E_ay}) introduce a coupling between them. Therefore, to implement 
(\ref{lim_E_ay}) we have to study  the equation of motion of $a_y$, written in 
(\ref{eom_massive_ay}). Near the horizon $\rho\approx 0$ this equation reduces to:
\beq
a_y''\,+\,{\Lambda^2\over \rho^{7-p}}\,a_y\,=\,0\,\,,
\label{ay_nh_eq}
\eeq
which is just the same as (\ref{chi_i_eq}). For $p<5$ the solution of (\ref{ay_nh_eq}) is given by the right-hand-side of (\ref{Hankel_nh}). Moreover, for $p<4$ this solution behaves for low frequencies as:
\beq
a_y(\rho)\,=\,C\,\rho+C\,c_p\,\Lambda^{{2\over 5-p}}\,+\,\cdots\,\,,
\qquad\qquad
(p<4)\,\,,
\label{ay_nh_low}
\eeq
with $C$ being a constant. We now perform the two limits in the opposite order. For low frequencies (\ref{eom_massive_ay}) reduces to:
\beq
\partial_{\rho}\,\Big[\sqrt{\rho^{\lambda}+d^2-c^2}\,a_y'\Big]\,=\,0\,\,,
\eeq
whose integration is straightforward:
\beq
a_y(\rho)\,=\,a_y^{(0)}\,-\,C_3\,{\cal J}_{3}(\rho)\,\,,
\label{a_y_lowfreq_zeroT}
\eeq
where $a_y^{(0)}=a_y(\rho\to\infty)$, $C_3$ is a constant of integration, and ${\cal J}_3(\rho)$ is the following integral (for $\lambda>2$):
\beq
{\cal J}_3(\rho)\,=\,\int_{\rho}^{\infty}\,
{d\bar\rho\over
(\bar\rho^{\lambda}+d^2-c^2)^{{1\over 2}}}\,=\,{2\over \lambda-2}\,\rho^{1-{\lambda\over 2}}\,
F\Big(\,{1\over 2}, {1\over 2}-{1\over\lambda}; {3\over 2}\,-\,{1\over\lambda};
- {d^2-c^2\over \rho^{\lambda}}\,\Big)\,\,.
\eeq
Let us now expand $a_y(\rho)$ in powers of $\rho$. First, one can check that, for small $\rho$, 
the integral ${\cal J}_3(\rho)$ can be approximated as:
\beq
{\cal J}_3(\rho)\,\approx\,{\mu\over d}\,-\,{\rho\over \sqrt{d^2-c^2}}\,\,,
\eeq
where $\mu$ is the chemical potential (\ref{chemical-pot-zeroT}). Therefore, for small $\rho$, $a_y$ can be approximated as:
\beq
a_y(\rho)\,\approx\,a_y^{(0)}\,-\,C_3\,{\mu\over d}\,+\,C_3\,{\rho\over \sqrt{d^2-c^2}}\,\,.
\label{ay_low_nh}
\eeq
Let us now match (\ref{ay_nh_low}) and (\ref{ay_low_nh}). From the linear terms, we get the following relation between the constants $C$ and $C_3$:
\beq
C\,=\,{C_3\over  \sqrt{d^2-c^2}}\,\,.
\eeq
Using this relation, and identifying the constant terms in  (\ref{ay_nh_low}) and (\ref{ay_low_nh}), we get the following relation between $a_y^{(0)}$ and $C_3$:
\beq
a_y^{(0)}\,=\,\Big[\,{\mu\over d}\,+\,{c_p\over  \sqrt{d^2-c^2}}\,\Lambda^{{2\over 5-p}}
\,\Big]\,C_3\,\,.
\eeq

Let us now rewrite the boundary conditions (\ref{lim_E_ay}) at low frequency and momentum. From the expressions of $E$ and $a_y$ in this regime (eqs. (\ref{E-xi-prime_zeroT}) and (\ref{a_y_lowfreq_zeroT})), we conclude that they  behave in the UV as:
\beq
E'\big|_{\rho\to\infty}\approx (\omega^2-k^2)\,\rho^{-{\lambda\over 2}}\,C_1\,\,,
\qquad\qquad
a_y'\big|_{\rho\to\infty}\approx \rho^{-{\lambda\over 2}}\,C_3\,\,.
\eeq
Taking this into account, we can recast the boundary conditions for the alternative quantization as a relation between the constants $E^{(0)}$, $a_y^{(0)}$, $C_2$, and 
$C_3$. Indeed, let us define  $E_{\textfrak{n}}^{(0)}$ and $a_{y,\textfrak{n}}^{(0)}$ as:
\beq
E_{\textfrak{n}}^{(0)}\,\equiv E^{(0)}\,+\,i\,\textfrak{n}\, C_3\,\,,
\qquad\qquad
a_{y,\textfrak{n}}^{(0)}\,=\,a_y^{(0)}\,-\,i\,\textfrak{n}\,C_1\,\,.
\eeq
Then, (\ref{lim_E_ay}) is  equivalent to the conditions:
\beq
E_{\textfrak{n}}^{(0)}\,=\,a_{y,\textfrak{n}}^{(0)}\,=\,0\,\,.
\label{En0_ay_n0_def}
\eeq
The UV values $E_{\textfrak{n}}^{(0)}$, $\xi^{(0)}$, and $a_{y,\textfrak{n}}^{(0)}$ can be related to the constants $C_1$, $C_2$, and $C_3$. In matrix form this relation becomes:
\beq
\begin{pmatrix}
E_{\textfrak{n}}^{(0)}\\ \\ \xi^{(0)}\\ \\ a_{y,\textfrak{n}}^{(0)}
\end{pmatrix}
\,=\,
\begin{pmatrix}
 \Lambda^{{2\over 5-p}}\,c_p\,a_1\,-\,b_1
 &&&
 \Lambda^{{2\over 5-p}}\,c_p\,a_2\,-\,b_2
 &&&
 i\,\textfrak{n}
  \\
  {} & {} \\
 \Lambda^{{2\over 5-p}}\,c_p\,\tilde a_1\,-\, \tilde  b_1 &&& 
  \Lambda^{{2\over 5-p}}\,c_p\, \tilde  a_2\,-\, \tilde  b_2
   &&&
0\\ {} & {} \\
-i\,\textfrak{n} &&& 0  &&& {\mu\over d}+{c_p\over  \sqrt{d^2-c^2}}\,\Lambda^{{2\over 5-p}}
 \end{pmatrix}\,
 \begin{pmatrix}
C_1\\ \\ C_2\\ \\ C_3
\end{pmatrix}\,\,,
\label{matrix_alternative_quantization}
\eeq
where $a_1,a_2,b_1,b_2$ and $\tilde a_1,\tilde a_2,\tilde b_1,\tilde b_2$ are given in (\ref{b_a_coeff}) and (\ref{tilde_b_a_coeff}), respectively.
To have a non-trivial solution of the condition $E_{\textfrak{n}}^{(0)}= \xi^{(0)}=a_{y,\textfrak{n}}^{(0)}=0$ we must require that the determinant of the matrix in (\ref{matrix_alternative_quantization}) be zero. This leads to:
\bear
&&\Big(\Big[ \Lambda^{{2\over 5-p}}\,c_p\,a_1\,-\,b_1\Big] 
\Big[ \Lambda^{{2\over 5-p}}\,c_p\, \tilde  a_2\,-\, \tilde  b_2\Big]-
\Big[\Lambda^{{2\over 5-p}}\,c_p\,a_2\,-\,b_2\Big]
\Big[ \Lambda^{{2\over 5-p}}\,c_p\,\tilde a_1\,-\, \tilde  b_1\Big]\Big)\rc\rc
&&\qquad\qquad
\times
\Big[{\mu\over d}+{c_p\over  \sqrt{d^2-c^2}}\,\Lambda^{{2\over 5-p}}\Big]+
\textfrak{n}^2\,\Big[\tilde b_2-\Lambda^{{2\over 5-p}}\,c_p\,\tilde a_2\Big]\,=\,0\,\,.
\label{alternative_dispersion}
\eear
At leading order in frequency and momentum  this equation simplifies as:
\beq
b_1\,\tilde b_2+b_2^2+{d\, \textfrak{n}^2\over \mu}\,\tilde b_2\,=\,0\,\,.
\label{alternative_dispersion_leading}
\eeq
Since:
\beq
b_1\,\tilde b_2+b_2^2\,=\,{\gamma^2\over \lambda}\,
\big(d^2-c^2\big)^{{2\over \lambda}-2}\,
\Big[2(d^2-c^2)\,k^2\,-\, (\lambda d^2-2 c^2)\,\omega_0^2\Big]\,\,,
\eeq
then (\ref{alternative_dispersion_leading}) implies the following gapped dispersion relation:
\beq
\omega^2\,=\,\omega^2_0\,=\,{2(d^2-c^2)\over \lambda\, d^2-2\,c^2}\,k^2+
\Big({d\,\textfrak{n}\over \mu}\Big)^2
\,\,.
\eeq
In terms of the reduced mass parameter ${\bf m}$, defined  in (\ref{reduced_mass}), we have
\beq
\omega^2_0\,=\,2\,\,{1-{\bf m}^2\over \lambda\, -2\,{\bf m}^2}\,k^2
+\Big({d\,\textfrak{n}\over \mu}\Big)^2\,\,.
\label{anyonic-speed}
\eeq
One can also calculate the next order term in the dispersion relation. Indeed, one can check that $\omega=\omega_0+\delta \omega$, where $\delta\omega$ is given by:
\beq
\delta\omega\,=\,-\,
2^{{p-3\over 2(5-p)}}\,c_p\,{\lambda\,\over \mu}\,
{(1-{\bf m}^2)^{{6-p\over 5-p}-{1\over 2}}\over (\lambda-2 {\bf m}^2)
^{{7-p\over 2(5-p)}+1}}
\,\,k^{{7-p\over 5-p}}\,\,
\,-\,
{\textfrak{n}^2\over \lambda-2}\,{c_p\over k\mu^3}\,
\Big({\gamma\over \mu}\Big)^{{\lambda\over 2}}\,
{\omega_0^{{p-3\over 5-p}}\over 
(1-{\bf m}^2)^{{1\over 2}+{\lambda\over 4}}}\,\,.
\eeq

\begin{figure}[ht]
\center
 \includegraphics[width=0.45\textwidth]{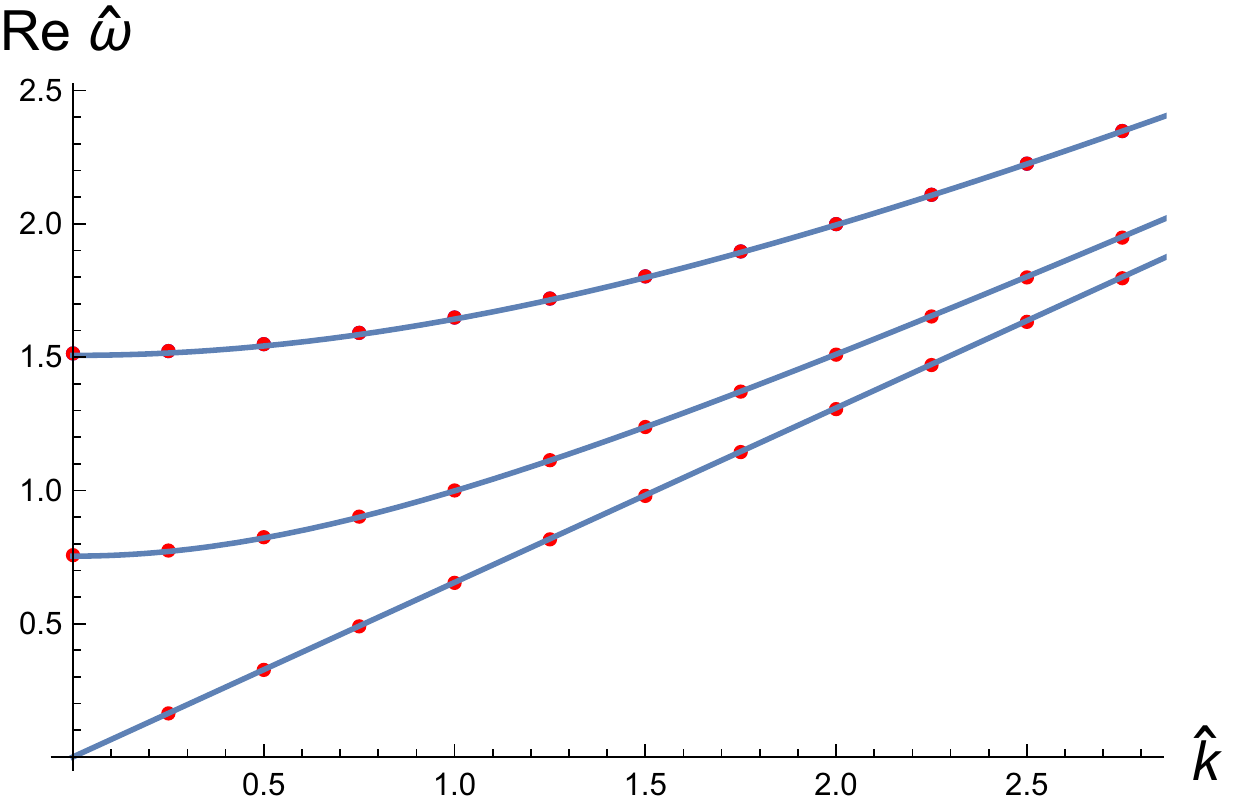}
 \includegraphics[width=0.45\textwidth]{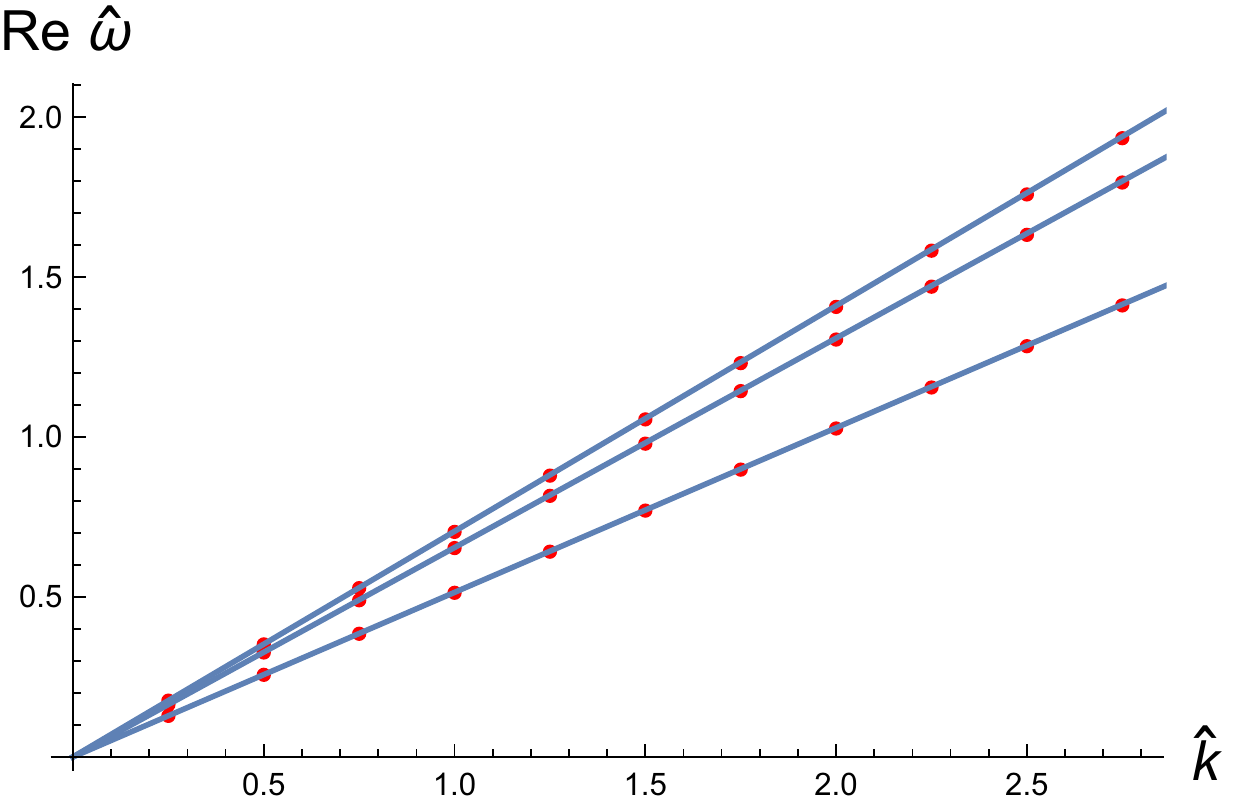}
 \caption{We plot the dispersions in the D3-D5 model $(p=3,\lambda=4)$ at $\hat d = 10^6$ and $\hat B=3\cdot 10^3$. In both plots the red points stand for numerical results whereas the blue curves are the analytic from (\ref{anyonic-speed-B}); we emphasize that the analytic result (\ref{anyonic-speed-B}) is an educated guess, but reproduces the numerics precisely. (left) We vary the quantization parameter $\textfrak{n}=0,\frac{1}{2}\textfrak{n}_{crit},\textfrak{n}_{crit}$ (top-down) at fixed $\frac{m}{\mu}=0.5$. (right) The quantization parameter is chosen to be critical $\textfrak{n}=\textfrak{n}_{crit}$. Different lines correspond to varying $\frac{m}{\mu}=0.1,0.5,0.8$ (top-down).}
\label{Zero_sound_velocity_anyon}
\end{figure}

It was noticed  in \cite{Jokela:2015aha} for the massless embeddings that the effect of the alternative quantization is equivalent to switching on a magnetic field  traversing the $x^1x^2$ plane. Actually, it was found in \cite{Jokela:2015aha} that the effect of a magnetic field $B$ effectively changes the parameter $\textfrak{n}$ as
\beq
\textfrak{n}\to \textfrak{n}-{B\over d}\,\,.
\label{substitution-rule}
\eeq
In the present massive case we cannot verify analytically the substitution rule (\ref{substitution-rule}) since the embedding function $z(\rho)$ is not a cyclic variable in the presence of a $B$ field. Therefore, we conjecture that the dispersion relation of the zero sound with general anyonic boundary conditions and magnetic field is given (at leading order) by:
\beq
\omega^2_0\,=\,2\,\,{1-{\bf m}^2\over \lambda\, -2\,{\bf m}^2}\,k^2
+{1\over \mu^2}\,\big(d\,\textfrak{n}-B\big)^2\,\,.
\label{anyonic-speed-B}
\eeq
Thus, the spectrum is generically gapped for non-vanishing $B$ and $\textfrak{n}$.  However, it can be made gapless by adjusting the alternative quantization parameter $\textfrak{n}$ to the critical value:
\beq
\textfrak {n}_{crit}\equiv {B\over d}\,\,.
\eeq
This particular case corresponds to one, where the anyonic fluid experiences zero net effective magnetic field, thus the resulting spectrum is also gapless.
In Fig.~\ref{Zero_sound_velocity_anyon} we compare the  results obtained from the  numerical integration of the fluctuation equations to our analytic formula (\ref{anyonic-speed-B}). We see that the agreement is very good and, in particular, the numerics confirm that the spectrum becomes gapless at $\textfrak{n}=\textfrak{n}_{crit}$.

\section{Finite temperature}
\label{finiteT}

Let us now consider the  D$p$-D$q$ intersections  $(n\, |\, p\perp q)$ at non-zero temperature and magnetic field. First, we introduce a more convenient system of coordinates. Let us represent the  different components of  the Cartesian  coordinates $\vec y$ transverse to the D$p$-brane  as:
\bear
&&y^{m}\,=\,r\,\cos\theta\,\eta^{m}\,\,,
\qquad\qquad
m=1,\cdots, q-n\,\,,\rc\rc
&&y^{l}\,=\,r\sin\theta\,\xi^{l}\,\,,
\qquad\qquad\,\,\,\,\,\,
l=q-n+1\,\cdots, 9-p\,\,,
\eear
where $\eta^{m}$ and $\xi^{l}$ satisfy:
\beq
\sum_{m=1}^{q-n}\big(\eta^{m}\big)^2\,=\,
\sum_{l=q-n+1}^{9-p}\big(\xi^{l}\big)^2\,=\,1\,\,.
\eeq
Clearly, the $\eta^{m}$ ($\xi^{l}$) are the coordinates of a $(q-n-1)$-sphere ($(8+n-p-q)$-sphere). As:
\beq
\sum_{l=q-n+1}^{9-p}\big(y^{l}\big)^2\,=\,r^2\,\sin^2\theta\,\,,
\qquad\qquad
\sum_{m=1}^{q-n}\big(y^{m}\big)^2\,=\,r^2\,\cos^2\theta\,\,,
\eeq
we  identify the coordinates $z$ and $\rho$ used so far with:
\beq
z\,=\,r\,\sin\theta\,\,,
\qquad\qquad
\rho\,=\,r\,\cos\theta\,\,.
\label{z-rho-r-theta}
\eeq
It is straightforward to check that
\beq
d\vec y\cdot d\vec y\,=\,
dr^2\,+\,r^2\,\big[d\theta^2\,+\,\cos^2\theta\,d\Omega^2_{\parallel}\,+\,
\sin^2\theta\,d\Omega^2_{\perp}\big]\,\,,
\eeq
where $d\Omega^2_{\parallel}=d\Omega^2_{q-n-1}$ is the line element of the $(q-n-1)$-sphere of the D$q$-brane worldvolume and  $d\Omega^2_{\perp}=d\Omega^2_{8+n-p-q}$ is the metric of the $(8+n-p-q)$-sphere transverse to the D$q$-brane. The ten-dimensional metric of a black D$p$-brane in these coordinates is:
\beq
ds^2_{10}=\Big({r\over R}\Big)^{{7-p\over 2}}\big[-f_p(r)\,dt^2\,+\, d \vec  x^{\,2}\big]+\Big({R\over r}\Big)^{{7-p\over 2}}\Big[{dr^2\over f_p(r)}\,+\,r^2\big(
d\theta^2+\cos^2\theta\,d\Omega^2_{\parallel}+
\sin^2\theta\,d\Omega^2_{\perp}\big)\Big]\,\,,
\eeq
where $R$ is a constant radius and the blackening factor $f_p$ is:
\beq
f_p(r)\,=\,1\,-\,\Big({r_h\over r}\Big)^{7-p}\,\,,
\label{fp}
\eeq
and $r_h$ is the horizon radius, that is related to the temperature as follows:
\beq
T\,=\, {7-p\over 4\pi }\,r_h^{{5-p\over 2}}\,\,.
\label{T_rh}
\eeq
Let us consider a  D$q$-brane probe extended along $t, x^1\ldots, x^n, r$ and the $(q-n-1)$-sphere. If the brane is at a fixed point in the transverse sphere and we take $\theta=\theta(r)$, the induced metric is (for $R=1$):
\beq
ds^2_{q+1}\,=\,r^{{7-p\over 2}}\,\big[-f_p\,dt^2+(dx^1)^2+\cdots +(dx^n)^2\big]\,+\,
r^{{p-7\over 2}}\,\big[(1+r^2\,f_p\,\dot\theta^2){dr^2\over f_p}+
r^2\cos\theta^2\,d\Omega^2_{\parallel}\big]\,\,,
\eeq
with $\dot\theta=d\theta/dr$.  In what follows we will take $n$, $p$ and $q$ to be related as in  (\ref{n_p_q}). Moreover,  we will  add a magnetic field in the $x^1\,x^2$ directions. The Ansatz for the worldvolume gauge field strength in this case becomes:
\beq
F\,=\,\dot A_t\,dr\wedge dt\,+\,B\,dx^1\wedge dx^2\,\,.
\label{F_charge_B}
\eeq
The DBI Lagrangian density for this Ansatz is:
\beq
{\cal L}\,=\,-{\cal N}\,\sqrt{H}\,(\cos\theta)^{{\lambda\over 2}}\,
\sqrt{1-\dot A_t^2\,+\,r^2\,f_p\,\dot\theta^2}\,\,,
\label{cal_L_theta_B}
\eeq
where $\lambda$ is given by (\ref{lambda_SUSY}) and  we have introduced a new function $H$, defined as:
\beq
H\,\equiv\,r^{\lambda}\,+\,r^{\lambda+p-7}\,B^2\,\,.
\label{H_def}
\eeq
In this Lagrangian $A_t$ is a cyclic variable. Its equation of motion can be integrated once to give:
\beq
{(\cos\theta)^{{\lambda\over 2}}\,\sqrt{H}\,\dot A_t\over 
\sqrt{1-\dot A_t^2\,+\,r^2\,f_p\,\dot\theta^2}}\,=\,d\,\,,
\eeq
with $d$ being an integration constant. From this last equation we obtain $\dot A_t$ as:
\beq
\dot A_t\,=\,d\,{\sqrt{1\,+\,r^2\,f_p\,\dot\theta^2}\over
\sqrt{d^2+H\,\big(\cos\theta\big)^{\lambda}}}\,\,.
\label{dot_At}
\eeq
After eliminating $A_t$, we find the following equation for the embedding function $\theta(r)$:
\beq
\partial_r\,\Bigg[r^2\,f_p\,
\sqrt{
{d^2+H\,(\cos\theta)^{\lambda}\over
1\,+\,r^2\,f_p\,\dot\theta^2}}\,\,\dot\theta\Bigg]+
{\lambda\over 2}\,H\,(\cos\theta)^{\lambda-1}\,\sin\theta\,
{\sqrt{1\,+\,r^2\,f_p\,\dot\theta^2}\over
\sqrt{d^2+H\,\big(\cos\theta\big)^{\lambda}}}\,=\,0\,\,.
\label{embbeding_equation}
\eeq
The equation of motion (\ref{embbeding_equation}) has explicit dependence on the blackening factor $f_p$, which has factors of the horizon radius $r_h$. This feeds in temperature dependence via (\ref{T_rh}).
The horizon radius $r_h$ can be scaled out by an appropriate change of variables, followed by a redefinition of the density $d$ and the magnetic field $B$. Indeed, let us define the reduced radial variable $\hat r$ as follows:
\beq
\hat r\,=\,{r\over r_h}\,\,.
\eeq
It is then straightforward to verify  that, in terms of $\hat r$, the embedding equation is just (\ref{embbeding_equation})
with $r_h=1$ and $d$ and $B$ substituted by the scaled quantities $\hat d$ and $\hat B$, defined as:
\beq
\hat d\,=\,{d\over r_h^{{\lambda\over 2}}}\,\,,
\qquad\qquad
\hat B\,=\,{B\over r_h^{{7-p\over 2}}}\,\,.
\label{hat_d_hat_B_def}
\eeq
We will integrate (\ref{embbeding_equation}) by imposing that the D$q$-brane intersects the horizon $r=r_h$  at some value $\theta_h\equiv\theta(r=r_h)$, \ie,  we will require that our embedding is a black hole embedding.\footnote{It is interesting to write the zero temperature results of section \ref{zero_temperature_embeddings} in terms of the $(r,\theta)$ variables used in this section. Let $\theta_*$ be the angle at the horizon when $T=0$, \ie, 
$\theta_*=\theta(r=0)$. Then, $\tan \theta_*\,=\,c/ \sqrt{d^2-c^2}$. Other useful relations at zero temperature are
$m=\gamma\,d^{{2\over \lambda}}\,\tan \theta_*\, (\cos\theta_*)^{{2\over \lambda}}$
 and
$\mu=\gamma\,d^{{2\over \lambda}}\,/(\cos\theta_*)^{1-{2\over \lambda}}$, which imply $\sin\theta_*\,=\,m/ \mu$.} At the UV $r\to\infty$ the function $\theta(r)$ behaves generically as:
\beq
\theta(r)\sim {m\over r}\,+\,{{\cal C}\over r^{{\lambda\over 2}}}\,+\,\cdots\,=\,
{\hat m\over \hat r}\,+\,{\hat {\cal C}\over \hat r^{{\lambda\over 2}}}\,+\,\cdots
\qquad\qquad
(r\to\infty)\,\,,
\label{theta_UV}
\eeq
where $m$ and ${\cal C}$ are related to the mass and condensate, respectively. 
Notice that we have introduced in (\ref{theta_UV}) the scaled quantities $\hat m$ and $\hat {\cal C}$, related to $m$ and ${\cal C}$ as:
\beq
\hat m\,=\,{m\over r_h}\,\,,
\qquad\qquad
{\hat {\cal C}}\,=\,{{\cal C}\over r_h^{{\lambda\over 2}}}\,\,.
\eeq
It is also interesting to write the  chemical potential $\mu$ in terms of the scaled quantities. We have:
\beq
\hat \mu\,=\,{ \mu\over r_h}\,\,,
\eeq
where $\hat \mu$ is given by the following integral:
\beq
\hat\mu\,=\,\hat d\,\int_1^{\infty} d\hat r
{\sqrt{1+\hat r^{p-5}\,(\hat r^{7-p}-1)\,\big({d\theta\over d\hat r}\big)^2}\over
\sqrt{\hat d^2+(\hat r^{\lambda}+\hat r^{\lambda+p-7}\hat B^2)(\cos\theta)^{\lambda}}}\,\,.
\eeq
Notice that $\hat m/\hat \mu=m/\mu$, \ie, the horizon radius $r_h$ drops out when one computes the mass/chemical potential ratio as both of the quantities have the same dimension.

\subsection{Charge susceptibility}
Let us consider  now the case $B=0$ and compute the charge susceptibility $\chi$, which is defined as:
\beq
\chi\,=\,{\partial\rho_{ch}\over \partial\mu}\,\,.
\eeq
Taking into account that the charge density $\rho_{ch}$ is related to $d$ as $\rho_{ch}\,=\,{\cal N}\,d$, we can rewrite the last expression as:
\beq
\chi^{-1}\,=\,{1\over {\cal N}}\,{\partial \mu\over \partial d}=\frac{1}{{\cal N}}\int_{r_h}^{\infty}\,
{\partial \dot A_t\over \partial d} \,dr\,\,.
\label{charge_sus_def}
\eeq
By a direct calculation using (\ref{dot_At}) for $B=0$  , we get:
\beq
{\partial \dot A_t\over \partial d}\,=\,\sqrt{\Delta}\,
{r^{{\lambda\over 2}}\,\big(\cos\theta\big)^{{\lambda\over 2}}\over
d^2+r^{\lambda}\,\big(\cos\theta\big)^{\lambda}}\,\Bigg[
1+d\Bigg({\lambda\over 2}\,\tan\theta\,{\partial \theta\over \partial d}\,+\,
{r^2\,f_p\,\dot\theta\over \Delta}\,
{\partial\dot \theta\over \partial d}\Bigg)\Bigg]\,\,,
\eeq
where $\Delta$ is defined as:
\beq
\Delta\,\equiv\,1-\dot A_t^{2}+r^2\,f_p\,\dot\theta^2\,=\,
r^{\lambda}\,(\cos\theta)^{\lambda}\,
{1+r^2\,\,f_p\,\dot\theta^2\over d^2+r^{\lambda}\,(\cos\theta)^{\lambda}}\,\,.
\eeq
Therefore, the charge susceptibility can be written as:
\beq
\chi^{-1}\,=\,{1\over {\cal N}}\,\int_{r_h}^{\infty}\,dr\,
\sqrt{\Delta}\,
{r^{{\lambda\over 2}}\,\big(\cos\theta\big)^{{\lambda\over 2}}\over
d^2+r^{\lambda}\,\big(\cos\theta\big)^{\lambda}}\,\Bigg[
1+d\Bigg({\lambda\over 2}\,\tan\theta\,{\partial \theta\over \partial d}\,+\,
{r^2\,f_p\,\dot\theta\over \Delta}\,
{\partial\dot \theta\over \partial d}\Bigg)\Bigg]\,\,.
\label{sus_non-zeroT}
\eeq
Let us consider some particular cases of (\ref{sus_non-zeroT}).  First of all, we consider the  massless case, in which  $\theta=0$ and the integral in (\ref{sus_non-zeroT})  can be performed explicitly. We get:
\beq
{\cal N}\,\chi^{-1}\,=\,{2\over \lambda-2}\,r_h^{1-{\lambda\over 2}}\,
F\Big({3\over 2}, {1\over 2}-{1\over \lambda};{3\over 2}\,-\,{1\over \lambda};-{d^2\over r_h^{\lambda}}\Big) \ , \ \ m = 0 \ .
\eeq
Another interesting limiting case is when $T=0$ . In this case we can obtain $\chi$ without using (\ref{sus_non-zeroT}). Indeed, we can compute the derivative of $\mu$ from the second equation in  (\ref{c_d_massive}). Computing $\partial d/\partial \mu$ for constant $m$, we get:
\beq
{\partial d\over \partial \mu}\,=\,
\gamma^{-{\lambda\over 2}}\,\big(\mu^2-m^2\big)^{{\lambda-6\over 4}}\,
\Big[{\lambda\over 2}\,\mu^2\,-\,m^2\Big] \ , \ \ T = 0 \ ,
\eeq
where $\gamma$ is the constant defined in (\ref{gamma_def}). 
Then, it follows that:
\beq
\chi\,=\,{\cal N}\,
\gamma^{-{\lambda\over 2}}\,
{{\lambda\over 2}\,\mu^2\,-\,m^2\over 
\big(\mu^2-m^2\big)^{{6-\lambda\over 4}}} \ \ , \ T = 0 \ .
\label{sus_zeroT_SUSY}
\eeq
Notice that (for  $\lambda<6$), the zero temperature susceptibility blows up when $m=\mu$.

\begin{figure}[ht]
\center
 \includegraphics[width=0.7\textwidth]{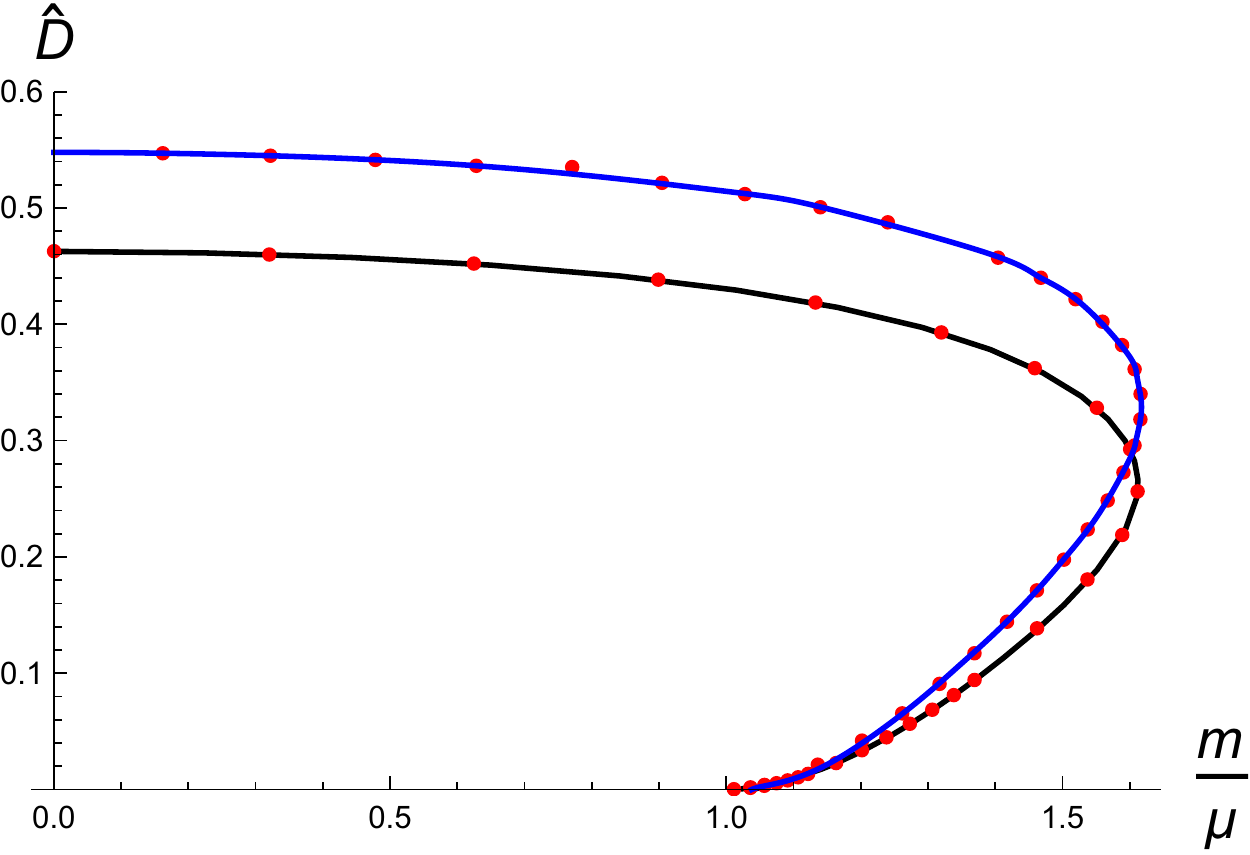}
 \caption{We plot the rescaled diffusion constant  $\hat D\,=\, r_h^{{5-p\over 2}}\,D$ for the D2-D6 model at $\hat d=1$. The solid blue curve is obtained purely from the background for $\hat B=0$, \emph{i.e.}, by evaluating the integral (\ref{Diffusion_Einstein}). The red points, on the other hand, are numerical results from the fluctuation analysis by solving the coupled equations of motion in (\ref{E_eom_momentum_B})-(\ref{ay_eom_momentum_B}) for zero magnetic field. We see that the two different methods agree perfectly.
 The lower  black curve is the result obtained from (\ref{Diffusion_Einstein_B}) for $\hat B=1/2$, while the red points are the result of the numerical integration of the fluctuation equations for this value of $\hat B$. Again, both methods agree perfectly. }
\label{diffusion_zero_B}
\end{figure}

\subsection{Einstein relation}
The diffusion constant $D$  can be related  to the charge susceptibility $\chi$ by means of the so-called Einstein relation, which reads:
\beq
D\,=\,\sigma\,\chi^{-{1}}\,\,,
\label{Einstein_rel}
\eeq
where $\sigma$ is the DC conductivity. The value of $\sigma$ can be extracted from the analysis of the two-point correlators of the transverse currents. This analysis is carried out in appendix \ref{appendixC}. The final result for $\sigma$ is written in (\ref{sigma}). Plugging this value of $\sigma$  and the susceptibility written in (\ref{sus_non-zeroT}) into 
(\ref{Einstein_rel}), we arrive at the following expression for $D$:
\bear
&&D\,=\,r_h^{{p-7\over 2}}\,
\sqrt{r_h^{\lambda}(\cos\theta_h)^{\lambda}+d^2}\,
\int_{r_h}^{\infty}\,dr\,
\sqrt{\Delta}\,
{r^{{\lambda\over 2}}\,\big(\cos\theta\big)^{{\lambda\over 2}}\over
d^2+r^{\lambda}\,\big(\cos\theta\big)^{\lambda}}\nonumber \\
&&\qquad\qquad\qquad\qquad\times
\Bigg[
1+d\Bigg({\lambda\over 2}\,\tan\theta\,{\partial \theta\over \partial d}\,+\,
{r^2\,f_p\,\dot\theta\over \Delta}\,
{\partial\dot \theta\over \partial d}\Bigg)\Bigg]\,\,.
\label{Diffusion_Einstein}
\eear
Let us now extract the low temperature behavior of $D$  by using the $T=0$ susceptibility written in (\ref{sus_zeroT_SUSY}). As 
$\sigma\sim {\cal N}\,d\,r_h^{{p-7\over 2}}$ for low $T$, we get:
\beq
D\approx 2\,\gamma^{{\lambda\over 2}}\,
{\big(\mu^2-m^2\big)^{{6-\lambda\over 4}}\over 
\lambda\,\mu^2-2\,m^2}\,\,d\,r_h^{{p-7\over 2}}\,\,\,,
\qquad\qquad\quad
(T\sim 0).
\label{Diffusion_Einstein_lowT}
\eeq
The expression (\ref{Diffusion_Einstein}) for $D$ can be compared with the values obtained by analyzing the spectrum of diffusive modes of the probe in the hydrodynamical regime (see section \ref{Fluctuations} below).
This comparison is shown in Fig.~\ref{diffusion_zero_B} for the D2-D6 intersection. We have obtained a very good agreement between the two methods in all intersections studied.

\subsection{Fluctuations}
\label{Fluctuations}

Let us now consider a fluctuation of the embedding angle and of the gauge field of the form:
\beq
\theta(x^{\mu}, r)\,=\,\theta_0(r)+\zeta(x^{\mu}, r)\,\,,
\qquad\qquad
A(x^{\mu}, r)\,=\,A^{(0)}(r)\,+\,a(x^{\mu}, r)\,\,,
\label{fluct_gauge_scalar}
\eeq
where $a(x^{\mu}, r)=a_{\nu}(x^{\mu}, r)dx^{\nu}$ and $A^{(0)}=A_\nu^{(0)}\,dx^{\nu}=A_t\,dt+B\,x^{1}\,dx^2$ is the one-form for the unperturbed gauge field. We will choose the gauge in which $a_r=0$ and we will consider fluctuation fields $a_\nu$ and $\zeta$ depending only on  $r$, $t$, and $x^1$. In this case it is possible to restrict to the case in which $a_{\nu}\not=0$ only when $\nu=t$, $x^{1}\equiv x$, and $x^2\equiv y$. In appendix \ref{appendixB} we obtain the Lagrangian density for the fluctuations and we perform a detailed analysis of the corresponding equations of motion. This analysis is performed in momentum space. Accordingly, let us Fourier transform $a_{\nu}$ and $\zeta$ as:
\bear
a_\nu(r, t, x) &  = & \int {d\omega\,dk\over (2\pi)^2}\,a_\nu(r, \omega, k)\,e^{-i\omega\,t\,+\,i k x}\nonumber \\
\zeta(r, t, x) & = & \int {d\omega\,dk\over (2\pi)^2}\,\zeta(r, \omega, k)\,e^{-i\omega\,t\,+\,i k x}\,\,.
\label{Fourier_a_zeta}
\eear
At very low temperature the numerical analysis of the coupled fluctuation equations (\ref{E_eom_momentum_B}), (\ref{zeta_eom_momentum_B}), and (\ref{ay_eom_momentum_B}) allows to find sound modes,  \ie, the zero sound. The corresponding dispersion relation is given in terms of the rescaled frequency and momentum $\hat \omega$ and $\hat k$, related to $\omega$ and $k$ as:
\beq
\hat \omega\,=\,{\omega\over  r_h^{{5-p\over 2}}}\,\,,
\qquad\qquad
\hat k\,=\,{k\over  r_h^{{5-p\over 2}}}\,\,.
\label{hat_omega_k}
\eeq
\begin{figure}[ht]
\center
 \includegraphics[width=0.80\textwidth]{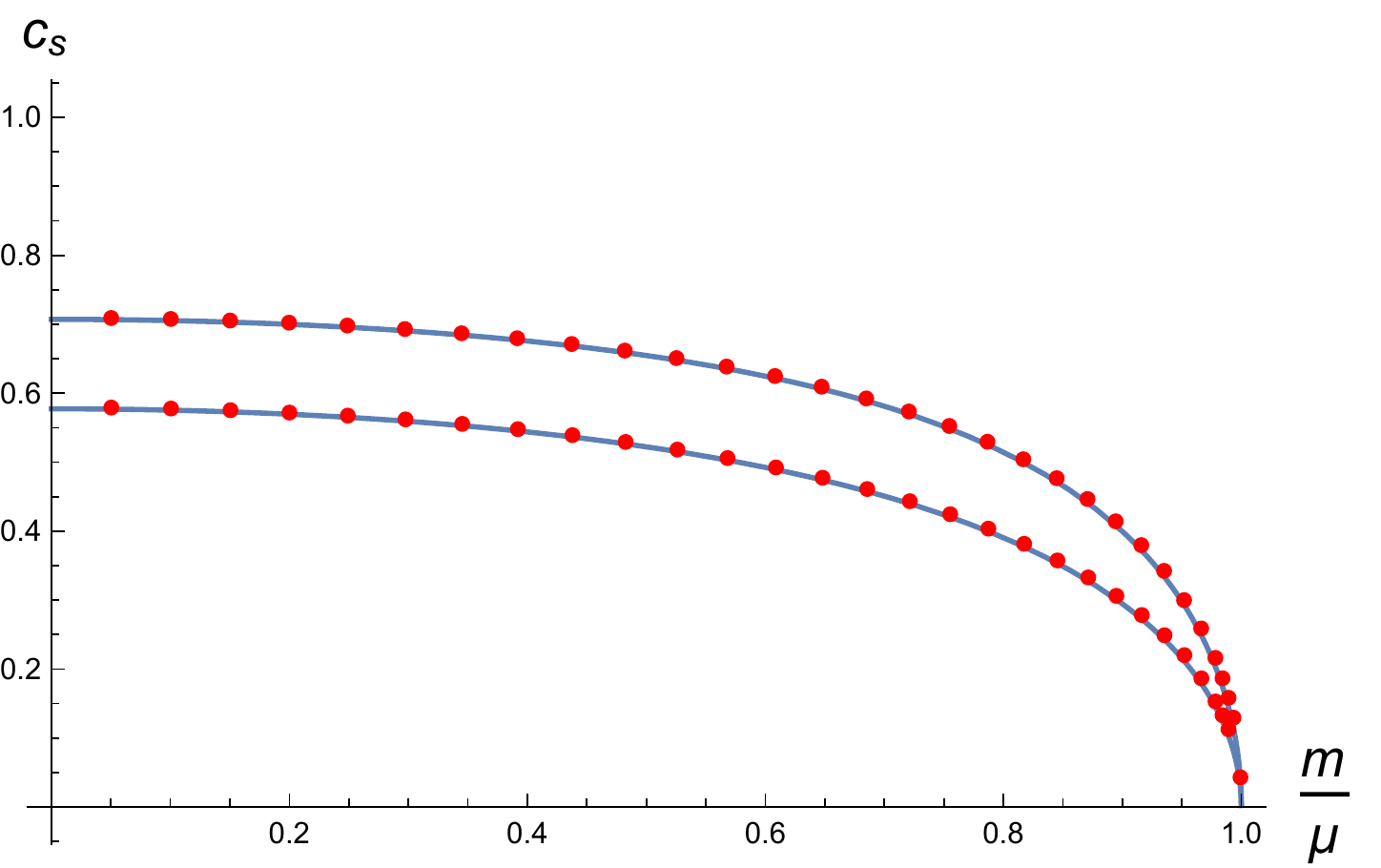}
 \caption{We demonstrate how well the analytics and numerics match also for the non-conformal brane intersections, by focusing on $p=2$ and at low temperature and vanishing magnetic field. We depict the speed of zero sound for two intersections D2-D4 ($\lambda=4$) and D2-D6 ($\lambda=6$) at $\hat d=10^\lambda$. The numerics are represented by red points whereas the analytics, following (\ref{speed_zero_sound}), are continuous curves. Lower dataset is for $\lambda=6$ and higher for $\lambda=4$. In extracting the slope we kept $\frac{\hat k}{\hat d^\frac{5-p}{\lambda}}=0.01$.}
\label{Nonconformal_Zero_sound_velocity}
\end{figure}
For vanishing magnetic field the numerical results are in very good agreement with the analytic equations of section \ref{zero_sound_section}, as it was illustrated already in Fig.~\ref{zero_sound} for the conformal D3-D5 intersection. This agreement is confirmed in Fig.~\ref{Nonconformal_Zero_sound_velocity} for the non-conformal cases D2-D4 and D2-D6.

\begin{figure}[ht]
\center
 \includegraphics[width=0.80\textwidth]{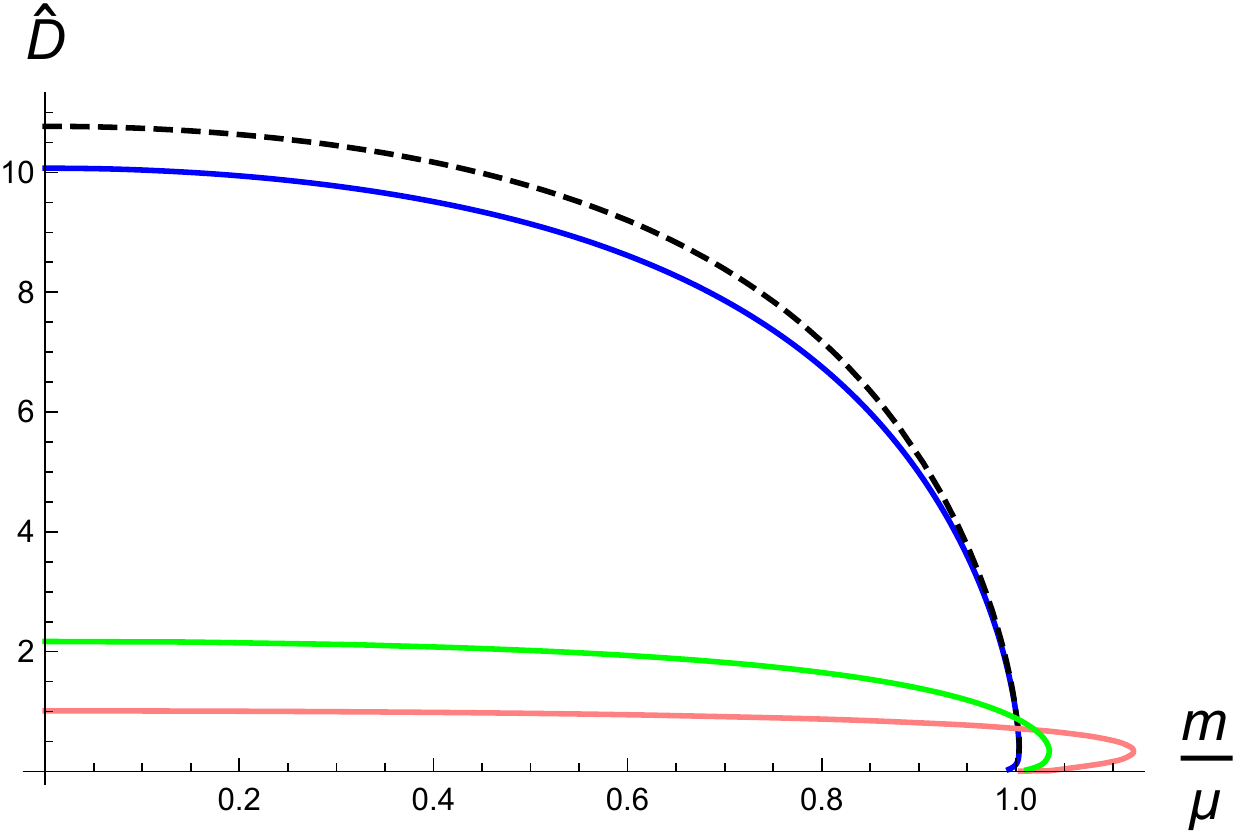}
 \caption{We depict the diffusion constant at various temperatures  and vanishing magnetic field for the D2-D6 model $(p=2,\lambda=6)$ as obtained by solving the fluctuation equations. The continuous curves correspond to $\hat d=10^1,10^2,10^4$ (bottom-up). As a reference we have also included the $T\to 0$ analytic result (\ref{hat_D_Einstein_lowT})  for $\hat d=10^4$,  depicted as a dashed black curve, showing how well the $\hat d=10^4$ numerical curve is converging to it. Higher values of $\hat d$ would be overlapping even more.}
\label{D2D6_diffusion}
\end{figure}

At higher temperatures the system is in a hydrodynamic regime, in which the dominant mode is a diffusion mode with purely imaginary frequency. The spectrum of these diffusion modes can be written in terms of the rescaled frequency and momentum defined in (\ref{hat_omega_k}) as:
\beq
\hat \omega\,=\,-i\,\hat D\,\hat k^{2}\,\,,
\eeq
where $\hat D$ is the rescaled diffusion constant, related to $D$ as:
\beq
\hat D\,=\, r_h^{{5-p\over 2}}\,D\,\,.
\label{hatD_def}
\eeq
The value of $\hat D$ predicted by Einstein relation can be straightforwardly obtained from (\ref{Diffusion_Einstein}). Indeed, one  must  simply take $r_h=1$ and change $d$ by $\hat d$ in (\ref{Diffusion_Einstein}).  The low temperature limit of $\hat D$ can also be obtained easily from (\ref{Diffusion_Einstein_lowT}). We get:
\beq
\hat D\approx 2\, \gamma^{{\lambda\over 2}}\,
{\big(\hat\mu^2-\hat m^2\big)^{{6-\lambda\over 4}}\over 
\lambda\,\hat\mu^2-2\,\hat m^2}\,\,\hat d\,\,\,,
\qquad\qquad\qquad
(T\sim 0).
\label{hat_D_Einstein_lowT}
\eeq
In Fig.~\ref{D2D6_diffusion} we show the temperature dependence of $\hat D$ for the D2-D6 model. The temperature is decreased by increasing $\hat d$. The results displayed in Fig.~\ref{D2D6_diffusion} indeed show that $\hat D$ approaches the value written in (\ref{hat_D_Einstein_lowT}) as $\hat d\to \infty$.

\begin{figure}[ht]
\center
 \includegraphics[width=0.45\textwidth]{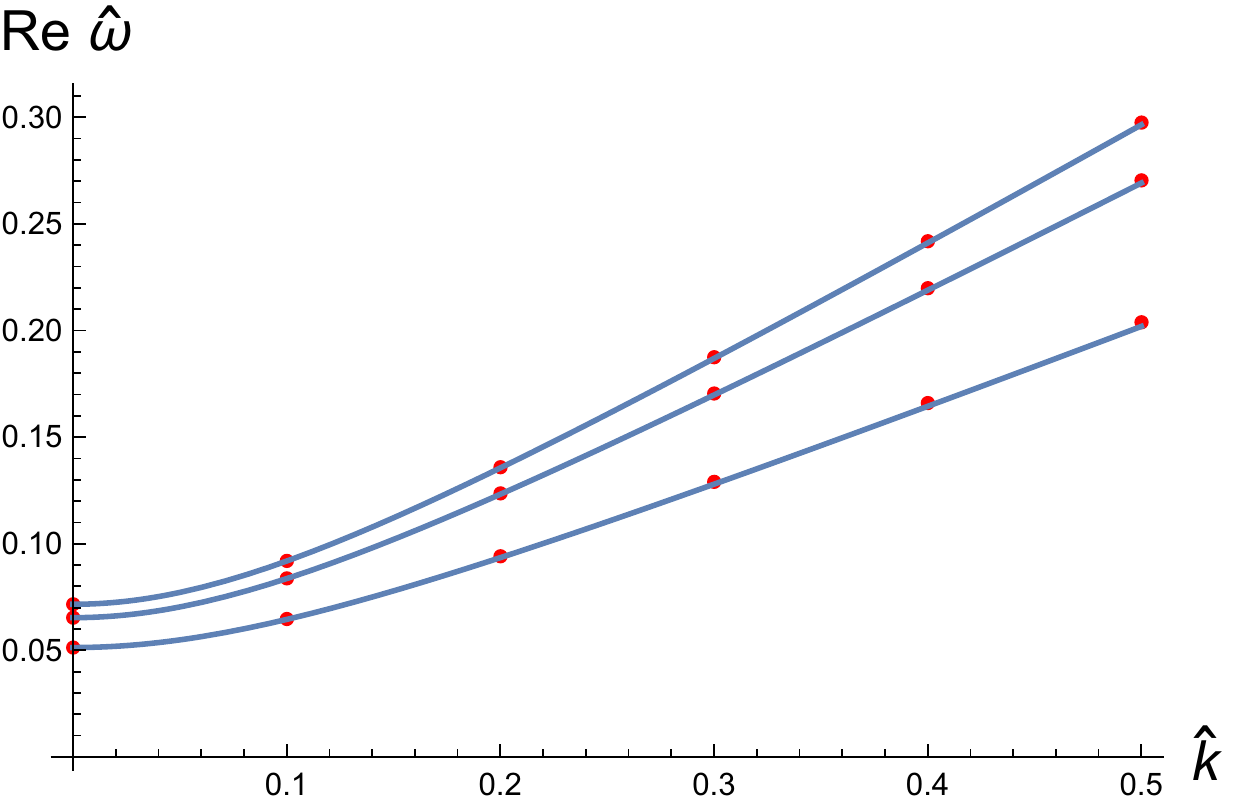}
 \includegraphics[width=0.45\textwidth]{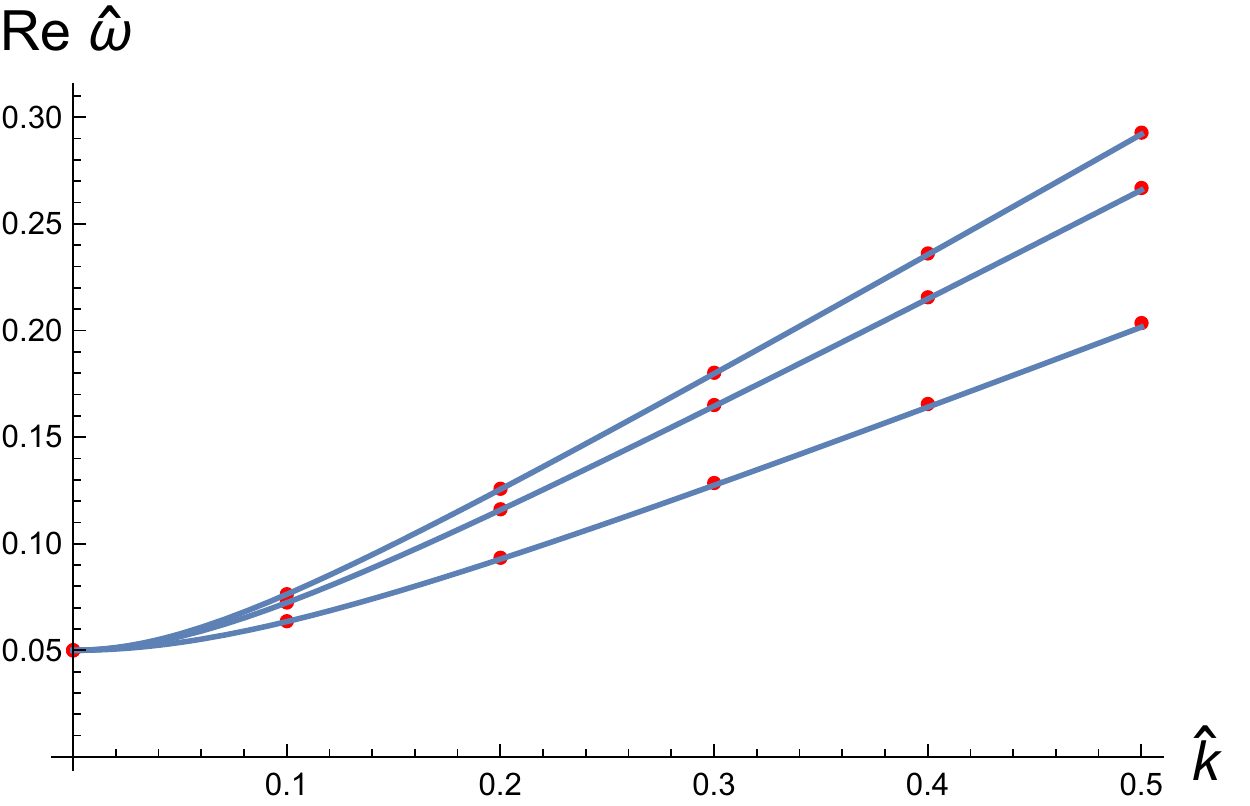}
 \caption{We present numerical evidence that our conjecture for the dispersions at low temperature (\ref{zero-sound-speed-B}) is supported at finite magnetic field strength. We demonstrate this in the case of the D2-D6 model ($p=2$,$\lambda=6$) at $\hat B=10$ for two different cases: in the canonical and in the grand canonical ensemble. From the latter case we clearly find that the mass gap of the zero sound is indeed independent of the mass ($\hat m$) of the fundamentals. Different curves in both panels correspond to $\frac{m}{\mu}=0.1,0.5,0.8$ (top-down). (Left) We keep the charge density fixed $\hat d=10^6$. (Right) We keep the chemical potential fixed $\hat\mu=200$.}
\label{Zero_sound_B}
\end{figure}

Let us now consider the dependence on the magnetic field. The results of section \ref{alternative_quantization} (and those of refs. \cite{Jokela:2012vn,Brattan:2012nb,Jokela:2015aha}) strongly suggest  that the spectrum of the zero sound is gapped and that the gap is just $B/\mu$. Therefore we are led to conjecture the following expression of the leading order dispersion relation of the zero sound:
\beq
\omega^2_0\,=\,2\,\,{1-{\bf m}^2\over \lambda\, -2\,{\bf m}^2}\,k^2
+{B^2\over \mu^2}\,\,,
\label{zero-sound-speed-B}
\eeq
where we have just added the gap to the gapless value of $\omega_0^2$. Notice that (\ref{zero-sound-speed-B}) implies that the gap is independent of the quark mass $m$ for fixed chemical potential $\mu$. We have explicitly verified this feature numerically in Fig.~\ref{Zero_sound_B} for the D2-D6 system.

Let us next analyze the dependence of the diffusion constant on the magnetic field $B$. In order to write the expression of $D$ which follows from the Einstein relation, we need to know the value of the DC conductivity $\sigma$ when $B\not=0$. In principle, this conductivity could be obtained from the analysis of the transverse correlators, as was done in appendix  \ref{appendixC} for $B=0$. However, the fluctuation equations couple the transverse and longitudinal modes when $B\not=0$ and and it is not clear to us how to deal with this coupling. For this reason we we have computed $\sigma$ by applying the method of ref. \cite{Karch:2007pd}. The details of this calculation are explained in appendix \ref{appendixD}. The final result for $\sigma$ is:
\beq
\sigma\,=\,{\cal N}\,{
\sqrt{r_h^{\lambda}(1+r_h^{p-7}\,B^2)\,(\cos\theta_h)^{\lambda}\,+d^2}\over r_h^{7-p}\,+\,B^2}\,\,
r_h^{{7-p\over 2}}\,\,.
\label{sigma_B}
\eeq
It is now straightforward to write down the expression of $D$ which follows from (\ref{Einstein_rel}). Indeed, 
let us define $\Delta_B$ as:
\beq
\Delta_B\,=\,{H\,(\cos\theta)^{\lambda}\,(1+r^2\,f_p\,\dot\theta^2)\over d^2+H\,(\cos\theta)^{\lambda}}\,\,,
\eeq
where $H$ is the quantity defined in (\ref{H_def}).  Then, the Einstein relation gives the following value of the diffusion constant:
\bear
&&D\,=\,{
\sqrt{r_h^{\lambda}(1+r_h^{p-7}\,B^2)\,(\cos\theta_h)^{\lambda}\,+d^2}\over r_h^{{7-p\over 2}}\,+\,r_h^{{p-7\over 2}}\,B^2}\,\,
\int_{r_h}^{\infty}\,dr\,
\sqrt{\Delta_B}\,
{H\,\big(\cos\theta\big)^{{\lambda\over 2}}\over
d^2+H\,\big(\cos\theta\big)^{\lambda}}\nonumber \\
&&\qquad\qquad\qquad\qquad\times
\Bigg[
1+d\Bigg({\lambda\over 2}\,\tan\theta\,{\partial \theta\over \partial d}\,+\,
{r^2\,f_p\,\dot\theta\over \Delta_B}\,
{\partial\dot \theta\over \partial d}\Bigg)\Bigg]\,\,.
\label{Diffusion_Einstein_B}
\eear
In Fig.~\ref{diffusion_zero_B}  we compare the predictions of (\ref{Diffusion_Einstein_B})   for the D2-D6 model and 
the numerical results obtained by direct integration of the coupled fluctuation equations (\ref{E_eom_momentum_B})-(\ref{ay_eom_momentum_B}). As can be appreciated in this figure, the agreement between the two methods is very good.

\section{Summary and conclusions}
\label{conclusions}

In this paper we studied the collective excitations of flavor D$q$-branes in the supergravity background generated by color D$p$-branes. The two set of branes are separated in their transverse directions, which corresponds to adding massive flavors  in the dual field theory. We first studied this D$p$-D$q$ model  at $T=0$ and $\mu\not=0$ in the quenched approximation. The non-zero chemical potential is generated by a suitable worldvolume gauge field on the probe. We then generalized these results for $T\not= 0$ and  non-vanishing magnetic field. 

At zero temperature and non-vanishing chemical potential the supersymmetric D$p$-D$q$  intersections with \#ND=4 can be studied  analytically. We obtained  their thermodynamics and  first and zero sound, generalizing previous results in the literature for the conformal cases with $p=3$.  These results allow to characterize the quantum phase transition that occurs when $\mu=m$ and $d=0$. In this point several thermodynamic quantities vanish and the system displays a non-relativistic scaling behavior with hyperscaling violation. We have been able to compute the corresponding critical exponents.

We also analyzed the massive flavor brane systems at non-zero temperature and magnetic field. We verified numerically that, when the magnetic field is non-vanishing, the zero sound spectrum becomes gapped, with the gap given by $B/\mu$. Moreover, when $T$ is large enough the system enters into a hydrodynamic regime,  which is dominated  by a diffusion mode.  We determined numerically the corresponding diffusion constant and verified the validity of the Einstein relation.  

When the intersection is $(2+1)$-dimensional we performed an alternative quantization of the fluctuations, which corresponds to adding degrees of freedom with fractional statistics (anyons). In those systems the zero sound is generically gapped, although it becomes gapless if the magnetic field is chosen appropriately. In fact, this choice corresponds to a fluid of anyons experiencing zero effective magnetic field, thus the occurrence of gapless mode was expected. Our understanding of the anyonic fluid is still lacking, though. In order to describe its properties better one would need to make a definite choice for the $SL(2,\mathbb{Z})$ transformation as this is needed to make an identification of the resulting charge density of the anyons. Moreover, as there is a residual gauge freedom in adding boundary terms to the action, the calculation of the free energy depends crucially on the chosen $SL(2,\mathbb{Z})$ transformation. The variational principle is still well-defined, which allowed us in the current analysis to investigate the transport properties and collective phenomena of the anyon fluid in terms of the statistics, proportional to the quantization parameter $\mathfrak{n}$.

There are several other open topics which deserve further investigation. The D$p$-brane metrics with $p\not =3$ violate hyperscaling \cite{Dong:2012se} with $\theta=-(p-3)^2/(5-p)$. It would be worth to explore the relation between this scaling of the background and the one found above for the probe. Another interesting problem for the  future would be the analysis of more general D$p$-D$q$ intersections. Contrary to the supersymmetric cases studied here, the massive embeddings of a general D$p$-D$q$ model are generically unstable and one must turn on fluxes on the worldvolume of the probe to stabilize them 
(see, for example \cite{Myers:2008me,Bergman:2010gm,Jokela:2011eb}). These additional worldvolume gauge fields give an important contribution to the Wess-Zumino term of the probe action.\footnote{An interesting alternative viewpoint without fluxes is discussed in \cite{Kutasov:2011fr,Mezzalira:2015vzn}. In this context too, however, one would need to take other Wess-Zumino terms into account (together with modifying the UV asymptotics) and our results are not directly applicable.} It would be very interesting to develop a general formalism for the collective excitations of the probe brane which could incorporate all the particular cases studied in the literature. 

It would also be interesting to analyze the systems in which the backgrounds are not generated by branes in flat space. Let us mention the cases of branes on the conifold (as in the Klebanov-Witten model \cite{Klebanov:1998hh}) and  the ABJM  model \cite{Aharony:2008ug}.  Since the massive embeddings depend on the particular model, it is expected that the results will not be completely universal.  It is interesting, however, to determine the features common to all the cases. 

The  collective excitations of brane intersections analyzed so far in the literature have been carried out in the probe approximation. Therefore, it is  quite natural   to explore the effects  on the results  of having dynamical quarks. In order to provide an answer to this problem  we need to have  supergravity backgrounds which include the backreaction of the flavor branes. By employing different approximations, these backgrounds can be found  for some systems. Let us mention the case of ABJM with smeared flavor branes \cite{Conde:2011sw,Jokela:2012dw,Bea:2013jxa,Bea:2014yda}, which are geometries free of pathologies, although they do not incorporate the effect of non-zero density.  This effect is included in the geometry  recently found in \cite{Faedo:2015urf}, which is dual to three-dimensional super Yang-Mills theory with compressible matter.  In the near future we intend to study the collective excitations of the flavor branes for some of  these systems.

\vspace{1.5cm}

{\bf \large Acknowledgments}
We thank Yago Bea and Carlos Hoyos for discussions and critical readings of the manuscript. 
N.J. is supported by the Academy of Finland grant no. 1268023. 
A.~V.~R.  and G.~I.  are funded by the Spanish grant FPA2014-52218-P, by the Consolider-Ingenio 2010 Programme CPAN (CSD2007-00042), by Xunta de Galicia (GRC2013-024), and by FEDER. 
G.~I. is also funded by FPA2012-35043-C02-02.

\appendix

\vskip 1cm
\renewcommand{\theequation}{\rm{A}.\arabic{equation}}
\setcounter{equation}{0}

\section{Fluctuation equations of motion}
In this appendix we obtain the Lagrangian density, and the corresponding equations of motion, for the fluctuations 
of the embedding scalar and the gauge fields at non-vanishing charge density $d\ne 0$ and magnetic field $B\ne 0$. As was the case for the background equations, it is useful to treat the analysis for $T=0$ and $T\ne 0$ using different parametrization.

\subsection{Fluctuations at zero temperature}
\label{appendixA}
In this subsection we focus on $T=0$ case. Let us consider a fluctuation of the gauge field and embedding as in (\ref{zeroT-fluct}) and (\ref{zeroT-fluct-F}).
The induced metric $g$ takes the form:
\beq
g\,=\,\bar g\,+\,\hat g\,\,,
\eeq
where $\bar g$ is the zeroth-order metric and $\hat g$ is the perturbation. Let us split $\hat g$ in the form:
\beq
\hat g\,=\,\hat g^{(1)}\,+\,\hat g^{(2)}\,\,.
\eeq
The non-zero elements of $\hat g^{(1)}$ are:
\beq
\hat g^{(1)}_{\rho x^{\mu}}\,=\,{z_0'\over r^{{7-p\over 2}}}\,\partial_{\mu}\xi\,\,,
\qquad\qquad
\hat g^{(1)}_{\rho \rho}\,=\,{2\,z_0'\over r^{{7-p\over 2}}}\,\partial_{\rho}\xi\,,
\eeq
whereas $\hat g^{(2)}$ has the form:
\beq
\hat g^{(2)}_{ab}\,=\,
{1\over r^{{7-p\over 2}}}
\,\partial_a\xi\,\partial_b\xi\,\,.
\eeq
(we are taking the radius $R=1$ in (\ref{Dp-metric_dilaton})). In order to expand the DBI D$q$-brane action we notice that the Born-Infeld  determinant can be written as:
\beq
\sqrt{-\det(g+F)}\,=\,\sqrt{-\det \big(\,\bar g \,+\,{
F^{(0)}}\,\big)}\,
\sqrt{\det(1+X)}\,\,,
\label{detX}
\eeq
where the matrix $X$ is given by:
\beq
X\,\equiv\,\big(\,\bar g\,+\,F^{(0)}\,\big)^{-1}\,\,
\big(\,\hat g\,+\,{f}\,\big)\,\,.
\label{X_def}
\eeq
To evaluate  the right-hand side of eq. (\ref{detX}), we shall use 
the expansion:
\beq
\sqrt{\det(1+X)}\,=\,1\,+\,{1\over 2}\,\tr X\,-\,{1\over 4}\,\tr X^2\,+\,
{1\over 8}\,\big(\tr X\big)^2\,+\,{\cal O}(X^3)\,\,.
\label{expansion}
\eeq
Moreover, in the inverse  matrix
$\big(\,\bar g\,+\,F^{(0)}\,\big)^{-1}$ we will separate the symmetric and antisymmetric parts:
\beq
\Big(\,\bar g\,+\,F^{(0)}\Big)^{-1}\,=\,{\cal G}^{-1}\,+\,{\cal J}\,\,,
\label{openmetric}
\eeq
where ${\cal J}$ is the antisymmetric component and the symmetric matrix ${\cal G}$ is the open string metric. The relevant components of ${\cal G}$ are:
\bear
{\cal G}^{tt} & = & -{\bar g_{rr}\,(1+z_0'^2)\over \bar g_{rr}\,|\bar g_{tt}|(1+z_0'^{2})\,-\,A_t^{(0)'2}}\,\,,\qquad\qquad {\cal G}^{x^i\,x^j}\,=\,{\delta^{ij}\over\bar  g_{xx}}\nonumber \\
{\cal G}^{\rho\rho} & = & -{\bar g_{tt}\over \bar g_{rr}\,|\bar g_{tt}|(1+z_0'^{2})\,-\,A_t^{(0)'2}}\,\,.
\eear
Using the fact that $\bar g_{rr}\,|\bar g_{tt}|=1$, and eliminating $z_0'$ and $A_t^{(0)'}$, we get:
\bear
{\cal G}^{tt} & = & -{\rho^{\lambda}+d^2\over |\bar g_{tt}|\,\rho^{\lambda}}\,=\,
-{\rho^{\lambda}+d^2\over (\rho^2+z_0^2)^{{7-p\over 4}}\,\rho^{\lambda}}\nonumber \\
{\cal G}^{\rho\rho} &  = & {\rho^{\lambda}+d^2-c^2\over \bar g_{rr}\,\rho^{\lambda}}\,=\,
 (\rho^2+z_0^2)^{{7-p\over 4}}\,\,
 {\rho^{\lambda}+d^2-c^2\over \rho^{\lambda}}\nonumber \\
 {\cal G}^{x^i\,x^j} &  = & {\delta^{ij}\over 
 (\rho^2+z_0^2)^{{7-p\over 4}} }\,\,,
\eear
which are just the components written in (\ref{os-metric-zeroT}). 
The elements of the antisymmetric matrix ${\cal J}$ are:
\beq
{\cal J}^{t\rho}\,=\,-{\cal J}^{\rho t}\,=\,-
{A_t^{(0)'}
\over \bar g_{rr}\,|\bar g_{tt}|(1+z_0'^{2})\,-\,A_t^{(0)'2}} = \,-{d\sqrt{\rho^{\lambda}+d^2-c^2}\over \rho^{\lambda}}\,\,.
\label{cal-J-zeroT}
\eeq
By explicit calculation one can verify that $\tr \,X$ is given by:
\beq
\tr \,X\,=\,2\,{z_0'\over r^{{7-p\over 2}}}\,{\cal G}^{\rho\rho}\,\partial_{\rho}\,\xi\,+\,
2\,{\cal J}^{t\rho}\,f_{\rho t}\,+\,{{\cal G}^{ab}\over r^{{7-p\over 2}}}\,
\partial_{a}\xi\,\partial_{b}\xi\,\,,
\eeq
while $\tr \,X^2$ is:
\bear
&&\tr \,X^2\,=\,-{\cal G}^{ac}\,{\cal G}^{bd}\,f_{cd}\,f_{ab}\,+\,
{\cal G}^{ac}\,{\cal G}^{bd}\,\hat g^{(1)}_{ab}\,\hat g^{(1)}_{cd}\nonumber \\
&&\qquad\qquad\qquad+\,
2 ({\cal J}^{t\rho})^2\,\Big[ (\hat g^{(1)}_{t\rho})^2\,+\,
(f_{t\rho})^2\Big]\,-\,
4{\cal J}^{t\rho}\,{\cal G}^{ab}\,\hat g^{(1)}_{\rho a}\,f_{tb}\,\,.
\eear
This last expression can be written more explicitly as:
\bear
&&\tr \,X^2\,=\,-{\cal G}^{ac}\,{\cal G}^{bd}\,f_{cd}\,f_{ab}\,+\,
2{(z_0')^2\over  r^{7-p}}\,{\cal G}^{\rho\rho}\,{\cal G}^{a b}\,
\partial_{a}\xi\,\partial_{b}\xi\,+2\,{(z_0')^2\over  r^{7-p}}\,({\cal G}^{\rho \rho})^2\,
(\partial_{\rho}\,\xi)^2 \\
&&\quad
+2 ({\cal J}^{t\rho})^2\,\Big[{ (z_0')^2\over  r^{7-p}}\,(\partial_{t}\xi)^2\,+\,
(f_{t\rho})^2\Big]\,-\,{4z_0'\over  r^{{7-p\over 2}}}
{\cal J}^{t\rho}{\cal G}^{ab}\partial_{a}\xi\,f_{t b}
-\,{4z_0'\over  r^{{7-p\over 2}}}
{\cal J}^{t\rho}{\cal G}^{\rho \rho}\partial_{\rho}\xi\,f_{t \rho}\,\,.\nonumber
\eear
From these expressions we get that:
\bear
&&{1\over 2}\,\tr X\,-\,{1\over 4}\,\tr X^2\,+\,
{1\over 8}\,\big(\tr X\big)^2\,=\,
{z_0'\over r^{{7-p\over 2}}}\,{\cal G}^{\rho\rho}\,\partial_{\rho}\,\xi\,+\,{\cal J}^{t\rho}\,f_{\rho t}\,+\,
{1\over 4}\,{\cal G}^{ac}\,{\cal G}^{bd}\,f_{cd}\,f_{ab}\nonumber \\
&&+
{{\cal G}^{ab}\over 2  r^{{7-p\over 2}}}\Big[1-
{(z_0')^2\,{\cal G}^{\rho\rho}\over  r^{{7-p\over 2}}}
\Big]
\partial_{a}\xi\,\partial_{b}\xi-
{(z_0')^2\over 2\, r^{7-p}} ({\cal J}^{t\rho})^2\,(\partial_{t}\xi)^2\,+\,
{z_0'\over  r^{{7-p\over 2}}}\,{\cal J}^{t\rho}
{\cal G}^{ab}\,\partial_{a}\xi\,f_{tb}\,\,.
\eear
Let us now obtain the Lagrangian density from these results. First of all, we can check that the first-order terms do not contribute to the equations of motion and, therefore, we just 
drop them. Moreover, in the second-order terms we can substitute $r$ by $r_0(\rho)$, given by:
\beq
r_0(\rho)\,=\,\sqrt{\rho^2+z_0(\rho)^2}\,\,.
\eeq
Taking into account the zeroth-order Lagrangian and that:
\beq
1-{(z_0')^2\,{\cal G}^{\rho\rho}\over  r_0^{{7-p\over 2}}}\,=\,
{1-A_t^{(0)'2}\over 1+(z_0')^2-A_t^{(0)'2}}\,\,,
\eeq
we get:
\bear
&&{\cal L}\,=\,-{\cal N}\,\rho^{{\lambda\over 2}}\,\sqrt{1+(z_0')^2-A_t^{(0)'2}}\,\Bigg[
{1\over 4}\,{\cal G}^{ac}\,{\cal G}^{bd}\,f_{cd}\,f_{ab}\,+\,
{1\over 2 r_0^{{7-p\over 2}}}\,
{1-A_t^{(0)'2}\over 1+(z_0')^2-A_t^{(0)'2}}\,{\cal G}^{a b}\,
\partial_{a}\xi\,\partial_{b}\xi\nonumber \\
&&\qquad\qquad\qquad\qquad-
{(z_0')^2\over 2\, r_0^{7-p}}\, ({\cal J}^{t\rho})^2\,(\partial_{t}\xi)^2\,+\,
{z_0'\over  r_0^{{7-p\over 2}}}\,{\cal J}^{t\rho}
{\cal G}^{ab}\,\partial_{a}\xi\,f_{tb}\,\,\Bigg]\,\,.\qquad
\eear
Substituting the values of  $z_0'$ and $A_t^{(0)'}$ (written in (\ref{zprime-Atprime})), the Lagrangian density for the fluctuations at zero temperature can be written as in (\ref{massive_Lagrangian}). 

\vskip 1cm

\subsection{Fluctuations at non-zero temperature}
\label{appendixB}

In this subsection we focus on $T\ne 0$ and $B\ne 0$, by fluctuating the scalar and the gauge fields (\ref{fluct_gauge_scalar}). First we compute the variation of the induced metric.  By using the expansions
\bear
&&d\theta^2\,=\,\dot\theta_0^2\,dr^2\,+\,2\dot\theta_0\,\partial_a\zeta\,dr dx^a\,+\,
\partial_a\,\zeta\,\partial_b\,\zeta\,dx^a\,dx^b\,+\,\cdots\nonumber \\
&& \cos^2\theta\,=\,\cos^2\theta_0\,-\,\sin(2\theta_0)\,\zeta\,-\,\cos(2\theta_0)\zeta^2\,+\,\cdots\,\,,
\eear
where $x^{a}=(x^{\mu}, r)=(t, x^{i}, r)$, 
we can represent  the induced metric $g$ in the form:
\beq
g\,=\,\bar g\,+\,\hat g\,\,,
\eeq
where $\bar g$ is the zeroth-order metric and $\hat g$ is the perturbation. We will expand $\hat g$ up to second order in the fluctuations. Accordingly,  let us split $\hat g$ in the form:
\beq
\hat g\,=\,\hat g^{(1)}\,+\,\hat g^{(2)}\,\,,
\eeq
where $\hat g^{(1)}$  ($\hat g^{(2)}$) are the first (second) order terms of $\hat g$. 
The non-zero elements of $\hat g^{(1)}$ are:
\beq
\hat g^{(1)}_{rr}\,=\,2\,r^{{p-3\over 2}}\,\dot\theta_0\,\dot\zeta\,\,,
\qquad
\hat g^{(1)}_{r x^{\mu}}\,=\,r^{{p-3\over 2}}\,
\dot\theta_0\,\partial_{\mu}\zeta\,\,,
\qquad
\hat g^{(1)}_{mn}\,=\,-r^{{p-3\over 2}}\,\sin(2\theta_0)\,\zeta\,\gamma_{mn}\,\,,
\label{first_order_metric}
\eeq
whereas those of $\hat g^{(2)}$ are:
\beq
\hat g^{(2)}_{ab}\,=\,r^{{p-3\over 2}}\,\partial_a\zeta\,\partial_b\zeta\,\,,
\qquad\qquad
\hat g^{(2)}_{mn}\,=\,-r^{{p-3\over 2}}\,\cos(2\theta_0)\,\zeta^2\,\gamma_{mn}\,\,,
\label{2nd_order_metric}
\eeq
where $m,n$ are indices along the internal $(q-n-1)$-sphere and $\gamma_{mn}$ is the metric of  a unit 
${\mathbb S}^{q-n-1}$.  Let us now define the open string metric ${\cal G}$  and the antisymmetric tensor ${\cal J}$  as in (\ref{openmetric}), 
with $F^{(0)}$ being the gauge field strength (\ref{F_charge_B}).  The components of the inverse of the open string metric in this case  are:
\bear
{\cal G}^{tt} & = &-\,{\bar g_{rr}\,(f_p^{-1}+r^2\,\dot\theta_0^2)\over
|\bar g_{tt}|\,\bar g_{rr}(1+r^2\,f_p\,\dot\theta_0^2)\,-\,\dot A_t^{(0)2}}\,\,,
\qquad
{\cal G}^{rr}\,=\,{|\bar g_{tt}|\,f_p\over
|\bar g_{tt}|\,\bar g_{rr}(1+r^2\,f_p\,\dot\theta_0^2)\,-\,\dot A_t^{(0)2}}\,\,,
\qquad\rc\rc\rc
{\cal G}^{x^1\,x^1} & = & {\cal G}^{x^2\,x^2}\,=\,{\bar  g_{xx}\over 
\bar  g_{xx}^2\,+\,B^2}\,,
\qquad\qquad\qquad
{\cal G}^{x^i\,x^j}\,=\,{\delta^{ij}\over \bar  g_{xx}}\,\,,
\qquad (i,j=3,4,\ldots)\,\,,
\rc\rc
{\cal G}^{mn} & = &{\gamma^{mn}\over r^2\,\bar g_{rr}\,\cos^2\theta_0}\,\,,
\eear
where  $A^{(0)}$ is the gauge potential for the field strength $F^{(0)}$. Using these explicit equations for the metric and eliminating $\dot A_t^{(0)}$, we get:
\bear
{\cal G}^{tt} & = &-{1\over r^{{7-p\over 2}}\,f_p}\,
\Big[1\,+\,{d^2\over H\,(\cos\theta_0)^{ \lambda}}\Big]\,\,,
\qquad
{\cal G}^{rr}\,=\,
{r^{{7-p\over 2}}\,f_p\over 1+r^2\,f_p\,\dot\theta_0^2}\,
\Big[1\,+\,{d^2\over H\,(\cos\theta_0)^{ \lambda}}\Big]\,\,,\rc\rc\rc
{\cal G}^{x^1\,x^1} & = & {\cal G}^{x^2\,x^2}\,=\,{\bar  g_{xx}\over 
\bar  g_{xx}^2\,+\,B^2}\,\equiv\,{\cal G}^{x\,x}\,\,,
\qquad\qquad
{\cal G}^{x^i\,x^j}\,=\,{\delta^{ij}\over \bar  g_{xx}}\,\,,
\qquad (i,j=3,4,\ldots)\,\,,
\rc\rc
{\cal G}^{mn} & = & {\gamma^{mn}\over r^2\,\bar g_{rr}\,\cos^2\theta_0}\,\,.
\label{open_srt_metric}
\eear
The only non-zero elements of the antisymmetric matrix ${\cal J}$ are:
\bear
{\cal J}^{tr} & = &-{\cal J}^{rt}\,=\,-
{\dot A_t^{(0)}\over 
|\bar g_{tt}|\,\bar g_{rr}(1+r^2\,f_p\,\dot\theta_0^2)\,-\,\dot A_t^{(0)2}}\nonumber \\
{\cal J}^{x^1\,x^2} & = &-{\cal J}^{x^2\,x^1}\,=\,-{B\over \bar g_{xx}^2+B^2}\,\,.
\eear
More explicitly:
\bear
{\cal J}^{tr} & = &-{\cal J}^{rt}\,=\,-
{d\over H\,(\cos\theta_0)^{ \lambda}}\,\,
{\sqrt{H\,(\cos\theta_0)^{ \lambda}+d^2}\over
\sqrt{1+r^2\,f_p\,\dot\theta_0^2}}\nonumber \\
{\cal J}^{x^1\,x^2} & = &-{\cal J}^{x^2\,x^1}\,=\,-{B\over \bar g_{xx}^2+B^2}\,\equiv {\cal J}^{x y}
\,\,.
\label{cal_J_B}
\eear

We next define the matrix $X$ as in (\ref{X_def}) and we perform the expansion (\ref{expansion}) of the DBI determinant. The traces of $X$ needed are:
\beq
\tr \,X\,=\,{\cal G}^{MN}\,\hat g_{MN}\,-\,{\cal J}^{MN}\,f_{MN}\,\,,
\eeq
and
\beq
\tr \,X^2\,=\,\big({\cal G}^{MN}\,{\cal G}^{PQ}\,-\,{\cal J}^{MN}\,{\cal J}^{PQ}\big)
\,(\hat g_{MP}\,\hat g_{NQ}\,-f_{MP}\, f_{NQ})\,-\,
4\,{\cal G}^{MN}\,{\cal J}^{PQ}\,\hat g_{MP}\,f_{NQ}\,\,.\qquad
\eeq
In these formulas the indices $M$, $N$, $P$, and $Q$ run over all worldvolume directions (including the angular ones). The Lagrangian  density for the fluctuations is given by:
\beq
{\cal L}\,=\,{\cal L}_0\,\Big[
1\,+\,{1\over 2}\,\tr X\,-\,{1\over 4}\,\tr X^2\,+\,
{1\over 8}\,\big(\tr X\big)^2\,+\,{\cal O}(X^3)\Big]\,\,,
\label{Lagrangian_fluct_trX}
\eeq
where ${\cal L}_0$ is the zeroth-order Lagrangian density,  given by:
\beq
{\cal L}_0\,=\,-{\cal N}\,H\,(\cos\theta_0)^{\lambda}\,
{\sqrt{1+r^2\,\,f_p\,\dot\theta_0^2}\over 
\sqrt{d^2+H\,(\cos\theta_0)^{\lambda}}} \ .
\eeq
Notice  that the equation for the embedding  $\theta_0(r)$ can be written as:
\beq
\partial_r\,\Big[{\cal L}_0\,r^2\,\bar g_{rr}\,{\cal G}^{rr}\,\dot\theta_0\Big]\,=\,-{\lambda\over 2}\,\tan\theta_0\,{\cal L}_0\,\,.
\label{embedding_L0}
\eeq
Let us now consider the first-order contributions to ${\cal L}$. They originate  from the $\tr \,X$ term in 
(\ref{Lagrangian_fluct_trX}).  Therefore:
\beq
{\cal L}^{(1)}\,=\,{\cal L}_0\,\Big[\,{1\over 2}\,{\cal G}^{MN}\,\,\hat g_{MN}^{(1)}\,-\,{1\over 2}\,{\cal J}^{MN}\,\, f_{MN}\,\Big]
\,\,.
\label{1st_order_Lagrangian}
\eeq
By using the values  of the first-order metric written in (\ref{first_order_metric}), we get that the first term in (\ref{1st_order_Lagrangian}) can be written as:
\beq
{{\cal L}_0\over 2}\,{\cal G}^{MN}\hat g_{MN}^{(1)}\,=\,{\cal L}_0\,
\Big[r^2\,\bar g_{rr}\,{\cal G}^{rr}\,\dot\theta_0\,\dot\zeta\,-\,{\lambda\over 2}\,\tan\theta_0\,\zeta\Big]\,\,.
\label{total_derivarive_1st_order}
\eeq
Integrating by parts the first term in (\ref{total_derivarive_1st_order})  and using (\ref{embedding_L0}) one can easily check that  (\ref{total_derivarive_1st_order})   reduces to a total derivative and, therefore,  can be dropped from the Lagrangian. Moreover, the second term in (\ref{1st_order_Lagrangian})  can be written as:
\beq
-{1\over 2}{\cal L}_0\,{\cal J}^{MN} f_{MN}\,=\,{\cal N}\,d\,f_{tr}\,+\,{\cal L}_0\,
{B\over \bar g_{xx}^2+B^2}\,f_{x^1 x^2}\,\,,
\eeq
and clearly does not contribute to the equations of motion of the fluctuations. Let us now concentrate on the second-order terms in ${\cal L}$. After some work, we get:
\bear
&&{\cal L}\,=\,{\cal L}_0\,\Big[
{1\over 4}\,\Big({\cal G}^{ab}{\cal G}^{cd}\,-\,{\cal J}^{ab}{\cal J}^{cd}\,+\,
{1\over 2}{\cal J}^{ac}{\cal J}^{bd}\Big)\,f_{ac}\,f_{bd}\,+\,
{r^2 \bar g_{rr}\over 2}
\big(1-r^2 \bar g_{rr}{\cal G}^{rr}\dot\theta_0^2\big)
{\cal G}^{ab}\partial_a\zeta\,\partial_b\zeta\nonumber \\
&&-\,{\lambda\over 4}\,
\Big(1+\big(1-{\lambda\over 2}\big)\tan^2\theta_0\Big)\,\zeta^2
-{\lambda\over 2}\,r^2 \bar g_{rr}\,{\cal G}^{rr}\,\tan\theta_0\,\dot\theta_0\,\zeta\,\dot\zeta
-{r^4 \bar g_{rr}^2\over 2}\,\big({\cal J}^{tr}\big)^2\,\dot\theta_0^2\,(\partial_t\zeta)^2\nonumber \\
&&
+{\lambda\over  4}\,\tan\theta_0\,{\cal J}^{ab}\,\zeta\,f_{ab}\,+\,
r^2 \bar g_{rr}\dot\theta_0\,\Big({\cal J}^{tr}{\cal G}^{ab}\partial_a\zeta f_{tb}\,+\,
{\cal J}^{ab}{\cal G}^{rr}\partial_a\zeta f_{rb}-{1\over 2}{\cal J}^{ab}{\cal G}^{rr}
\partial_r\zeta f_{ab}\Big)\Big]\,\,.\qquad\qquad
\label{Lagrangian_fluct_raw}
\eear
Let us integrate by parts the $\zeta\dot\zeta$ term on the second line of  (\ref{Lagrangian_fluct_raw}). In this process we generate the following contribution to ${\cal L}$:
\beq
{\lambda\over 4}\partial_r\big[{\cal L}_0\,r^2 \bar g_{rr} \,{\cal G}^{rr}\tan\theta_0\,\dot\theta_0\big]\,\zeta^2\,=\,
-{\lambda^2\over 8}\,{\cal L}_0\,\big(\tan\theta_0\big)^2\,\zeta^2\,+\,
{\lambda\over 4}\,{\cal L}_0\,r^2 \bar g_{rr}{\cal G}^{rr}{\dot\theta_0^2\over \cos^2\theta_0}\,\zeta^2\,\,,
\eeq
where we have used the embedding equation (\ref{embedding_L0}).  Plugging this result into 
(\ref{Lagrangian_fluct_raw}) we get the final form of the  Lagrangian for the fluctuations, which is given by:
\bear
&&{\cal L}={\cal L}_0\,\Big[
{1\over 4}\,\Big({\cal G}^{ab}{\cal G}^{cd}\,-\,{\cal J}^{ab}{\cal J}^{cd}\,+\,
{1\over 2}{\cal J}^{ac}{\cal J}^{bd}\Big)\,f_{ac}\,f_{bd} \\
&&
+\Big(1-r^2 \bar g_{rr}{\cal G}^{rr}\dot\theta_0^2\Big)\Big(
{r^2 \bar g_{rr}\over 2}{\cal G}^{ab}\partial_a\zeta\,\partial_b\zeta\,-\,
{\lambda\over 4 \cos^2\theta_0}\,\zeta^2\Big)
-{r^4 \bar g_{rr}^2\over 2}\,\big({\cal J}^{tr}\big)^2\,\dot\theta_0^2\,(\partial_t\zeta)^2\nonumber \\
&&
+{\lambda\over  4}\,\tan\theta_0\,{\cal J}^{ab}\,\zeta\,f_{ab}\,+\,
r^2 \bar g_{rr}\dot\theta_0\,\Big({\cal J}^{tr}{\cal G}^{ab}\partial_a\zeta f_{tb}\,+\,
{\cal J}^{ab}{\cal G}^{rr}\partial_a\zeta f_{rb}-{1\over 2}{\cal J}^{ab}{\cal G}^{rr}
\partial_r\zeta f_{ab}\Big)\Big]\,\,. \nonumber
\eear
Let us now work out the equations of motion derived from this Lagrangian density. We will   assume  that all fields only depend on $t$, $r$ and one of the Cartesian coordinates (say $x$). First of all, we write the equation of $a_r$ in the $a_r=0$ gauge.  We get the following Gauss' law:
\beq
{\cal G}^{tt}\,\partial_t\,\dot a_t\,+\,{\cal G}^{xx}\,\partial_i\,\dot a_i\,=\,r^2\,\bar g_{rr}\,{\cal J}^{tr}\,\dot\theta_0\,
\partial_t\,\dot\zeta\,+\,{\bar\lambda\over 2}\,{{\cal J}^{tr}\over {\cal G}^{rr}}\,\tan\theta_0\,\partial_t\zeta\,\,.
\label{Gauss_law}
\eeq
The equation for $a_t$ becomes:
\bear
&&\partial_r\Big[{\cal L}_0\,{\cal G}^{rr}\Big({\cal G}^{tt}\,\dot a_t\,-\,r^2\,\bar g_{rr}\,{\cal J}^{tr}\,\dot\theta_0\,\dot\zeta\Big)\,-\,
{\cal L}_0\,{\cal J}^{tr}\,\Big({ \lambda\over 2}\,\tan\theta_0\,\zeta\,
+\,{\cal J}^{x y}\,\partial_x a_y\Big)
\Big]\nonumber \\
&&\qquad\qquad
+{\cal L}_0\,{\cal G}^{x\,x}\,\big[{\cal G}^{tt}\,\partial_x\,f_{xt}\,-\,
r^2\,\bar g_{rr}\,{\cal J}^{tr}\,\dot\theta_0\,\partial_x^2\zeta\big]\,+\,{\cal L}_0\,{\cal J}^{tr}\,{\cal J}^{xy}\,\partial_x\,\dot a_y
=\,0\,\,.
\label{at_eom_B}
\eear
The equation of $a_x$ is:
\bear
&&\partial_r\,\Big[{\cal L}_0\,\Big({\cal G}^{rr}\,{\cal G}^{xx}\,\dot a_x\,+\,
{\cal J}^{tr}\,{\cal J}^{x y}\,\partial_t\,a_y\Big)\Big]\,
+\,{\cal L}_0\,{\cal G}^{tt}\,{\cal G}^{xx}\,\partial_{t}\,f_{tx}\nonumber \\
&&
\qquad\qquad+
{\cal L}_0\,{\cal G}^{xx}\,r^2\,\bar g_{rr}\,{\cal J}^{tr}\,\dot\theta_0\,
\partial_t\,\partial_x\,\zeta\,-\,{\cal L}_0\,{\cal J}^{tr}\,{\cal J}^{xy}\,\partial_t\,\dot a_y\,=\,0\,\,.
\label{ax_eom_B}
\eear
Taking into account that ${\cal L}_0\,{\cal J}^{tr}={\rm constant}$, this last equation can be rewritten as:
\bear
\partial_r\big[{\cal L}_0{\cal G}^{rr}{\cal G}^{xx}\dot a_x\big]+{\cal L}_0{\cal G}^{tt}{\cal G}^{xx}\partial_{t}f_{tx}+
{\cal L}_0{\cal G}^{xx}r^2\bar g_{rr}{\cal J}^{tr}\dot\theta_0\partial_t\partial_x\zeta=-{\cal L}_0{\cal J}^{tr}\partial_r\big({\cal J}^{xy}\big)\partial_t a_y\,\,.
\label{ax_eom_B_simple}
\eear
Moreover, after some simplifications, the equation of motion of $a_y$ can be written as:
\bear
&&\partial_r\,\big[{\cal L}_0\,{\cal G}^{rr}\,{\cal G}^{xx}\,f_{ry}\big]\,+\,
{\cal L}_0\,{\cal G}^{xx}\,
\big({\cal G}^{tt}\,\partial_t\,f_{ty}\,+\,{\cal G}^{xx}\,\partial_x\,f_{xy}\,\big)\nonumber \\
&&\qquad\qquad\qquad\qquad
=\,{\cal L}_0\,\partial_r\big({\cal J}^{x y}\big)\,\Big(
{\cal J}^{tr}\,f_{tx}\,-\,
r^2\,\bar g_{rr}\,{\cal G}^{rr}\,\dot\theta_0\,\partial_x\zeta\Big)\,\,.
\eear
Finally, let us write the equation of motion of the scalar fluctuations.  We get:
\bear
&&\partial_r\Big[{\cal L}_0 r^2\bar g_{rr}{\cal G}^{rr}\Big(\big(1-r^2 \bar g_{rr}{\cal G}^{rr}\dot\theta_0^2\big)\dot \zeta-{\cal J}^{tr}\dot\theta_0\dot a_t\Big)\Big]+{\lambda\over 2\cos^2\theta_0}{\cal L}_0\big(1-r^2 \bar g_{rr}{\cal G}^{rr}\dot\theta_0^2\big)\zeta\nonumber \\
&&\
+{\lambda\over 2}\tan\theta_0{\cal L}_0{\cal J}^{tr}\dot a_t+
{\cal L}_0 r^2\bar g_{rr}\big(1-r^2 \bar g_{rr}{\cal G}^{rr}\dot\theta_0^2\big)
\big({\cal G}^{tt}\partial^2_t\zeta+{\cal G}^{xx}\partial^2_x\zeta\big)\nonumber \\
&&
-{\cal L}_0 r^4\bar g_{rr}^2({\cal J}^{tr})^2\dot\theta_0^2\partial^2_t\zeta +{\cal L}_0 r^2\bar g_{rr}{\cal J}^{tr}{\cal G}^{xx}\dot\theta_0\partial_xf_{tx}={\cal L}_0\partial_r\big({\cal J}^{x y}\big)
r^2\bar g_{rr}\dot\theta_0{\cal G}^{rr}f_{xy}\,\,.
\label{zeta_eom_B}
\eear
Let us  next Fourier transform  the gauge field  and the scalar to momentum space as in (\ref{Fourier_a_zeta})
and let us  define the electric field $E$ as the gauge-invariant combination:
\beq
E\,=\,k\,a_t\,+\,\omega\,a_x\,\,.
\label{E_at_ax_r}
\eeq
In momentum space the Gauss law (\ref{Gauss_law}) becomes:
\beq
\omega\,{\cal G}^{tt}\,\dot a_t\,-\,k\,{\cal G}^{xx}\,\dot a_x\,=\,\omega\,r^2\,\bar g_{rr} {\cal J}^{tr}\,\dot\theta_0\,\dot\zeta\,+\,
{\bar\lambda\over 2}\,\omega\,{{\cal J}^{tr}\over {\cal G}^{rr}}\,\tan\theta_0\,\zeta\,\,.
\label{Gauss_law_momentum}
\eeq
We can combine (\ref{Gauss_law_momentum}) and (\ref{E_at_ax_r}) to get $\dot a_t$ and $\dot a_x$ in terms of the gauge-invariant combination $E$ and the scalar field $\zeta$:
\bear
\dot a_t & = & {{\cal G}^{xx}\,k \dot E\,+\,
\omega^2\,r^2\,\bar g_{rr} {\cal J}^{tr}\,\dot\theta_0\,\dot\zeta\,+\,\omega^2\,
{\bar\lambda\over 2}\,{{\cal J}^{tr}\over {\cal G}^{rr}}\,\tan\theta_0\,\zeta\over
{\cal G}^{tt}\omega^2+{\cal G}^{xx}k^2}\nonumber \\
\dot a_x & = &
{{\cal G}^{tt}\,\omega \dot E\,-\,
k\,\omega\,r^2\,\bar g_{rr} {\cal J}^{tr}\,\dot\theta_0\,\dot\zeta\,-\,
k\,\omega\,{\bar\lambda\over 2}\,{{\cal J}^{tr}\over {\cal G}^{rr}}\,\tan\theta_0\,\zeta\over
{\cal G}^{tt}\omega^2+{\cal G}^{xx}k^2}\,\,.
\label{at_ax_zeta}
\eear
Moreover,  using (\ref{at_ax_zeta})  one can demonstrate that (\ref{at_eom_B}) and (\ref{ax_eom_B_simple})  are equivalent to the following 
equation for the electric field $E$:
\bear
&&\partial_r\,\Bigg[{{\cal L}_0\,{\cal G}^{rr}\,{\cal G}^{xx}\over {\cal G}^{tt}\omega^2+{\cal G}^{xx}k^2}
\Big({\cal G}^{tt}\,\dot E\,-\,k\,r^2\,\bar g_{rr} {\cal J}^{tr}\,\dot\theta_0\,\dot\zeta\,-\,
k\,{\lambda\over 2}\,{{\cal J}^{tr}\over {\cal G}^{rr}}\,\tan\theta_0\,\zeta\Big)\Bigg]\nonumber \\
&&\qquad\qquad
-{\cal L}_0\,{\cal G}^{tt}\,{\cal G}^{xx}\,E\,+\,k\,{\cal L}_0\,{\cal G}^{xx}\,
\,r^2\,\bar g_{rr} {\cal J}^{tr}\,\dot\theta_0\,\zeta\,=\,
i\,{\cal L}_0\,{\cal J}^{tr}\,\,\partial_r\,({\cal J}^{xy})\,a_y
\,\,,\qquad
\label{E_eom_momentum_B}
\eear
where ${\cal G}^{xx}$ has been defined in (\ref{open_srt_metric}).  Similarly, we can work out the equation for the scalar $\zeta$ in terms of $E$. In momentum space this equation becomes:
\bear
&&\partial_r\Big[{\cal L}_0r^2\bar g_{rr}{\cal G}^{rr}\Big(\big(1-r^2 \bar g_{rr}{\cal G}^{rr}\dot\theta_0^2\big)\dot \zeta-{\cal J}^{tr}\dot\theta_0\dot a_t\Big)\Big]+
{ \lambda\over 2\cos^2\theta_0}{\cal L}_0\big(1-r^2 \bar g_{rr}{\cal G}^{rr}\dot\theta_0^2\big)\zeta\nonumber \\
&&\qquad\qquad
+{\lambda\over 2}\tan\theta_0{\cal L}_0{\cal J}^{tr}\dot a_t-{\cal L}_0 r^2\bar g_{rr}\big(1-r^2 \bar g_{rr}{\cal G}^{rr}\dot\theta_0^2\big)\big({\cal G}^{tt}\omega^2+{\cal G}^{xx}k^2\big)\zeta\nonumber \\
&&
+{\cal L}_0 r^4\bar g_{rr}^2({\cal J}^{tr})^2\dot\theta_0^2\omega^2\zeta +{\cal L}_0 r^2\bar g_{rr}{\cal J}^{tr}{\cal G}^{xx}\dot\theta_0 k E=
ik{\cal L}_0\dot\theta_0 r^2\bar g_{rr}{\cal G}^{rr}\partial_r({\cal J}^{x y})a_y\,,
\label{zeta_eom_momentum_B}
\eear
where it should be understood that $\dot a_t$ is given by the first equation in (\ref{at_ax_zeta}). 
Finally,  the equation of motion of the transverse fluctuation $a_y$ is:
\bear
&&\partial_r\Big[{\cal L}_0{\cal G}^{rr}{\cal G}^{xx}\dot a_y\Big]-{\cal L}_0{\cal G}^{xx}\big({\cal G}^{tt} \omega^2+{\cal G}^{xx}k^2\big)a_y=\rc\rc
&&\qquad\qquad
-i{\cal L}_0{\cal J}^{tr}\partial_r({\cal J}^{x y})E-ik{\cal L}_0\dot\theta_0 r^2\bar g_{rr}{\cal G}^{rr}\partial_r({\cal J}^{x y})\zeta\,\,.
\label{ay_eom_momentum_B}
\eear

\vskip 1cm
\renewcommand{\theequation}{\rm{B}.\arabic{equation}}
\setcounter{equation}{0}

\section{Transverse correlators and  the conductivity }
\label{appendixC}

Let us consider the case in which the magnetic field vanishes, $B=0$. In this case, 
the equation of motion (\ref{ay_eom_momentum_B}) for the transverse fluctuation $a_y$ is:
\beq
\partial_r\Big[{\cal L}_0\,{\cal G}^{rr}\,{\cal G}^{xx}\,\dot a_y\Big]\,-\,{\cal L}_0\,{\cal G}^{xx}\,
\big({\cal G}^{tt} \omega^2\,+\,{\cal G}^{xx}\,k^2\big)\,a_y\,=\,0\,\,.
\eeq 
This equation can be rewritten as:
\beq
\ddot a_y\,+\,\partial_r\log\Big[{\cal L}_0\,{\cal G}^{rr}\,{\cal G}^{xx}\Big]\,\dot a_y\,-\,
{{\cal G}^{tt} \omega^2\,+\,{\cal G}^{xx}\,k^2\over {\cal G}^{rr}}\,a_y\,=\,0\,\,.
\eeq
More explicitly, the equation of motion for $a_y$ is:
\bear
&&\ddot a_y\,+\,\partial_r\log\Bigg[
{\sqrt{d^2+ r^{\lambda}\,\big(\cos\theta_0\big)^{\lambda}}\over
\sqrt{1+ r^2\,f_p\dot\theta_0^{2}}}\,f_p\Bigg]\,\dot a_y\nonumber \\
&&\qquad\qquad\qquad\qquad
+{1+ r^2\,f_p\dot\theta_0^{2}\over r^{7-p}\,f_p^2}\,
{(\omega^2-f_p\,k^2)\,r^{\lambda}\,(\cos\theta_0)^{\lambda}\,+\,\omega^2\,d^2\over 
d^2+ r^{\lambda}\,\big(\cos\theta_0\big)^{\lambda}}\,
a_y\,=\,0\,\,.
\label{transverse_fluct_eq}
\eear
We now study the equation of motion (\ref{transverse_fluct_eq}) for $a_y$ in the low frequency regime in which $k\sim {\mathcal O}(\epsilon)$ and $\omega\sim {\mathcal O}(\epsilon^2)$.
Let us  first study (\ref{transverse_fluct_eq}) near the horizon $r=r_h$. With this purpose we expand $\theta_0(r)$ near $r=r_h$:
\beq
\theta_0(r)\approx \theta_h\,-\,
{\lambda \over 2(7-p)}\,{r_h^{\lambda-1}\,\big(\cos\theta_h\big)^{\lambda}\,\tan\theta_h\over
d^2+r_h^{\lambda}\,\big(\cos\theta_h\big)^{\lambda}}\,(r-r_h)\,+\,\cdots \ .
\eeq
We also expand the coefficients of the  equation of the transverse fluctuations:
\bear
\partial_r\log\Bigg[
{\sqrt{d^2+ r^{\lambda}\,\big(\cos\theta_0\big)^{\lambda}}\over
\sqrt{1+ r^2\,f_p\dot\theta_0^{2}}}\,f_p\Bigg] & = & {1\over r-r_h}+d_1\,+\,\cdots\nonumber \\
{1+ r^2\,f_p\dot\theta_0^{2}\over r^{7-p}\,f_p^2}\,
{(\omega^2-f_p\,k^2)\,r^{\lambda}\,(\cos\theta_0)^{\lambda}\,+\,\omega^2\,d^2\over 
d^2+ r^{\lambda}\,\big(\cos\theta_0\big)^{\lambda}} & = & {A\over (r-r_h)^2}\,+\,{c_2\over r-r_h}\,+\,\cdots\,\,,
\eear
where $A$, $d_1$, and $c_2$ are given by:
\bear
A & = &{\omega^2\over (7-p)^2\,r_h^{5-p}}\nonumber \\
d_1 & = & {1\over 2 r_h}\,{(p-8)\,d^2\,+\,(p+\lambda-8)\,r_h^{\lambda}\,\big(\cos\theta_h\big)^{\lambda}\over
d^2+r_h^{\lambda}\,\big(\cos\theta_h\big)^{\lambda}}\,+\,
{\lambda^2\over 8(7-p)}{r_h^{2\lambda-1}\,\big(\cos\theta_h\big)^{2\lambda}\over
\big[d^2+r_h^{\lambda}\,\big(\cos\theta_h\big)^{\lambda}\big]^2}\,\tan^2\theta_h\nonumber \\
c_2 & = & -{1\over 7-p}\,{r_h^{p+\lambda-6}\,\,\big(\cos\theta_h\big)^{\lambda}\over
d^2+r_h^{\lambda}\,\big(\cos\theta_h\big)^{\lambda}}\,k^2\,+\,{1\over (7-p)^2\,r_h^{6-p}}\,\omega^2\nonumber \\
&&
+\,{\lambda^2\over 4(7-p)^3}\,{r_h^{p+2\lambda-6}\,\,\big(\cos\theta_h\big)^{2\lambda}\,
\tan^2\theta_h\over 
\big[d^2+r_h^{\lambda}\,\big(\cos\theta_h\big)^{\lambda}\big]^2}\,\omega^2 \ .
\label{nh_coeff}
\eear

Let us now solve for $a_y$ in Frobenius series around $r=r_h$:
\beq
a_y(r)\,=\,(r-r_h)^{\alpha}\,(1+\beta\,(r-r_h)+\ldots)\,\,,
\label{ay_nh}
\eeq
where the exponents $\alpha$ and $\beta$, at order $\epsilon^2$,  are given by:
\beq
\alpha\,=\,-{i\omega\over (7-p)\,r_h^{{5-p\over 2}}}\,\,,
\qquad\qquad
\beta\approx -(\alpha\,d_1+\,c_2)\,\,.
\label{exponents_alpha_beta}
\eeq
From the expressions of $d_1$ and $c_2$ written in (\ref{nh_coeff}) we find that $\beta$ is given by:
\bear
&&\beta\,=\,i\Bigg[{1\over 2(7-p)r_h^{{7-p\over 2}}}\,
{(p-8)\,d^2\,+\,(p+\lambda-8)\,r_h^{\lambda}\,\big(\cos\theta_h\big)^{\lambda}\over
d^2+r_h^{\lambda}\,\big(\cos\theta_h\big)^{\lambda}}\nonumber \\
&&\qquad\qquad\qquad\qquad
+{\lambda^2\over 8(7-p)^2\,r_h^{{5-p\over 2}}}\,
{r_h^{2\lambda-1}\,\big(\cos\theta_h\big)^{2\lambda}\over
\big[d^2+r_h^{\lambda}\,\big(\cos\theta_h\big)^{\lambda}\big]^2}\,\tan^2\theta_h\,\Bigg]\,\omega\nonumber \\
&&\qquad\qquad\qquad\qquad\qquad\qquad
+{1\over 7-p}\,
{r_h^{p+\lambda-6}\,\,\big(\cos\theta_h\big)^{\lambda}\over
d^2+r_h^{\lambda}\,\big(\cos\theta_h\big)^{\lambda}}\,k^2\,\,.
\eear
Let us now take the near-horizon and low frequency limits in opposite order. First, we write (\ref{transverse_fluct_eq}) as:
\beq
\ddot a_y\,+\,{\dot G\over G}\,\dot a_y\,+\,Q\,a_y\,=\,0 \ ,
\label{a_y_eq_G_Q}
\eeq
where $G(r)$ is given by:
\beq
G(r)\,=\,{\sqrt{d^2+ r^{\lambda}\,\big(\cos\theta_0\big)^{\lambda}}\over 
\sqrt{1+ r^2\,f_p\dot\theta_0^{2}}}\,f_p
\,\,.
\eeq
Moreover, the expression of $Q(r)$ at order $\epsilon^2$ is:
\beq
Q(r)\approx -{1+ r^2\,f_p\dot\theta_0^{2}\over f_p}\,
{r^{\lambda+p-7}\,\big(\cos\theta_0\big)^{\lambda}
\over d^2+ r^{\lambda}\,\big(\cos\theta_0\big)^{\lambda}}\,\,k^2\,\,.
\eeq
Let us redefine $a_y(r)$ as:
\beq
a_y(r)\,=\,F(r)\,\alpha_y(r)\,\,,
\label{ay_alphay}
\eeq
where $\alpha_y(r)$ should be regular at $r=r_h$ and $F(r)$ is given by:
\beq
F(r)=(r-r_h)^{\alpha}\,\,.
\label{F_value}
\eeq
The resulting equation for $\alpha_y$ is:
\beq
\ddot\alpha_y\,+\,\Big({\dot G\over G}\,+\,2\,{\dot F\over F}\,\epsilon^2\Big)\,\dot \alpha_y\,+\,
\epsilon^2\,(P+Q)\, \alpha_y\,=\,0\,\,,
\label{alpha_y_eq}
\eeq
where we have explicitly introduced the powers of $\epsilon$ to keep track of the low frequency expansion and we have defined the new function $P(\rho)$ as:
\beq
P(r)\equiv {\ddot F\over F}\,+\,{\dot G\over G}\,{\dot F\over F}\,\,.
\label{P_def}
\eeq
We will solve (\ref{alpha_y_eq}) order by order in  a series expansion in $\epsilon$ of the form:
\beq
\alpha_y\,=\,\alpha_0\,+\,\epsilon^2\,\alpha_1\,+\,\ldots\,\,.
\label{epsilon_expansion_alphay}
\eeq
As in the massless case, $\dot\alpha_0=0$ if we impose regularity at $r=r_h$. 
Furthermore, without loss of generality we can take:
\beq
\alpha_0=1\,\,.
\eeq
The equation for $\alpha_1$ is
\beq
\ddot \alpha_1\,+\,{\dot G\over G}\,\dot \alpha_1\,=\,-P\,-\,Q \ .
\label{alpha1_eq}
\eeq
This equation can be solved by variation of constants. We put:
\beq
\dot \alpha_1(r)\,=\,
{A(r)\over G(r)}\,\,,
\eeq
where $A(r)$ is a function to be determined. By direct substitution into (\ref{alpha1_eq}) we get that $A(r)$ must satisfy:
\beq
\dot A\,=\,-G\,(P+Q)\,\,.
\eeq
The solution of this equation for $A$ at leading order in $\epsilon$ is:
\beq
A(r)\,=\,-G\, {\dot F\over F}\,-\,c\,-\int_{r_h}^{r}\,G(\bar r)\,Q(\bar r)\,d\bar r\,\,,
\eeq
where $c$ is a constant to be determined. Let us next define the integral ${\cal I}(r)$ as:
\beq
k^2\,{\cal I}(r)\,\equiv\,-\int_{r_h}^{r}\,G(\bar r)\,Q(\bar r)\,d\bar r\,\,,
\eeq
or, more explicitly:
\beq
{\cal I}(r)\,=\,\int_{r_h}^{r}\,
{\sqrt{1+ \bar r^2\,f_p(\bar r)\,\dot\theta_0^{2}(\bar r)}\over 
\sqrt{d^2+ \bar r^{\lambda}\,\big(\cos\theta_0(\bar r)\big)^{\lambda}}}\,\,
\bar r^{\lambda+p-7}\,(\cos\theta_0(\bar r))^{ \lambda}\,\,d\bar r\,\,.
\eeq
Therefore, $\dot \alpha_1$ can be written as:
\beq
\dot \alpha_1\,=\,-\Bigg[c\,
{\sqrt{1+  r^2\,f_p\,\dot\theta_0^{2}}\over 
\sqrt{d^2+  r^{\lambda}\,\big(\cos\theta_0\big)^{\lambda}}\,\,f_p}
\,+\,
{\alpha\over r-r_h}\,\Bigg]\,+\,{{\cal I}(r)\,k^2\over G(r)}\,\,.
\eeq
The constant $c$ is determined by requiring that $\dot\alpha_1$ be regular at $r=r_h$. We get:
\beq
c\,=\,i\,{\sqrt{r_h^{\lambda}\,
(\cos\theta_h)^{\lambda}+d^2}\over r_h^{{7-p\over 2}}}
\,\,\omega\,\,.
\eeq
As $\dot\alpha_y=\dot\alpha_1+{\cal O}(\epsilon^4)$, we get:
\beq
\dot \alpha_y\,=\,-{i\over (7-p)\,r_h^{{5-p\over 2}}}
\Bigg[{7-p\over r_h}\,\sqrt{1+  r^2\,f_p\,\dot\theta_0^{2}}\,
{\sqrt{r_h^{\lambda}\,
(\cos\theta_h)^{\lambda}+d^2}\over
\sqrt{r^{\lambda}\,
(\cos\theta_0)^{\lambda}+d^2}}\,{1\over f_p}\,-\,{1\over r-r_h}\,\Bigg]\,\omega\,+\,
{{\cal I}(r)\,k^2\over G(r)}\,\,.
\eeq
This solution should match (\ref{ay_nh}). One can check that this is indeed the case since 
$\dot\alpha_y(r=r_h)=\beta$. Moreover:
\beq
\dot a_y\,=\, \dot\alpha_1\,+\,{\alpha \over r-r_h}\,+\,{\mathcal O}(\epsilon^4)\,=\,
\dot\alpha_y\,+\,{\alpha\over  r-r_h}\,+\,{\mathcal O}(\epsilon^4)\,\,.
\eeq
Thus, we can write:
\beq
\dot a_y\,=\,-{1\over G(r)}\,
\Big[ i{\sqrt{r_h^{\lambda}\,
(\cos\theta_h)^{\lambda}+d^2}\over r_h^{{7-p\over 2}}}\,\omega\,-\,{\cal I}(r)\,k^2
\Big]\,\,.
\eeq

In order to obtain the $\left\langle J_y\,J_y\right\rangle $ correlator from these results, let us point out that
the term depending on $a_y$ of the Lagrangian density is of the form:
\beq
{\cal L}(a_y)\,=\,{\cal F}\,(f_{y\,r})^2\,=\,{\cal F}\, (\dot a_y)^2\,\,,
\eeq
where ${\cal F}$ is given by:
\beq
{\cal F}\,=\,-{\cal N}\,r^{{\lambda\over 2}}\,\big(\cos\theta_0\big)^{{\lambda\over 2}}\,
\sqrt{\Delta}\,{\cal G}^{yy}\,{\cal G}^{rr}\,=\,-{\cal N}\,G\,\,.
\eeq
Then, the  $\left\langle J_y\,J_y\right\rangle $ correlator takes the form:
\beq
\left\langle J_y(p)\,J_y(-p)\right\rangle\,=\,{\cal N}\,\Big[\,\Gamma_{\omega}\,i\omega\,+\,\Gamma_k\,k^2\,\Big]\,\,,
\label{transv_corr_Gammas}
\eeq
where the coefficients $\Gamma_{\omega}$  and $\Gamma_k$ are:
\bear
\Gamma_{\omega} & = & {\sqrt{r_h^{\lambda}(\cos\theta_h)^{\lambda}+d^2}\over r_h^{{7-p\over 2}}}\nonumber \\
\Gamma_k & = & -\int_{r_h}^{\infty}\,d\bar r
{\sqrt{1+\bar r^2\,f_p(\bar r)\,\dot\theta_0^{2}(\bar r)}\over
\sqrt{d^2+\bar r^{\lambda}\,\big(\cos\theta_0(\bar r)\big)^{\lambda}}}\,\,
\bar r^{\lambda+p-7}\,\big(\cos\theta_0(\bar r)\big)^{\lambda}\,\,.
\eear
Notice that the DC conductivity $\sigma$ is given by $\sigma={\cal N}\,\Gamma_{\omega}$. Therefore:
\beq
\sigma\,=\,{\cal N}\,
{\sqrt{r_h^{\lambda}(\cos\theta_h)^{\lambda}+d^2}\over r_h^{{7-p\over 2}}}\,\,.
\label{sigma}
\eeq
In terms of $\hat d\,=\,d/{r_h^{{\lambda\over 2}}}$, the conductivity can be written as:
\beq
\sigma\,=\, {\cal N}\,r_h^{{p+\lambda-7\over 2}}\,
\sqrt{(\cos\theta_h)^{\lambda}+\hat d^{\,2}}\,\,.
\label{sigma_hat_d}
\eeq

\vskip 1cm
\renewcommand{\theequation}{\rm{C}.\arabic{equation}}
\setcounter{equation}{0}

\section{Conductivity  by the Karch-O'Bannon method}
\label{appendixD}

In this appendix we evaluate the conductivity of the probe by using the method developed in \cite{Karch:2007pd}.  Let us work in the $(r,\theta)$ variables of section \ref{finiteT} and consider a D$q$-brane probe with the following worldvolume gauge field:
\beq
A\,=\,A_t\,dt\,+\,(Et+a_x(r))dx\,+\,(Bx+a_y(r))dy\,\,,
\label{A_general}
\eeq
where $x$ and $y$ are two spatial Minkowski directions along the brane. The field strength corresponding to 
(\ref{A_general}) is:
\beq
F\,=\,\dot A_t\,dr\wedge dt\,+\,B dx\wedge dy\,+\,E\,dt\wedge dx\,+\,
\dot a_x\,dr\wedge dx\,+\,\dot a_y\,dr\wedge dy\,\,.
\label{F_general}
\eeq
Notice that, as in the main text, the field $A_t$ is dual to the charge density, whereas $a_x$ and $a_y$ are dual to the components of the current along the directions $x$ and $y$, respectively. Notice also that we have switched on an electric field $E$ in the $x$ direction and a magnetic field $B$ across the $xy$ plane.  The DBI Lagrangian density for this configuration is given by:
\beq
{\cal L}\,=\,-{\cal N}\,(\cos\theta)^{{\lambda\over 2}}\,
\sqrt{\Sigma}\,\,,
\label{cal_L_KOB}
\eeq
where ${\cal N}$ is the normalization constant (\ref{calN}) and $\Sigma$ is defined as:
\beq
\Sigma=(1\,+\,r^2\,f_p\,\dot\theta^2)\,(H\,-\,r^{\lambda+p-7}f_p^{-1}\,E^2)-
r^{\lambda}\dot A_t^2+
r^{\lambda}\,f_p\,(\dot a_x^2+\dot a_y^2)-r^{\lambda+p-7}(E\dot a_y+B \dot A_t)^2\,\,,
\label{Sigma}
\eeq
with $H=H(r)$ being the function introduced in (\ref{H_def}). 
It follows from (\ref{cal_L_KOB}) and (\ref{Sigma}) that $A_t$, $a_x$, and $a_y$ are cyclic variables. Let $d$, $j_x$, and $j_y$ be the corresponding conserved canonical momenta. They are given by:
\bear
 d & = & (\cos\theta)^{{\lambda\over 2}}\,{H\,\dot A_t\,+\,r^{\lambda+p-7}\,E\,B\,\dot a_y\over \sqrt{\Sigma}}\nonumber \\
 -j_x & = & (\cos\theta)^{{\lambda\over 2}}\,{r^{\lambda}f_p\,\dot a_x\over  \sqrt{\Sigma}}\nonumber \\
-j_y & = & (\cos\theta)^{{\lambda\over 2}}\,{(r^{\lambda}\,f_p\,-\,r^{\lambda+p-7}\,E^2)\,\dot a_y
-r^{\lambda+p-7}\,E B\,\dot A_t\over \sqrt{\Sigma}} \ .
\label{canonical_momenta}
\eear
Notice that  we have absorbed the normalization constant ${\cal N}$ in the definitions (\ref{canonical_momenta}). 
Let us now solve for $\dot A_t$, $\dot a_x$, and $\dot a_y$. First of all, we define the quantity $X$ as:
\beq
X\equiv \Big[d^2\,f_p+r^{\lambda}\,f_p\,(\cos\theta)^{\lambda}\,-\,j_x^2\,-\,j_y^2\Big]\,
\big[\,f_p(r^{7-p}+B^2)-E^2)\big]\,-\,\big(d\, B\,f_p\,-\,E\,j_y)^2\,\,.
\eeq
Then, after some algebra, one can verify that:
\bear
 \dot A_t & = & {\sqrt{1\,+\,r^2\,f_p\,\dot\theta^2}\over r^{{7-p\over 2}}\,\sqrt{X}}\,\,
\Big[\big(E^2-r^{7-p}\,f_p)d\,-\,E\,B\,j_y\Big]\nonumber \\
 \dot a_x & = & {\sqrt{1\,+\,r^2\,f_p\,\dot\theta^2}\over r^{{7-p\over 2}}\,f_p\,\sqrt{X}}\,\,
\Big[E^2\,-\,f_p\,(r^{7-p}+B^2)\Big]\,j_x\nonumber \\
 \dot a_y & = & -{\sqrt{1\,+\,r^2\,f_p\,\dot\theta^2}\over r^{{7-p\over 2}}\,\sqrt{X}}\,\,
\Big[E\,B\,d\,-\,(r^{7-p}\,+\,B^2)\,j_y\Big]\,\,.
\eear
Following closely the arguments in \cite{Karch:2007pd}, let us determine the position $r=r_*$ of the pseudohorizon by imposing the 
three conditions at  $r=r_*$:
\bear
 f_p^*\,(r_*^{7-p}+B^2) & = & E^2\nonumber \\
j_x^2+j_y^2 & = & f_p^*\big[r_*^{\lambda}\,(\cos\theta_*)^{\lambda}\,+\,d^2\big]\nonumber \\
 E\,j_y & = & B\, f_p^*\,d\,\,,
\label{pseudohorizon}
\eear
where we have denoted $\theta_*\equiv \theta(r=r_*)$ and $f_p^*\equiv f_p(r=r_*)$. From the first equation in (\ref{pseudohorizon}) we can determine $r_*$ in terms of $r_h$, $E$, and $B$. Indeed, we have:
\beq
r_*^{7-p}\,=\,{1\over 2}\Big[E^2-B^2+r_h^{7-p}\,+\,
\sqrt{(E^2-B^2+r_h^{7-p})^2+4r_h^{7-p}\,B^2}\,\Big]\,\,.
\label{r*_rh}
\eeq
Notice that $r_*=r_h$ if the electric field $E$ vanishes. From now on we will assume that $E$ is small. Then, it follows from (\ref{r*_rh}) that:
\beq
r_*\,=\,r_h\,+\,{1\over 7-p}\,{r_h\over r_h^{7-p}\,+\,B^2}\,E^2\,+\,{\cal O}(E^4)\,\,,
\qquad\qquad
f_p^{*}\,=\,{E^2\over  r_h^{7-p}\,+\,B^2}\,+\,{\cal O}(E^4)\,\,.
\eeq
We can use these expressions in the last two equations in (\ref{pseudohorizon}) to get $j_x$ and $j_y$. 
At leading order in $E$, we get:
\bear
j_x & = & {r_h^{{7-p\over 2}}\,\sqrt{r_h^{\lambda}(1+r_h^{p-7}\,B^2)\,(\cos\theta_h)^{\lambda}\,+d^2}\over
r_h^{7-p}\,+\,B^2}\,E\nonumber \\
 j_y & = & {B\,d\over r_h^{7-p}\,+\,B^2}\,E\,\,.
\eear
Therefore, the longitudinal and transverse conductivities are given by:
\bear
\sigma_{xx} & = & {{\cal N}\,j_x\over E}\,=\,{\cal N}\,
{r_h^{{7-p\over 2}}\,\sqrt{r_h^{\lambda}(1+r_h^{p-7}\,B^2)\,(\cos\theta_h)^{\lambda}\,+d^2}\over
r_h^{7-p}\,+\,B^2}\nonumber \\
 \sigma_{xy} & = & {{\cal N}\,j_y\over E}\,=\,{\cal N}\,{B\,d\over r_h^{7-p}\,+\,B^2}\,\,,
\label{sigma_xx_xy}
\eear
where we have reintroduced the normalization factor ${\cal N}$. Notice that $\sigma_{xx}$  in (\ref{sigma_xx_xy}) coincides with the value written in (\ref{sigma_B}). The same value of $\sigma_{xx}$  can be found from the analysis of the transverse fluctuations (as the one in  appendix \ref{appendixC} for $B=0$) if we neglect the coupling between the fluctuation equations.



\begin{thebibliography}{99}



\bibitem{AdS_CFT_reviews} For reviews see: 
 J.~Casalderrey-Solana, H.~Liu, D.~Mateos, K.~Rajagopal and U.~A.~Wiedemann,
  ``Gauge/String Duality, Hot QCD and Heavy Ion Collisions,''
  arXiv:1101.0618 [hep-th];
 J.~McGreevy,
  ``Holographic duality with a view toward many-body physics,''
  Adv.\ High Energy Phys.\  {\bf 2010}, 723105 (2010)
  [arXiv:0909.0518 [hep-th]];
    A.~V.~Ramallo,
  ``Introduction to the AdS/CFT correspondence,''
  Springer Proc.\ Phys.\  {\bf 161} (2015) 411
  [arXiv:1310.4319 [hep-th]].
 
\bibitem{Karch:2002sh}
  A.~Karch and E.~Katz,
  ``Adding flavor to AdS / CFT,''
  JHEP {\bf 0206} (2002) 043
  [hep-th/0205236].
  
\bibitem{Kobayashi:2006sb}
  S.~Kobayashi, D.~Mateos, S.~Matsuura, R.~C.~Myers and R.~M.~Thomson,
  ``Holographic phase transitions at finite baryon density,''
  JHEP {\bf 0702} (2007) 016
  [hep-th/0611099].
  
  
  
\bibitem{Jokela:2015aha} 
  N.~Jokela and A.~V.~Ramallo,
  ``Universal properties of cold holographic matter,''
  Phys.\ Rev.\ D {\bf 92}, no. 2, 026004 (2015)
  doi:10.1103/PhysRevD.92.026004
  [arXiv:1503.04327 [hep-th]].

  
  
\bibitem{Karch:2008fa}
  A.~Karch, D.~T.~Son and A.~O.~Starinets,
  ``Zero Sound from Holography,''
  arXiv:0806.3796 [hep-th].
 
\bibitem{Karch:2009zz}
  A.~Karch, D.~T.~Son and A.~O.~Starinets,
  ``Holographic Quantum Liquid,''
  Phys.\ Rev.\ Lett.\  {\bf 102} (2009) 051602.


\bibitem{Bergman:2011rf}
  O.~Bergman, N.~Jokela, G.~Lifschytz and M.~Lippert,
  ``Striped instability of a holographic Fermi-like liquid,''
  JHEP {\bf 1110} (2011) 034
  [arXiv:1106.3883 [hep-th]].
  
  
\bibitem{Davison:2011ek}
  R.~A.~Davison and A.~O.~Starinets,
  ``Holographic zero sound at finite temperature,''
  Phys.\ Rev.\ D {\bf 85} (2012) 026004
  [arXiv:1109.6343 [hep-th]].

  
  
  
  
\bibitem{Jokela:2012vn}
  N.~Jokela, G.~Lifschytz and M.~Lippert,
  ``Magnetic effects in a holographic Fermi-like liquid,''
  JHEP {\bf 1205} (2012) 105
  [arXiv:1204.3914 [hep-th]].
  

\bibitem{Brattan:2012nb}
  D.~K.~Brattan, R.~A.~Davison, S.~A.~Gentle and A.~O'Bannon,
  ``Collective Excitations of Holographic Quantum Liquids in a Magnetic Field,''
  JHEP {\bf 1211} (2012) 084
  [arXiv:1209.0009 [hep-th]].


  
 

  
   

 
\bibitem{Kulaxizi:2008kv}
  M.~Kulaxizi and A.~Parnachev,
  ``Comments on Fermi Liquid from Holography,''
  Phys.\ Rev.\ D {\bf 78} (2008) 086004
  [arXiv:0808.3953 [hep-th]].




\bibitem{Kulaxizi:2008jx} 
  M.~Kulaxizi and A.~Parnachev,
  ``Holographic Responses of Fermion Matter,''
  Nucl.\ Phys.\ B {\bf 815}, 125 (2009)
  [arXiv:0811.2262 [hep-th]].
  
  
\bibitem{Kim:2008bv}
  K.~Y.~Kim and I.~Zahed,
  ``Baryonic Response of Dense Holographic QCD,''
  JHEP {\bf 0812} (2008) 075
  [arXiv:0811.0184 [hep-th]].

 
\bibitem{Hung:2009qk}
  L.~Y.~Hung and A.~Sinha,
  ``Holographic quantum liquids in 1+1 dimensions,''
  JHEP {\bf 1001} (2010) 114
  [arXiv:0909.3526 [hep-th]].
 



\bibitem{Edalati:2010pn}
  M.~Edalati, J.~I.~Jottar and R.~G.~Leigh,
  ``Holography and the sound of criticality,''
  JHEP {\bf 1010} (2010) 058
  [arXiv:1005.4075 [hep-th]].
 
\bibitem{Lee:2010uy}
  B.~H.~Lee and D.~W.~Pang,
  ``Notes on Properties of Holographic Strange Metals,''
  Phys.\ Rev.\ D {\bf 82} (2010) 104011
  [arXiv:1006.4915 [hep-th]].
    
\bibitem{HoyosBadajoz:2010kd}
  C.~Hoyos-Badajoz, A.~O'Bannon and J.~M.~S.~Wu,
  ``Zero Sound in Strange Metallic Holography,''
  JHEP {\bf 1009} (2010) 086
  [arXiv:1007.0590 [hep-th]].

\bibitem{Lee:2010ez}
  B.~H.~Lee, D.~W.~Pang and C.~Park,
  ``Zero Sound in Effective Holographic Theories,''
  JHEP {\bf 1011} (2010) 120
  [arXiv:1009.3966 [hep-th]].

\bibitem{Ammon:2011hz}
  M.~Ammon, J.~Erdmenger, S.~Lin, S.~M\"uller, A.~O'Bannon and J.~P.~Shock,
  ``On Stability and Transport of Cold Holographic Matter,''
  JHEP {\bf 1109} (2011) 030
  [arXiv:1108.1798 [hep-th]].

\bibitem{Goykhman:2012vy}
  M.~Goykhman, A.~Parnachev and J.~Zaanen,
  ``Fluctuations in finite density holographic quantum liquids,''
  JHEP {\bf 1210} (2012) 045
  [arXiv:1204.6232 [hep-th]].
  
\bibitem{Gorsky:2012gi}
  A.~Gorsky and A.~V.~Zayakin,
  ``Anomalous Zero Sound,''
  JHEP {\bf 1302} (2013) 124
  [arXiv:1206.4725 [hep-th]].



\bibitem{Jokela:2012se}
  N.~Jokela, M.~J\"arvinen and M.~Lippert,
  ``Fluctuations and instabilities of a holographic metal,''
  JHEP {\bf 1302} (2013) 007
  [arXiv:1211.1381 [hep-th]].

\bibitem{Davison:2013bxa}
  R.~A.~Davison and A.~Parnachev,
  ``Hydrodynamics of cold holographic matter,''
  JHEP {\bf 1306} (2013) 100
  [arXiv:1303.6334 [hep-th]].


\bibitem{Dey:2013vja}
  P.~Dey and S.~Roy,
  ``Zero sound in strange metals with hyperscaling violation from holography,''
  Phys.\ Rev.\ D {\bf 88} (2013) 046010
  [arXiv:1307.0195 [hep-th]].
  

\bibitem{Edalati:2013tma} 
  M.~Edalati and J.~F.~Pedraza,
  ``Aspects of Current Correlators in Holographic Theories with Hyperscaling Violation,''
  Phys.\ Rev.\ D {\bf 88}, 086004 (2013)
  [arXiv:1307.0808 [hep-th]].

  
  
\bibitem{Davison:2013uha}
  R.~A.~Davison, M.~Goykhman and A.~Parnachev,
  ``AdS/CFT and Landau Fermi liquids,''
  JHEP {\bf 1407} (2014) 109
  [arXiv:1312.0463 [hep-th]].
  
\bibitem{DiNunno:2014bxa}
  B.~S.~DiNunno, M.~Ihl, N.~Jokela and J.~F.~Pedraza,
  ``Holographic zero sound at finite temperature in the Sakai-Sugimoto model,''
  JHEP {\bf 1404} (2014) 149
  [arXiv:1403.1827 [hep-th]].
   

  
\bibitem{Itsios:2015kja} 
  G.~Itsios, N.~Jokela and A.~V.~Ramallo,
 ``Cold holographic matter in the Higgs branch,''
  Phys.\ Lett.\ B {\bf 747}, 229 (2015)
  doi:10.1016/j.physletb.2015.05.071
  [arXiv:1505.02629 [hep-th]].
 

\bibitem{Karch:2007br}
  A.~Karch and A.~O'Bannon,
  ``Holographic thermodynamics at finite baryon density: Some exact results,''
  JHEP {\bf 0711} (2007) 074
  [arXiv:0709.0570 [hep-th]].


  
\bibitem{Ammon:2012je}
  M.~Ammon, M.~Kaminski and A.~Karch,
  ``Hyperscaling-Violation on Probe D-Branes,''
  JHEP {\bf 1211} (2012) 028
  [arXiv:1207.1726 [hep-th]].
  
  
  
\bibitem{Jokela:2013hta}
  N.~Jokela, G.~Lifschytz and M.~Lippert,
  ``Holographic anyonic superfluidity,''
  JHEP {\bf 1310} (2013) 014
  [arXiv:1307.6336 [hep-th]].

 
 
\bibitem{Jokela:2014wsa}
  N.~Jokela, G.~Lifschytz and M.~Lippert,
  ``Flowing holographic anyonic superfluid,''
  JHEP {\bf 1410} (2014) 21
  [arXiv:1407.3794 [hep-th]].

  
\bibitem{Brattan:2013wya}
  D.~K.~Brattan and G.~Lifschytz,
  ``Holographic plasma and anyonic fluids,''
  JHEP {\bf 1402} (2014) 090
  [arXiv:1310.2610 [hep-th]].


\bibitem{Brattan:2014moa} 
  D.~K.~Brattan,
  ``A strongly coupled anyon material,''
  JHEP {\bf 1511}, 214 (2015)
  doi:10.1007/JHEP11(2015)214
  [arXiv:1412.1489 [hep-th]].
  
  
  
  
  
\bibitem{Karch:2005ms}
  A.~Karch, A.~O'Bannon and K.~Skenderis,
  ``Holographic renormalization of probe D-branes in AdS/CFT,''
  JHEP {\bf 0604} (2006) 015
  doi:10.1088/1126-6708/2006/04/015
  [hep-th/0512125].
  
  
\bibitem{Benincasa:2009ze}
  P.~Benincasa,
  ``A Note on Holographic Renormalization of Probe D-Branes,''
  arXiv:0903.4356 [hep-th].
  
\bibitem{Benincasa:2009be}
  P.~Benincasa,
  ``Universality of Holographic Phase Transitions and Holographic Quantum Liquids,''
  arXiv:0911.0075 [hep-th].
  
  
\bibitem{Karch:2009eb}
  A.~Karch, M.~Kulaxizi and A.~Parnachev,
  ``Notes on Properties of Holographic Matter,''
  JHEP {\bf 0911} (2009) 017
  [arXiv:0908.3493 [hep-th]].
  



  
  
  
 
 
 
\bibitem{Witten:2003ya}
  E.~Witten,
  ``SL(2,Z) action on three-dimensional conformal field theories with Abelian symmetry,''
  In *Shifman, M. (ed.) et al.: From fields to strings, vol. 2* 1173-1200
  [hep-th/0307041].

\bibitem{Yee:2004ju}
  H.~U.~Yee,
  ``A Note on AdS / CFT dual of SL(2,Z) action on 3-D conformal field theories with U(1) symmetry,''
  Phys.\ Lett.\ B {\bf 598} (2004) 139
  [hep-th/0402115].

 
\bibitem{Karch:2007pd}
  A.~Karch and A.~O'Bannon,
  ``Metallic AdS/CFT,''
  JHEP {\bf 0709} (2007) 024
  doi:10.1088/1126-6708/2007/09/024
  [arXiv:0705.3870 [hep-th]].

\bibitem{Dong:2012se}
  X.~Dong, S.~Harrison, S.~Kachru, G.~Torroba and H.~Wang,
  ``Aspects of holography for theories with hyperscaling violation,''
  JHEP {\bf 1206} (2012) 041
  doi:10.1007/JHEP06(2012)041
  [arXiv:1201.1905 [hep-th]].
  


\bibitem{Myers:2008me}
  R.~C.~Myers and M.~C.~Wapler,
  ``Transport Properties of Holographic Defects,''
  JHEP {\bf 0812} (2008) 115
  doi:10.1088/1126-6708/2008/12/115
  [arXiv:0811.0480 [hep-th]].
 
 
 
\bibitem{Bergman:2010gm}
  O.~Bergman, N.~Jokela, G.~Lifschytz and M.~Lippert,
  ``Quantum Hall Effect in a Holographic Model,''
  JHEP {\bf 1010} (2010) 063
  doi:10.1007/JHEP10(2010)063
  [arXiv:1003.4965 [hep-th]].
 
\bibitem{Jokela:2011eb}
  N.~Jokela, M.~J\"arvinen and M.~Lippert,
  ``A holographic quantum Hall model at integer filling,''
  JHEP {\bf 1105} (2011) 101
  doi:10.1007/JHEP05(2011)101
  [arXiv:1101.3329 [hep-th]].
 
 
\bibitem{Kutasov:2011fr}
  D.~Kutasov, J.~Lin and A.~Parnachev,
  ``Conformal Phase Transitions at Weak and Strong Coupling,''
  Nucl.\ Phys.\ B {\bf 858} (2012) 155
  doi:10.1016/j.nuclphysb.2012.01.004
  [arXiv:1107.2324 [hep-th]].

\bibitem{Mezzalira:2015vzn}
  A.~Mezzalira and A.~Parnachev,
  ``A Holographic Model of Quantum Hall Transition,''
  Nucl.\ Phys.\ B {\bf 904} (2016) 448
  doi:10.1016/j.nuclphysb.2016.01.022
  [arXiv:1512.06052 [hep-th]].
 
\bibitem{Klebanov:1998hh}
  I.~R.~Klebanov and E.~Witten,
  ``Superconformal field theory on three-branes at a Calabi-Yau singularity,''
  Nucl.\ Phys.\ B {\bf 536} (1998) 199
  doi:10.1016/S0550-3213(98)00654-3
  [hep-th/9807080].
 
 
\bibitem{Aharony:2008ug}
  O.~Aharony, O.~Bergman, D.~L.~Jafferis and J.~Maldacena,
  ``N=6 superconformal Chern-Simons-matter theories, M2-branes and their gravity duals,''
  JHEP {\bf 0810} (2008) 091
  doi:10.1088/1126-6708/2008/10/091
  [arXiv:0806.1218 [hep-th]].
  
 
 
\bibitem{Conde:2011sw}
  E.~Conde and A.~V.~Ramallo,
  ``On the gravity dual of Chern-Simons-matter theories with unquenched flavor,''
  JHEP {\bf 1107} (2011) 099
  doi:10.1007/JHEP07(2011)099
  [arXiv:1105.6045 [hep-th]].
  
\bibitem{Jokela:2012dw}
  N.~Jokela, J.~Mas, A.~V.~Ramallo and D.~Zoakos,
  ``Thermodynamics of the brane in Chern-Simons matter theories with flavor,''
  JHEP {\bf 1302} (2013) 144
  doi:10.1007/JHEP02(2013)144
  [arXiv:1211.0630 [hep-th]].
  
\bibitem{Bea:2013jxa} 
  Y.~Bea, E.~Conde, N.~Jokela and A.~V.~Ramallo,
  ``Unquenched massive flavors and flows in Chern-Simons matter theories,''
  JHEP {\bf 1312}, 033 (2013)
  doi:10.1007/JHEP12(2013)033
  [arXiv:1309.4453 [hep-th]].

\bibitem{Bea:2014yda}
  Y.~Bea, N.~Jokela, M.~Lippert, A.~V.~Ramallo and D.~Zoakos,
  ``Flux and Hall states in ABJM with dynamical flavors,''
  JHEP {\bf 1503} (2015) 009
  doi:10.1007/JHEP03(2015)009
  [arXiv:1411.3335 [hep-th]].
  
\bibitem{Faedo:2015urf} 
  A.~F.~Faedo, A.~Kundu, D.~Mateos, C.~Pantelidou and J.~Tarrio,
  ``Three-dimensional super Yang-Mills with compressible quark matter,''
  arXiv:1511.05484 [hep-th].


\end{thebibliography}
\end{document}